\documentclass[%
 reprint,
 amsmath,amssymb,
 aps,
]{revtex4-2}

\usepackage{graphicx}
\usepackage{dcolumn}
\usepackage{bm}
\usepackage[utf8]{inputenc}
\usepackage{subcaption}
\usepackage{mhchem}
\usepackage{multirow}
\usepackage{makecell}
\usepackage{hyperref}
\usepackage{color,soul}
\usepackage{chemformula}
\usepackage{comment}
\usepackage{textcmds}
\usepackage{caption}
\usepackage{chemscheme}
\usepackage{amsfonts}
\usepackage{csquotes}
\usepackage{fancyhdr}
\usepackage{float}

\begin{document}

\preprint{APS/123-QED}

\title{Materials-Discovery Workflows Guided by Symbolic Regression: Identifying Acid-Stable Oxides for Electrocatalysis}%

\author{Akhil S. Nair*}
\author{Lucas Foppa*}
\author{Matthias Scheffler}%
 \affiliation{The NOMAD Laboratory at the Fritz Haber Institute 
of the Max Planck Society, Faradayweg 4-6, D-14195 Berlin, Germany}

\date{\today}%

\begin{abstract}
The efficiency of active learning (AL) approaches to identify materials with desired properties relies on the knowledge of a few parameters describing the property. However, these parameters are unknown if the property is governed by a high intricacy of many atomistic processes. Here, we develop an AL workflow based on the sure-independence screening and sparsifying operator (SISSO) symbolic-regression approach. SISSO identifies the few, key parameters correlated with a given materials property via analytical expressions, out of many offered primary features. Crucially, we train ensembles of SISSO models in order to quantify mean predictions and their uncertainty, enabling the use of SISSO in AL. By combining bootstrap sampling to obtain training datasets with Monte-Carlo feature dropout, the high prediction errors observed by a single SISSO model are improved. Besides, the feature dropout procedure alleviates the overconfidence issues observed in the widely used bagging approach. We demonstrate the SISSO-guided AL workflow by identifying acid-stable oxides for water splitting using high-quality DFT-HSE06 calculations. From a pool of 1470 materials, 12 acid-stable materials are identified in only 30 AL iterations. The materials-property maps provided by SISSO along with the uncertainty estimates reduce the risk of missing promising portions of the materials space that were overlooked in the initial, possibly biased dataset.
\end{abstract}

\maketitle
The discovery of improved materials is critical for addressing global challenges such as the transition to renewable energies \cite{tabor2018accelerating,ludwig2019discovery}. However, the space of possible materials is practically infinite and the materials that exhibit desired properties are often very few. Thus, direct, high-throughput screening materials discovery is impractical. Artificial intelligence (AI) can accelerate materials discovery by identifying complex, nonlinear relationships between materials' parameters and certain properties of interest \cite{pyzer2022accelerating,merchant2023scaling}. In particular, AI can guide workflows that perform complex tasks such as the synthesis of materials or the evaluation of their properties \cite{montoya2020autonomous, lookman2019active}. An Active-learning (AL) workflow, for instance, can cast materials discovery as an optimization problem {\cite{boley2023prediction}}. In such AL frameworks, the AI model for the property of interest is retrained (updated) iteratively with more and more data acquired by the workflow. A data-acquisition strategy informed by the AI  model is defined in order to identify the materials that present the desired behavior, e.g., presenting a materials-property value below or above a certain threshold, in an efficient manner. Here, efficiency means that the number of property evaluations is kept at a manageable level, for instance, by intelligently selecting the materials that should be studied. The acquisition strategy might not only target materials with desired behavior (exploitation), but it might also include data associated with high prediction uncertainty (exploration), assuming that regions of the materials space that were overlooked in the training data are associated with uncertain predictions \cite{hwang2024overcoming,qian2023knowledge}. Despite their demonstrated success \cite{ye2022novel,kusne2020fly,montoya2024ai}, the efficiency of AL workflows based on widely used machine-learning approaches often relies on the knowledge of a few input parameters or features, i.e., on a low-dimensional representation. This might be a drawback in materials science because the high intricacy of many atomistic processes governing certain materials properties or functions implies that the key parameters required to describe them are typically unknown. 

\begin{scheme*}[ht]
 \includegraphics[width=\linewidth]{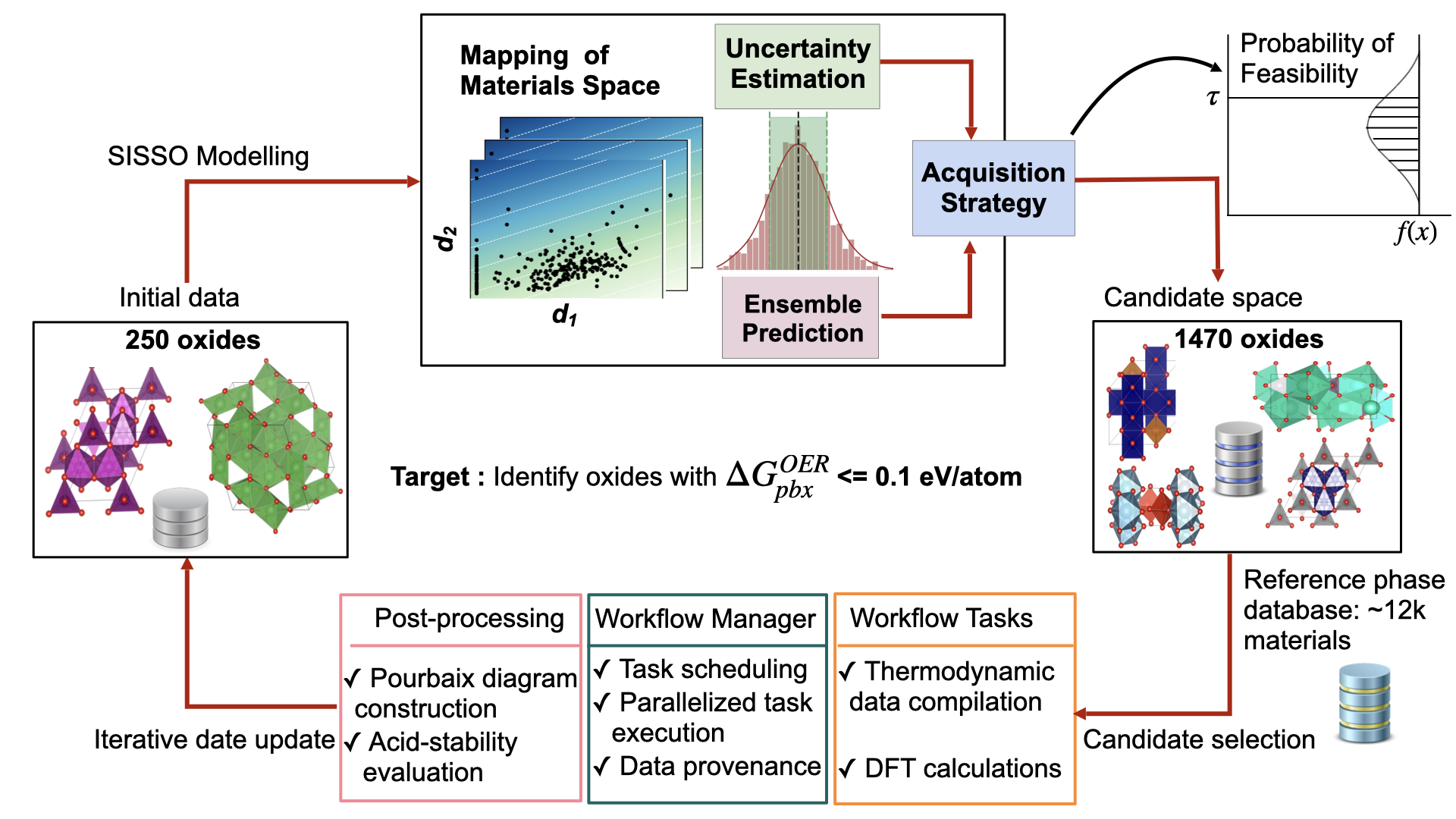}
\captionsetup{justification=raggedright,singlelinecheck=false}
  \caption{Starting from an initial dataset of oxides and many offered primary features, SISSO learns a representation and a model for the acid stability of oxides obtained from Pourbaix analysis. $d_1$ and $d_2$ in the materials-property map are the descriptors obtained from the SISSO model and the black circles represent the initial data in the descriptor space. Ensembles of SISSO models  are used to obtain predictions and their uncertainty estimates. The SISSO-model-informed acquisition strategy selects the region(s) of materials space that should be addressed next. A computational workflow expedites a large number of material calculations and enables the acid stability evaluation of oxides. The selected oxides along with computed acid stability are added to the dataset. The SISSO models are retrained and the process restarts.} 
  \label{Scheme 1}
\end{scheme*}

Here, we address this challenge by developing an AL workflow for materials discovery based on the sure-independence screening and sparsifying operator (SISSO) symbolic-regression approach \cite{ouyang2018sisso,foppa2021materials, purcell2023recent}. SISSO identifies models for a given target property of interest in the form of analytical expressions. These expressions depend only on a few key physical parameters, out of many offered primary features. By constructing models with physical parameters, SISSO might better capture the relationship between materials' parameters and the property of interest across the materials space compared to interpolation schemes \cite{wang2019symbolic, muckley2023interpretable}. Crucially, we construct \textit{ensembles} of SISSO models to obtain their mean predictions and uncertainty estimates, which are used to steer data acquisition from the parts of the materials space that are not covered by the training data.

The SISSO-guided AL workflow is demonstrated for the computational discovery of acid-stable oxides for
electrocatalytic water splitting, a key process for sustainable hydrogen production (Scheme 1). In particular, earth-abundant materials that withstand the harsh conditions of the oxygen evolution reaction (OER) are urgently required. However, accurately evaluating the thermodynamic stability of an oxide under electrochemical conditions (voltage and pH) requires the consideration of many competing phases that could co-exist in the Pourbaix diagram such as hydrides, hydroxides, elemental solids, etc. \cite{pourbaix1966atlas}. This severely limits the number of oxides that can be calculated using reliable methods such as hybrid density functional theory (DFT).   

Here, SISSO is used to intelligently select the materials that should be evaluated by high-quality DFT-HSE06 calculations. As a result, 12 acid-stable oxides are identified out of a space of 1470 oxides within only 30 iterations of the AL workflow. Many of these oxides have not been previously identified by widely used DFT calculations under the generalized gradient approximation (GGA).

\section*{Results}
The SISSO analysis starts with the collection of its input parameters termed primary features. By iteratively applying mathematical operators such as addition, multiplication, etc. to these primary features and also to the previously generated expressions $q$ times, SISSO generates $\sim$$\mathrm{10^8}$ analytical functions. Then, a few, $D$ expressions, typically 2 or 3, components of a descriptor vector $d_1$,$d_2$,..$d_D$, are selected (see Methods for more details). These components, combined with weighting coefficients, are those that best correlate with the target property for the training dataset. $q$ and $D$ are hyperparameters of SISSO and are generally determined by resampling methods such as nested cross-validation (NCV). 

In this work, the target property modeled by SISSO is the thermodynamic stability of oxide under acidic OER conditions: applied voltage of 1.23 V and pH=0, denoted $\Delta G_{pbx}^{\mathrm{OER}}$ hereafter. This quantity corresponds to the Pourbaix decomposition free energy and it is calculated with DFT using the Heyd–Scuseria–Ernzerhof (HSE06) exchange-correlation functional \cite{heyd2003hybrid} as efficiently implemented in the FHI-aims code \cite{kokott2024efficient}. A low $\Delta G_{pbx}^{\mathrm{OER}}$ indicates high stability of the material.  We stress that the stability under operation is an overlooked yet crucial criterion for the design of OER catalysts \cite{li2024stability,ju2024two} in addition to surface reactivity \cite{man2011universality,xu2021rational}. The significant improvements of DFT-HSE06 with respect to DFT-GGA for the description of oxide formation energy and $\Delta G_{pbx}^{\mathrm{OER}}$ are discussed in detail in Supplementary sections I-III. 
In order to model $\Delta G_{pbx}^{\mathrm{OER}}$ with SISSO, 14 primary features are offered (Table S2). These are elemental and compositional properties of free-atoms and the oxide, respectively such as orbital radii and oxidation state distribution. A  training dataset containing 250 oxides was created by computing $\Delta G_{pbx}^{\mathrm{OER}}$ for these materials. Among these, the well-known water splitting catalysts are found, such as \ce{IrO2} ($\Delta G_{pbx}^{\mathrm{OER}}$= -0.62 eV/atom) and \ce{RuO2} ($\Delta G_{pbx}^{\mathrm{OER}}$= -0.26 eV/atom). The $\Delta G_{pbx}^{\mathrm{OER}}$ distribution over these 250 materials has a mean value of 0.61 eV/atom (Figure S8).

\begin{figure}[h]
\centering
\captionsetup{justification=justified, singlelinecheck=false}
  \begin{subfigure}{\textwidth}
    \centering
    \includegraphics[width=1.0\linewidth]{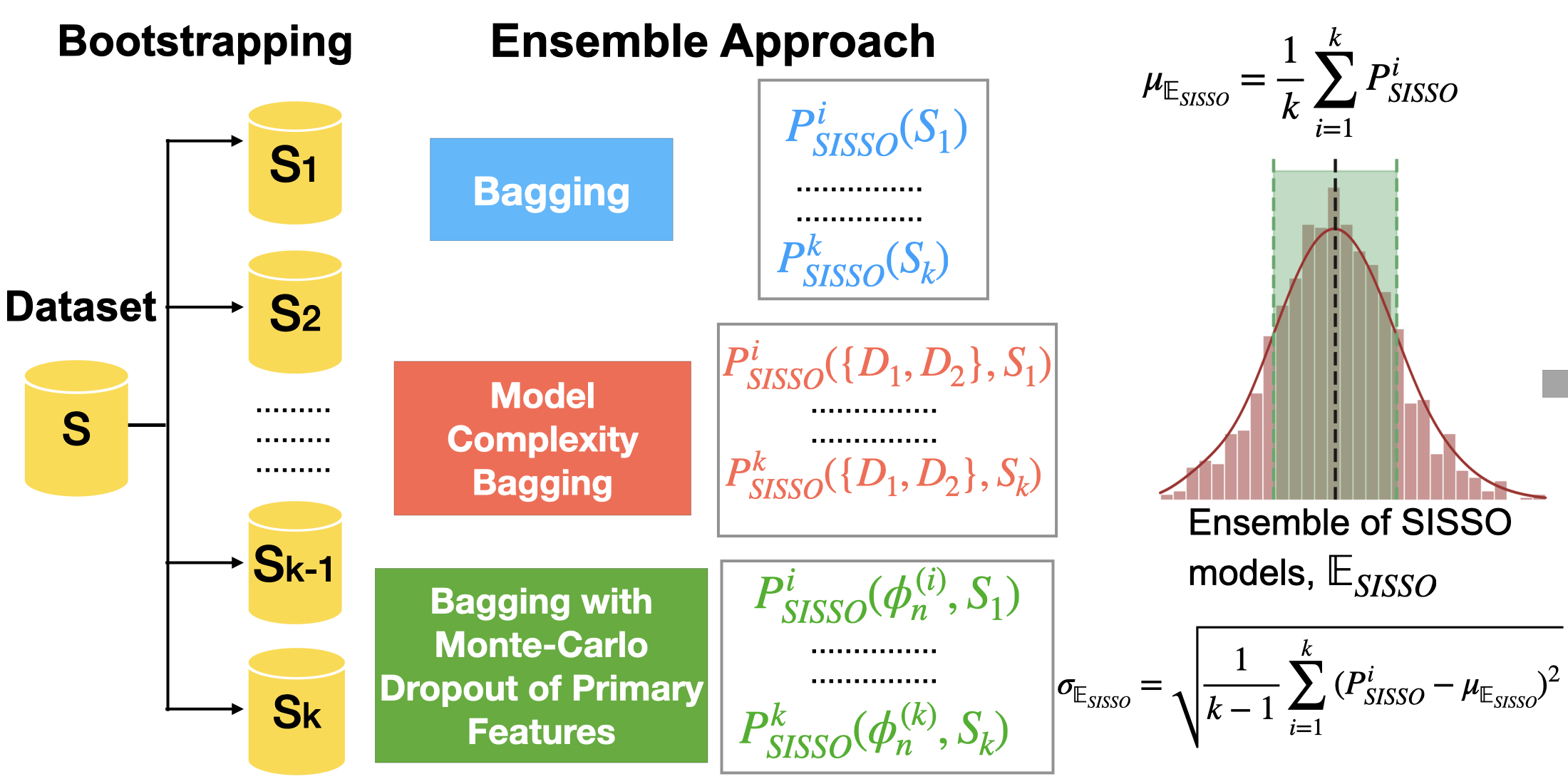}
    \caption*{(a)}
  \end{subfigure}

  \vspace{1pt} 

  \begin{subfigure}{\textwidth} 
    \centering
    \includegraphics[width=\linewidth]{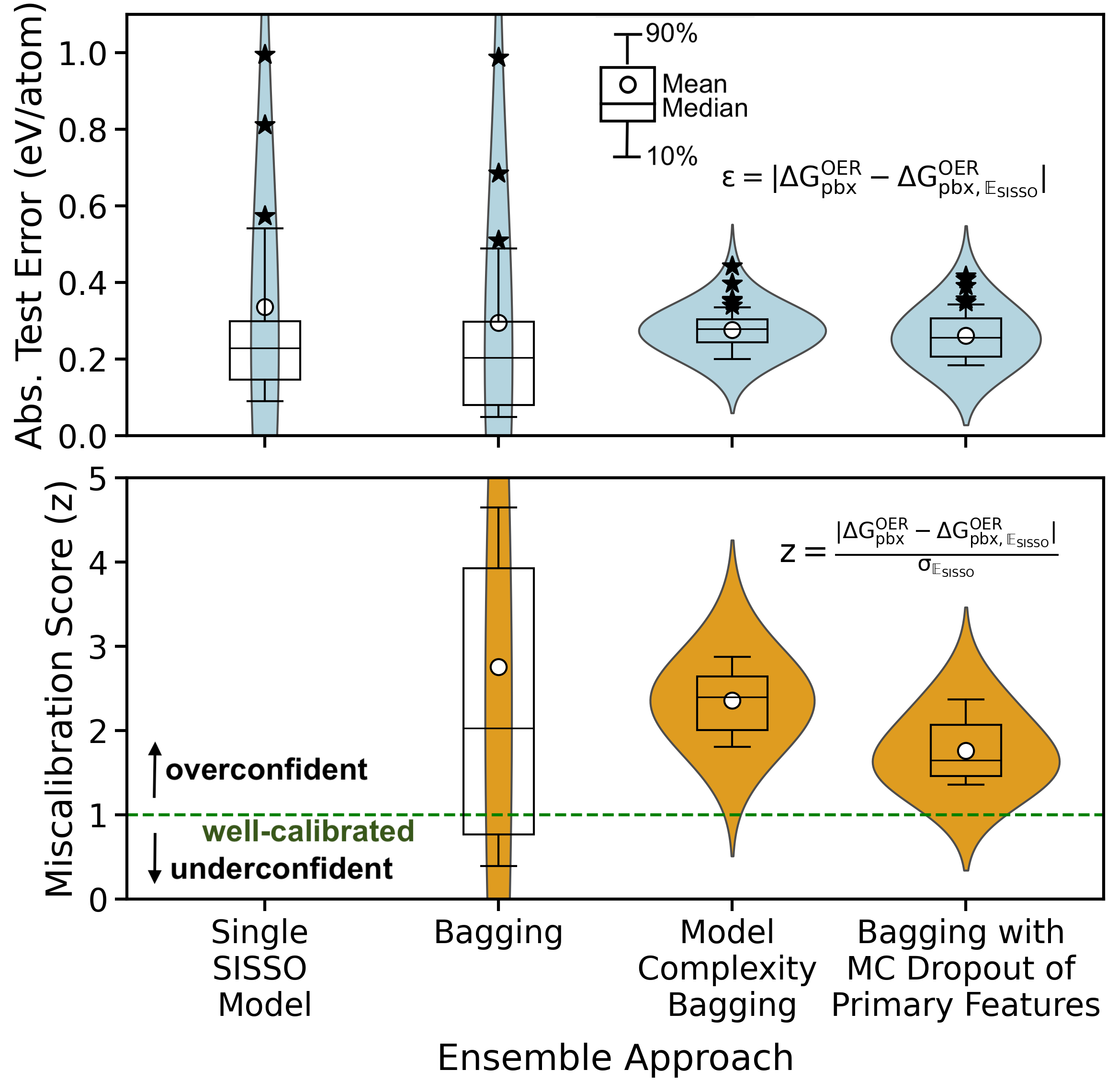}
    \caption*{(b)}
  \end{subfigure}
  \caption{(a) Schematic representation of different ensemble methods. S$_1$, ..., S$_k$ represent the subsets of the original dataset S obtained through bootstrapping. $P^{i}_{SISSO}$ is the SISSO model trained on the $i$\textsuperscript{th} bootstrap sample, and $\phi^{i}_{_n}$ represents the set of primary features retained for that sample (b) comparison of absolute prediction errors (top panel) and miscalibration scores (bottom panel) across different ensemble methods. The violin plots are constructed with errors and miscalibration scores obtained by 30 independent trials with $k=10$. Some of the high error predictions are indicated with star markers.}
  \label{Fig:1}
\end{figure}

Three strategies are investigated for constructing ensembles of SISSO models: bagging, model complexity bagging, and bagging with Monte-Carlo dropout of primary features (Figure \ref{Fig:1}a). All three begin by generating $k$ distinct training sets from the original dataset through randomly sampling data subsets with replacement (bootstrapping). This is followed by aggregating the predictions from SISSO models trained on these subsets. In the bagging approach, a SISSO model is trained on each of these $k$ bootstrapped datasets. Model complexity bagging is a modification of the bagging approach involving training two models for each training set: one with $D$ = 1 and another with $D$=2. This results in 2$k$ models in the ensemble. For bagging with Monte-Carlo dropout of primary features, each model is trained using a subset (here 20\%) of primary features, created by Monte Carlo dropout from the bootstrapped subsets. Different $k$ values are analyzed and $k=10$ is selected based on the convergence of prediction errors and uncertainty estimates (see Figure S11). For all the ensemble methods, an optimal model complexity ($q$=2, $D$=2) identified by NCV (see Supplementary Section IV) is used unless otherwise mentioned (as in model complexity bagging). We define $\Delta G_{pbx,\mathbb{E}_{SISSO}}^{\mathrm{OER}}$ as the mean prediction of a given ensemble. The standard deviation of the predictions of the different SISSO models ($\sigma_{\mathbb{E}_{SISSO}}$) serves as an estimate of the prediction uncertainty. Note that the uncertainty quantification by the ensemble does not necessarily reflect the uncertainty related to lack of data \cite{lu2023uncertainty,bauer2024roadmap}. 

We apply these three ensemble approaches to a training set of 200 materials, randomly drawn from the dataset of 250 materials. The performance of different ensemble approaches is compared using two metrics which are the absolute prediction errors ($\epsilon$) and the miscalibration scores ($z$)\cite{pernot2023calibration}. $\epsilon$ is evaluated as the difference in predicted and DFT calculated $\Delta G_{pbx}^{\mathrm{OER}}$ values. $z$ measures how calibrated the uncertainty estimates are \cite{palmer2022calibration}. $z=1$ indicates well calibrated uncertainty estimates, while $z > 1$ suggests overconfidence, and $z < 1$ signifies underconfident predictions. The $\epsilon$ and $z$ values are evaluated for the 50 materials of the dataset that were not used for training. To evaluate the performance of the ensemble approaches, we also consider a single SISSO model obtained by training on the same 200 data points used to train the ensembles but without bootstrapping.

The distributions of $\epsilon$ and $z$ are shown in Figure \ref{Fig:1}b for the three different ensemble approaches. The single SISSO model achieves a mean absolute error (MAE) of 0.34 eV/atom. In comparison, the bagging approach shows a reduced MAE of 0.29 eV/atom. Both model complexity bagging and bagging with Monte Carlo dropout of primary features further reduce the MAE to 0.29 and 0.26 eV/atom, respectively. The 90\%-iles of the  $\epsilon$ distributions for the three ensemble approaches (0.85, 0.37, and 0.35 eV/atom) also demonstrate improvement compared to a single SISSO model.

The analysis of $z$ shows a relatively more distinct performance across the three approaches. The bagging method exhibits significantly high 
$z$-values, with a mean of 2.76, indicating overconfident predictions. This issue is moderately addressed by the model complexity bagging approach, which lowers the mean $z$-value to 2.36. Bagging with Monte Carlo dropout of primary features achieves the most balanced uncertainty estimates, with a mean $z$-value of 1.76, marking a clear improvement over the other methods. A similar trend is observed in the  90\%-iles of the $z$-value distributions (5.62, 3.23, 2.44), further highlighting the superior performance of the Monte Carlo dropout method. Based on these findings, bagging with Monte Carlo dropout of primary features is identified as the most effective ensemble approach among the considered ones for uncertainty estimation in SISSO models. Consequently, this approach is adopted in the AL workflow discussed in the next section.

The dataset containing 250 oxides, is employed as the initial dataset for AL. A SISSO model trained on this dataset has an RMSE of 0.38 eV/atom and R$^2$=0.87. The candidate space, i.e., the set of new materials that could be queried during AL, consists of 1470 oxides (see Supplementary section V). The majority of materials in the candidate space are ternary oxides. We note that ternary oxides are less considered by DFT-HSE06 studies compared to binary oxides due to their typically larger unit-cell sizes, which makes the calculations expensive \cite{liu2024high,huang2015electrochemical}. These 1470 materials together with the competing phases needed to evaluate the $\Delta G_{pbx,SISSO}^{\mathrm{OER}}$ correspond to a total number of $\sim$ 12 thousand materials. 

Our goal is to identify acid-stable oxides, i.e., materials with very small $\Delta G_{pbx}^{\mathrm{OER}}$. To achieve this goal, we consider the probability of feasibility (POF ) as the data-acquisition strategy. POF quantifies the probability that an oxide is acid-stable. It is defined as;

\begin{equation}
\text{POF} = \mathbb{P}(\Delta G_{pbx}^{\mathrm{OER}} \leq \tau) = F\left( \frac{\tau - \Delta G_{pbx,\mathbb{E}_{SISSO}}^{\mathrm{OER}}}{\sigma_{\mathbb{E}_{SISSO}}} \right)
\end{equation}

where $\mathbb{P}$ is the probability that $\Delta G_{pbx}^{\mathrm{OER}}$ is less than $\tau$=0.00 and $F$ is the cumulative distribution function  of the standard normal distribution. POF is informed by both mean predictions ($\Delta G_{pbx,\mathbb{E}_{SISSO}}^{\mathrm{OER}}$) and the standard deviations of the predictions ($\sigma_{\mathbb{E}_{SISSO}}$), both obtained by the bagging with Monte-Carlo dropout of primary features ensemble approach.  In each AL iteration, we train ensembles of SISSO models (vide supra). The material associated with the highest POF value is selected and the corresponding $\Delta G_{pbx}^{\mathrm{OER}}$ is calculated. This is followed by ranking the candidate materials based on the values of the POF. We also consider \textit{random selection} (RS) as a baseline data-acquisition strategy that selects material from the candidate space randomly. The selected material is moved from the candidate space to the training dataset, and the process is repeated. 

\begin{figure}[ht]
 \includegraphics[width=\linewidth]{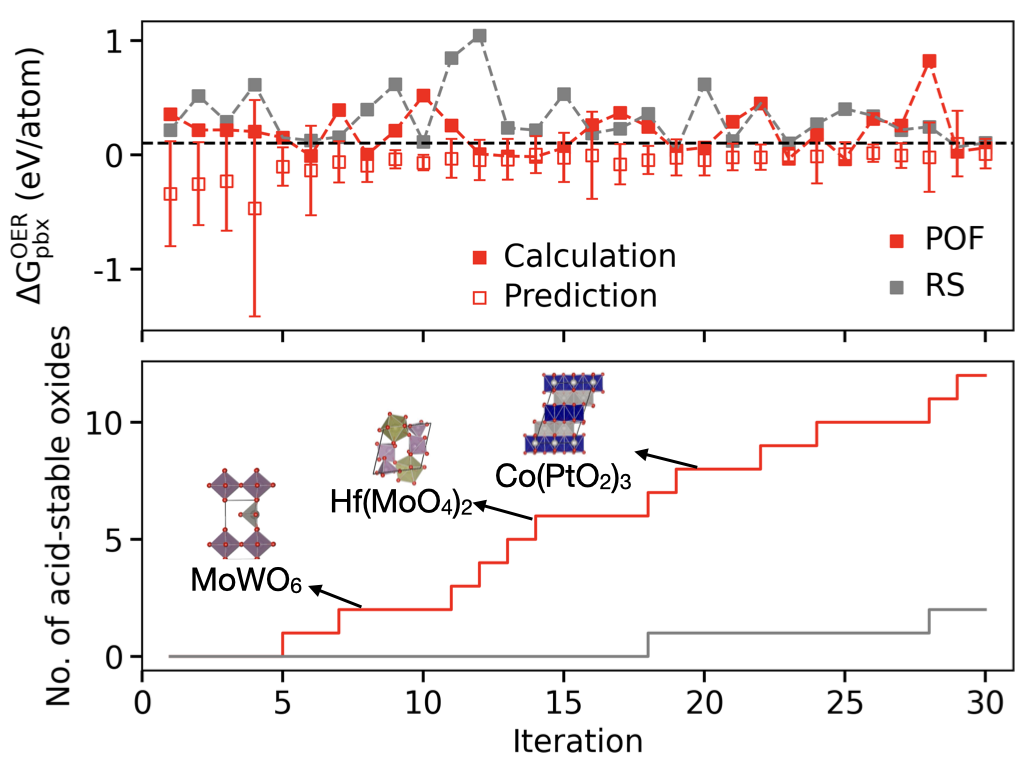}
\captionsetup{justification=justified,singlelinecheck=false}
  \caption{$\Delta G_{pbx}^{\mathrm{OER}}$ change (top panel) and number of acid-stable oxides identified (bottom panel) across 30 AL iterations with the probability of feasibility (POF) and random selection (RS) acquisition strategies The filled and open square marks indicate the DFT-HSE06 calculated $\Delta G_{pbx}^{\mathrm{OER}}$ and mean prediction of ensembles of SISSO models for the oxide selected at each iteration ($\Delta G_{pbx,\mathbb{E}_{SISSO}}^{OER}$), respectively. The error bars represent the corresponding uncertainty estimates ($\sigma_{\mathbb{E}_{SISSO}}$). The stability threshold of $\Delta G_{pbx}^{OER}=$ 0.1 eV/atom is indicated with the black dashed line. The formulae and structures of acid-stable oxides identified in iterations 8,14,20 are shown in the inset of the figure.}
  \label{Fig:2}
\end{figure}

The results of the SISSO-based AL campaigns are shown in Figure \ref{Fig:2} for 30 iterations. Materials selected and evaluated based on RS and POF are displayed in grey and red colors, respectively. We use a stability threshold of $\Delta G_{pbx}^{\mathrm{OER}} \leq$0.1 eV/atom to label an oxide as acid-stable as materials with small positive $\Delta G_{pbx}^{\mathrm{OER}}$ can retain their stability under acidic conditions due to self-passivation and formation of more stable phases \cite{singh2017electrochemical,park2024data,wang2020acid}. Previous high-throughput studies have used higher stability thresholds (e.g., 0.2 \cite{back2020discovery} or 0.5 \cite{gunasooriya2020analysis} eV/atom), but we adopt a more conservative criterion as we use higher-accuracy HSE06 functional. In Figure \ref{Fig:2}a, the filled square markers correspond to $\Delta G_{pbx}^{\mathrm{OER}}$ values obtained by DFT-HSE06 calculations. The open square markers show $\Delta G_{pbx,\mathbb{E}_{SISSO}}^{\mathrm{OER}}$, and the error bars reflect $\sigma_{\mathbb{E}_{SISSO}}$. Oxides selected by RS have $\Delta G_{pbx}^{\mathrm{OER}}$ in the range [0.04, 1.04] eV/atom and only two of them are acid-stable. This indicates that $\Delta G_{pbx}^{\mathrm{OER}}$ for the materials in the candidate space are concentrated in a rather narrow range with $\mathrm{\Delta G_{pbx}^{OER} \ge 0.1}$ eV/atom, corresponding to unstable materials. The situation is similar to the training dataset (Figure S8). However, in contrast with a random search,  materials suggested by POF have $\Delta G_{pbx}^{\mathrm{OER}}$ in the range [-0.04, 0.82] eV/atom.  This shows that by using the mean predictions and uncertainty estimates from SISSO ensembles, we are able to identify the stable materials in the candidate space much more efficiently than a random selection.  An overall reduction in uncertainty estimates is visible with the error bars becoming narrower as the iterations progress. While overconfidence is occasionally observed (e.g., iterations 10,22,26), the error bars generally capture the true values, highlighting the reliability of the uncertainty estimates. Within 30 iterations,  12 acid-stable materials are successfully identified (Figure \ref{Fig:2}b). The number of iterations can be extended to identify more  acid-stable materials.

\begin{figure}[ht]
    \includegraphics[width=\linewidth]{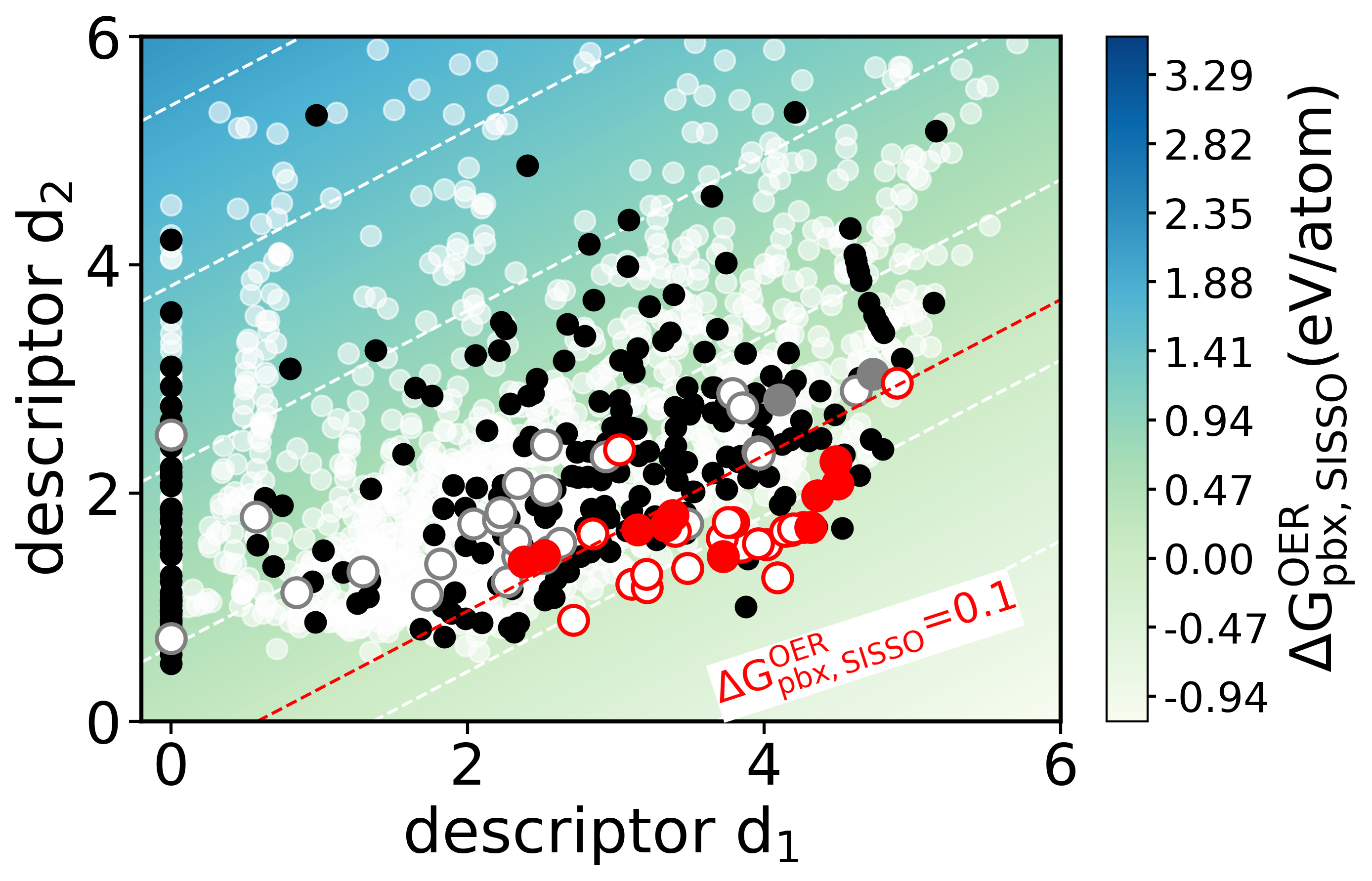}
  \caption{SISSO-descriptor-based material maps of oxide stability with materials in the initial training data (black), the candidate space of 1470 oxides  (white) and materials selected during AL campaigns with probability of feasibility (red) and random selection (grey). The $x$ and $y$ axes represent the descriptors obtained from the SISSO model. The filled (hollow) circles indicate materials suggested by AL which are acid-stable (unstable) from DFT-HSE06 calculations. The  $y$-axis range is limited for an enlarged view of the region of interest.} 
  \label{Fig:3}
\end{figure}

The descriptors identified by SISSO can be used to construct materials-property maps. In Figure \ref{Fig:3}, we show such a map for the property $\Delta G_{pbx,\mathbb{E}_{SISSO}}^{OER}$, constructed based on the initial training set of 250 materials. In this map, we plot the materials in this training set and in the candidate space as black and white circles, respectively. We also show the oxides selected via AL using POF and RS as red and grey circles, respectively. The filled and hollow red and grey circles indicate materials that turned out to be stable and unstable according to the DFT-HSE06 calculations, respectively. Only a small fraction of the 250 oxides in the initial training data are acid-stable, which is well captured by the SISSO model. The oxides chosen by RS predominantly fall within high-$\Delta G_{pbx,\mathbb{E}_{SISSO}}^{OER}$ regions of the descriptor space. In contrast, many oxides selected by POF are concentrated in the low-$\Delta G_{pbx,\mathbb{E}_{SISSO}}^{OER}$ region predicted by the SISSO model. This indicates that this initial SISSO model trained on 250 oxides is valid for most of the discovered materials. However, the oxides \ce{Ni(PtO2)3}, \ce{Co(PtO2)3},  identified to be acid-stable by POF, were incorrectly predicted to be unstable by the initial SISSO model. This reflects that the performance of the SISSO model for the stable materials of interest is improved by using the ensemble approach and also by the inclusion of more training data suggested by the AL workflow. The Pourbaix diagrams of all the oxides identified by AL as acid-stable  are given in Figure S12.

\begin{table}{Table I: $\Delta G_{pbx}^{\mathrm{OER}}$ values (in eV/atom) calculated using PBE and HSE06 for 12 acid-stable oxides identified by the workflow. Previous reports which identified some of the oxides are given in parentheses.}
\centering
\captionsetup{justification=raggedright, singlelinecheck=false}
\renewcommand{\arraystretch}{1.5} 
\label{Table_1}
\begin{tabular}{lcc}
\hline
Oxide & $\Delta G_{pbx,\mathrm{PBE}}^{\mathrm{OER}}$ & $\Delta G_{pbx,\mathrm{HSE06}}^{\mathrm{OER}}$  \\
\hline
\ce{MoW11O36}   &  0.004  &  0.007  \\
\ce{MoWO6} (\cite{gunasooriya2020analysis,wang2020acid})      &  0.001  &  0.002  \\
\ce{Ta2Mo2O11}  & -0.023  & -0.012  \\
\ce{Te(WO4)3}   &  0.062  &  0.061  \\
\ce{Ta12MoO33}  & -0.018  & -0.010  \\
\ce{Hf(MoO4)2}  &  0.052  & -0.017  \\
\ce{Fe2(MoO4)3} (\cite{gunasooriya2020analysis,wang2020acid}) &  0.130  & -0.034  \\
\ce{Ni(PtO2)3}  &  0.230  &  0.036  \\ 
\ce{Co(PtO2)3}  &  0.212  &  0.063  \\
\ce{NiMoO4}     &  0.273  & -0.037  \\
\ce{Ti(GeO3)2} (\cite{gunasooriya2020analysis,wang2020acid})  &  0.015  &  0.027  \\
\ce{CoWO4}      &  0.290  &  0.061  \\
\hline
\end{tabular}
\end{table}
\vspace{-5pt}
Some of the materials identified from the AL approach were also observed in previous high-throughput screening studies based on DFT-PBE(+U) \cite{wang2020acid, gunasooriya2020analysis} such as \ce{MoWO6} and \ce{Fe2(MoO4)3}. Overall, the presence of elements such as Mo, Ta, W are observed to favour acid-stability. We compare the stability of the oxides suggested by our AL framework as described by HSE06 and PBE in Table I. Five out of these twelve oxides are predicted to be acid-unstable based on PBE. For instance, \ce{NiMoO4} has $\Delta G_{pbx,\mathrm{PBE}}^{\mathrm{OER}}$ =0.27 eV/atom whereas $\Delta G_{pbx,\mathrm{HSE06}}^{\mathrm{OER}}$ =-0.04 eV/atom. This underscores the importance of combining an AI-driven approach with high-quality data for an accurate yet efficient screening of the vast materials space.

High-quality materials data is often scarce and expensive to obtain, posing significant challenges to the application of AI in materials discovery. AL  frameworks mitigate this limitation by guiding models toward unexplored, high-potential regions of the materials space. The approach presented here addresses key limitations of interpolation-based AI methods commonly used in AL workflows by eliminating reliance on pre-assumed representations, incorporating explainability, and enabling effective mapping of the target materials space. This advancement holds promise for expanding the application of symbolic regression methods in closed-loop materials discovery workflows. While challenges remain, such as the computational expense of generating large candidate feature spaces, recent developments, including GPU-accelerated SISSO++ implementations \cite{purcell2023recent}, can partially address these issues. Furthermore, exploring additional uncertainty quantification techniques, such as mathematical operator-based or distance-based approaches, could further alleviate overconfidence concerns. Overall, this work constitutes a step forward in the design of  efficient AL-based workflows for AI-driven materials discovery.


\section*{Methods}

\subsection*{FHI-aims Calculations}
All-electron DFT calculations were performed with the FHI-aims (version 230629) electronic structure package with numeric atom-centered-orbital basis sets \cite{blum2009ab}. The standard “light” basis sets are used with zero-order regular approximation to account for the relativistic effects. A k-grid density of 5 $\mathrm{\AA}^{-1}$ is applied to divide an evenly split k-points grid along the reciprocal lattice vectors. A tight self-consistency cycle convergence criteria of $\mathrm{10^{-5}}$ $\mathrm{e{\AA}^{3}}$ for charge density, $\mathrm{10^{-6}}$ eV for total energy and  $\mathrm{10^{-3}}$ eV for sum of eigenvalues is used. For geometry relaxations, forces are converged within $\mathrm{10^{-2}}$ $\mathrm{eV/\AA}$. 

\subsection*{Computational Workflow}
A computational workflow is devised to execute various tasks within the AL framework. The Atomic Simulation Recipes (ASR) \cite{gjerding2021atomic} framework is used with an interface to FHI-aims. In this framework,  DFT calculations are categorized as dependent or independent tasks. Then, these tasks are distributed across multiple batches. The submission of a large number of tasks is handled by workers from a MyQueue front-end deployed on a SLURM-based high-performance computing  system \cite{mortensen2020myqueue}. The workflow provenance is stored in a registry and monitored for error handling. At each iteration of AL, one oxide from the candidate space is selected by the (AI-model-informed) acquisition strategy. The workflow initiates a sequence of actions, including the selection of competing phases, the preparation of input files (geometry optimization at PBE level and self-consistent-field energy evaluation at the HSE06 level), the collection of thermodynamic data, and the construction of Pourbaix diagrams. The input and output files of the calculations as well as the data associated with each iteration of AL (SISSO models, acquisition scores, and selected material) are recorded. Finally, the workflow closes the loop by updating the initial database with all newly acquired data.

\subsection*{SISSO}
Sure-Independence-Screening and Sparsifying Operator (SISSO) is a symbolic regression method to identify descriptors governing material properties or functions \cite{ouyang2018sisso}. Given a set of primary features (\(\phi_{1}, \phi_{2}, \ldots, \phi_{n} \in \Phi\)) and a set of unary and binary mathematical operators (\(\omega_{1}, \omega_{2}, \ldots, \omega_{m} \in \Omega\)), SISSO generates descriptors by iteratively applying the operators over the primary features up to the assigned rung value (\(q\)). A subspace of the constructed feature space (\(\Phi_N\)) is selected by sure-independence screening (SIS), where descriptors are ranked by projection score on the target property. The \(N_{SIS}\) descriptors with the highest projection scores to the target property are used to create the best one-dimensional ($1D$) model. A second round of SIS is applied where the target property is now replaced by \(N_{res}\) residuals of the 1D model, followed by an \(\ell_0\)-regularized optimization to generate the best two-dimensional (2D) model. This process is repeated \(D\) times to generate the best $D$-dimensional SISSO model.
 In this work, the following set of mathematical operations are considered;
\begin{equation}
\begin{split}
\Omega = & \big\{ \phi_1 + \phi_2, \phi_1 - \phi_2, |\phi_1 - \phi_2|, \exp(\phi_1), \exp(-\phi_2), \\
& |\phi_1|, \phi_1 \cdot \phi_2, \frac{\phi_1}{\phi_2}, \ln(\phi_1), \phi_1^{-1},\phi_1^{2}, \phi_1^{3},\sqrt{\phi_1}, \sqrt[3]{\phi_1} \big\}
\label{eqn:features}
\end{split}
\end{equation}



\vspace{-20pt}
\subsection*{Data Availability}
The electronic structure data can be accessed at the NOMAD archive at \url{https://nomad-lab.eu/prod/v1/gui/user/datasets/dataset/id/ON0mTt4eRr67PE3h9K6l_Q}

\subsection*{Code Availability}
The electronic structure package FHI-aims is freely available for academic use with a voluntary donation. SISSO++ \cite{scheffler2022sisso++} is used for all the SISSO analysis. The source code of the active learning workflow along with all the processed data and supporting scripts are available at Gitlab repository \url{https://gitlab.com/akhilsnair/workflow_sisso}. A tutorial demonstrating the approach is provided at \url{https://gitlab.com/FHI-aims-club/tutorials/tutorial-sl-sisso}.
\vspace{-5pt}
\subsection*{Acknowledgement}
A.S.N thanks Stefano Americo, Kristian Sommer Thygssen, and Ask Hjorth Larsen for valuable discussions. A.S.N acknowledges Wahib Aggoune for providing access to computing resources for a brief duration. This work was funded by the NOMAD Center of Excellence (European Union’s Horizon 2020 research and innovation program, grant agreement N 951786), the ERC Advanced Grant TEC1p (European Research Council, grant agreement N 740233). This work used computing resources from the Max Planck Computing and Data Facility and North German Supercomputing Alliance (project ID 20814).

\bibliography{apssamp}

\end{document}


\preprint{AIP/123-QED}

\title{Materials-Discovery Workflows Guided by Symbolic Regression: Identifying Acid-Stable Oxides for Electrocatalysis}

\author{Akhil S. Nair*} 
\author{Lucas Foppa*}
\author{Matthias Scheffler}%
\affiliation{ The NOMAD Laboratory at the Fritz Haber Institute of the Max Planck Society, Faradayweg 4-6, D-14195 Berlin, Germany}
%
\maketitle

\begin{figure*}[t]
\captionsetup{justification=raggedright, singlelinecheck=false}
    \includegraphics[width=\linewidth]{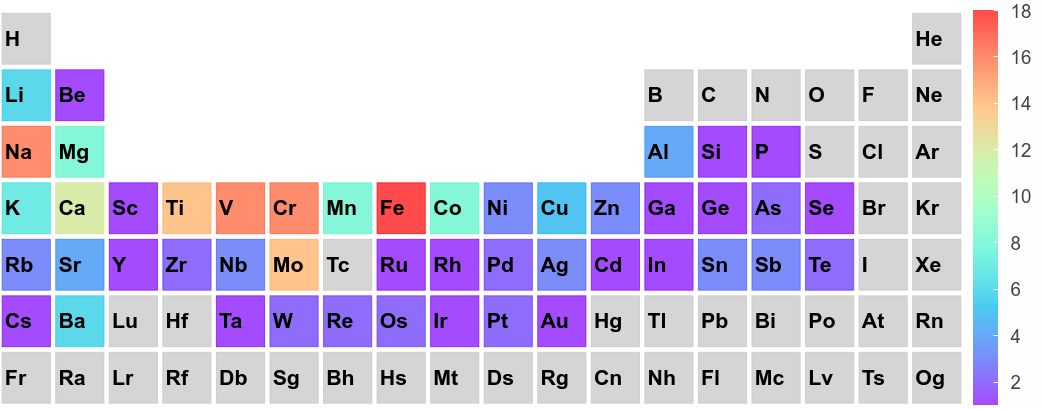}
    \caption*{\textbf{Figure S1.} Heat map illustrating the number of oxides containing each element (excluding oxygen) across a dataset of 153 oxides, used for benchmarking formation energies.}
    \label{Fig:S1}
\end{figure*}


\newpage
\pagestyle{plain}
\section{Benchmarking formation energies of oxides}

In order to quantify the improvement of the oxide description with DFT-HSE06 compared to DFT-PBE, we first calculated formation energies ($\Delta H_f$) of 153 oxides (with binary and ternary composition) with PBE and HSE06 and compared them with experimental data \cite{wang2021framework}.  The initial structures were obtained from the Materials Project (MP) database as of 01/08/2023 and relaxed with FHI-aims code using PBE. HSE06 is selected for the choice for hybrid functional as it is well-known to provide accurate bulk properties of solids \cite{ zhang2018performance}.  These oxides contain diverse elements such as alkali metals, alkaline-earth metals, transition metals, as well as main-group elements (Figure S1). 

The formation energy of an oxide $\mathrm{A_mO_n}$ was calculated as;
\begin{equation}
    \Delta H_{f}(A_mO_n)=E(A_mO_n)-m\mu A-n \mu O
\end{equation}
where $E(A_mO_n)$ is the energy of oxide and $\mu A$ and $\mu O$ are the chemical potentials of element $A$ and O, respectively. The most stable elemental solid (according to MP) and molecular oxygen (\ce{O2}) are considered as the reference for chemical potentials of A and O, respectively. The effect of zero-point energy and entropy contributions has been reported to be two orders of magnitude lower than the $\Delta H_{f}$ values \cite{bartel2019role}. Thus, we neglect these quantities for the calculation of formation energies.

In Figure S2 (left panel), the distribution of errors in the $\Delta H_f$ calculated with PBE (blue color) and HSE06 (red color) are shown. PBE underestimates $\Delta H_f$ and presents a mean absolute error (MAE) of 0.25 eV/atom. HSE06 provides more accurate $\Delta H_f$, with a MAE of 0.17 eV/atom. The underestimation of $\Delta H_f$ with PBE compared to the experiment can be explained by the overestimation of the $\mathrm{O_2}$ molecule binding energy by PBE (-6.07 vs. experimental value of -5.12 eV \cite{pople1989gaussian}) and also by the inadequate treatment of self-interaction error (SIE) in the oxides. HSE06 notably corrects these by providing accurate $\mathrm{O_2}$ binding energy (-5.16 eV) and reduced SIE, resulting in improved $\Delta H_f$ results of oxides. 


\begin{figure}[ht]
\captionsetup{justification=justified,singlelinecheck=false}
\includegraphics[width=0.8\linewidth]{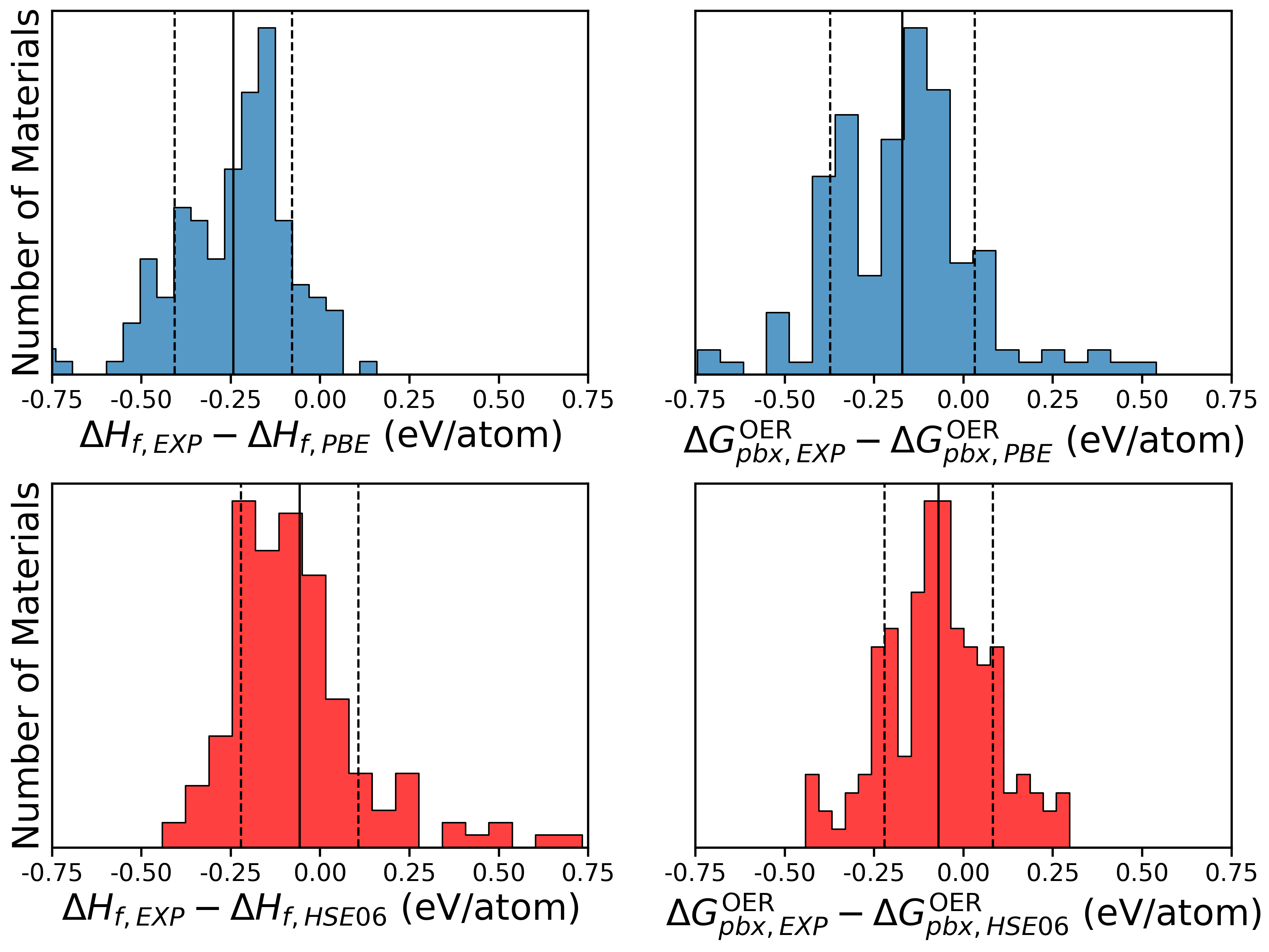}
\caption*{\textbf{Figure S2}: Distribution of errors in $\Delta H_f$ (left panel) and $\Delta G_{pbx}^{\mathrm{OER}}$ (right panel) calculated with PBE (top, blue color) and HSE06 (bottom, red color) with respect to experimental values for 153 oxides \cite{wang2021framework}. The solid lines in the histogram indicate the mean values and the dashed lines indicate the values of the mean +/- one standard deviation.}
\label{Fig:S2}
\end{figure}

\begin{figure*}[!ht]
\centering
\captionsetup{justification=raggedright, singlelinecheck=false}
  \begin{subfigure}{0.36\textwidth}
    \includegraphics[width=\linewidth]{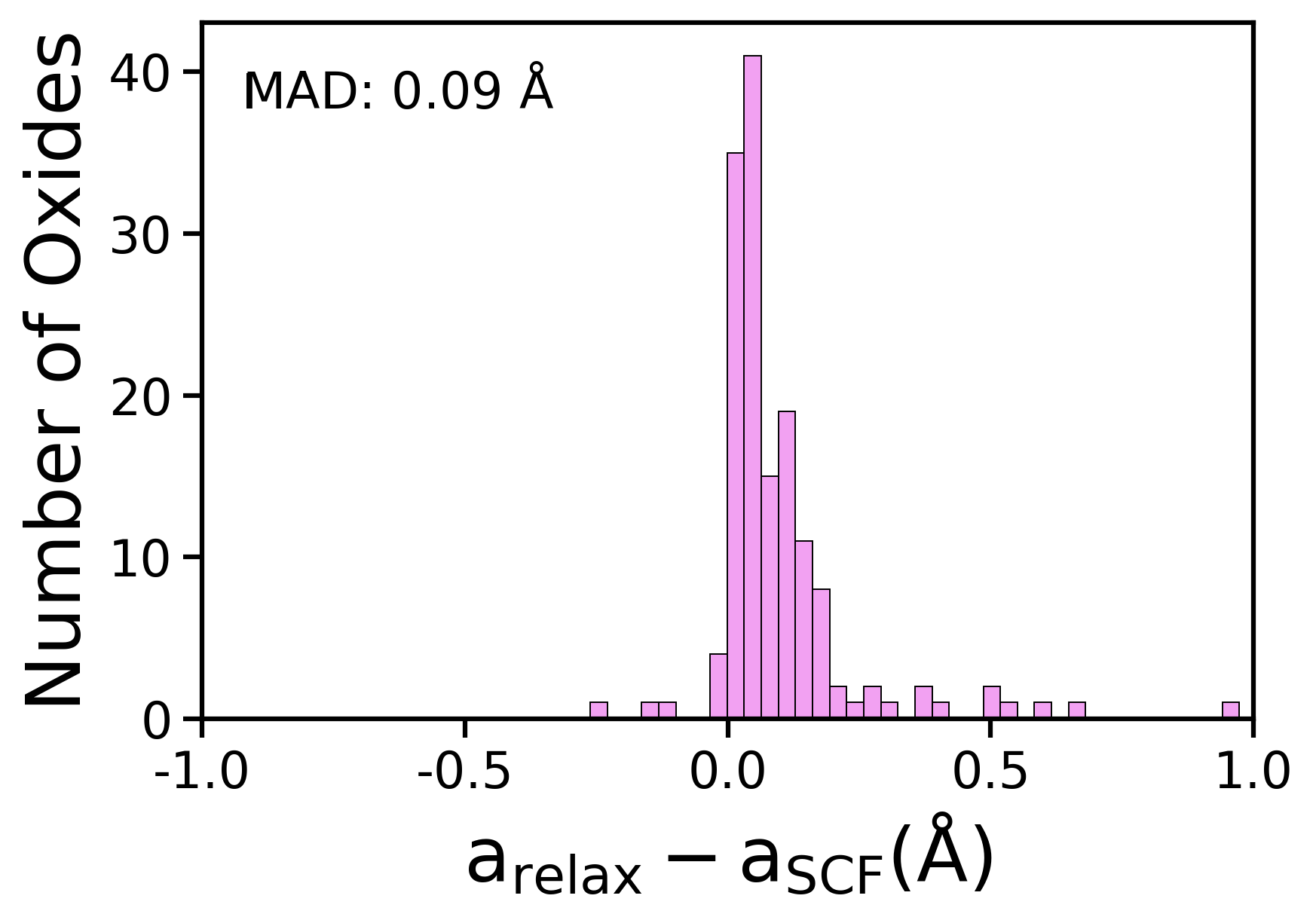}
    \caption*{}
  \end{subfigure}%
  \hspace*{\fill}   
  \begin{subfigure}{0.33\textwidth}
    \includegraphics[width=\linewidth]{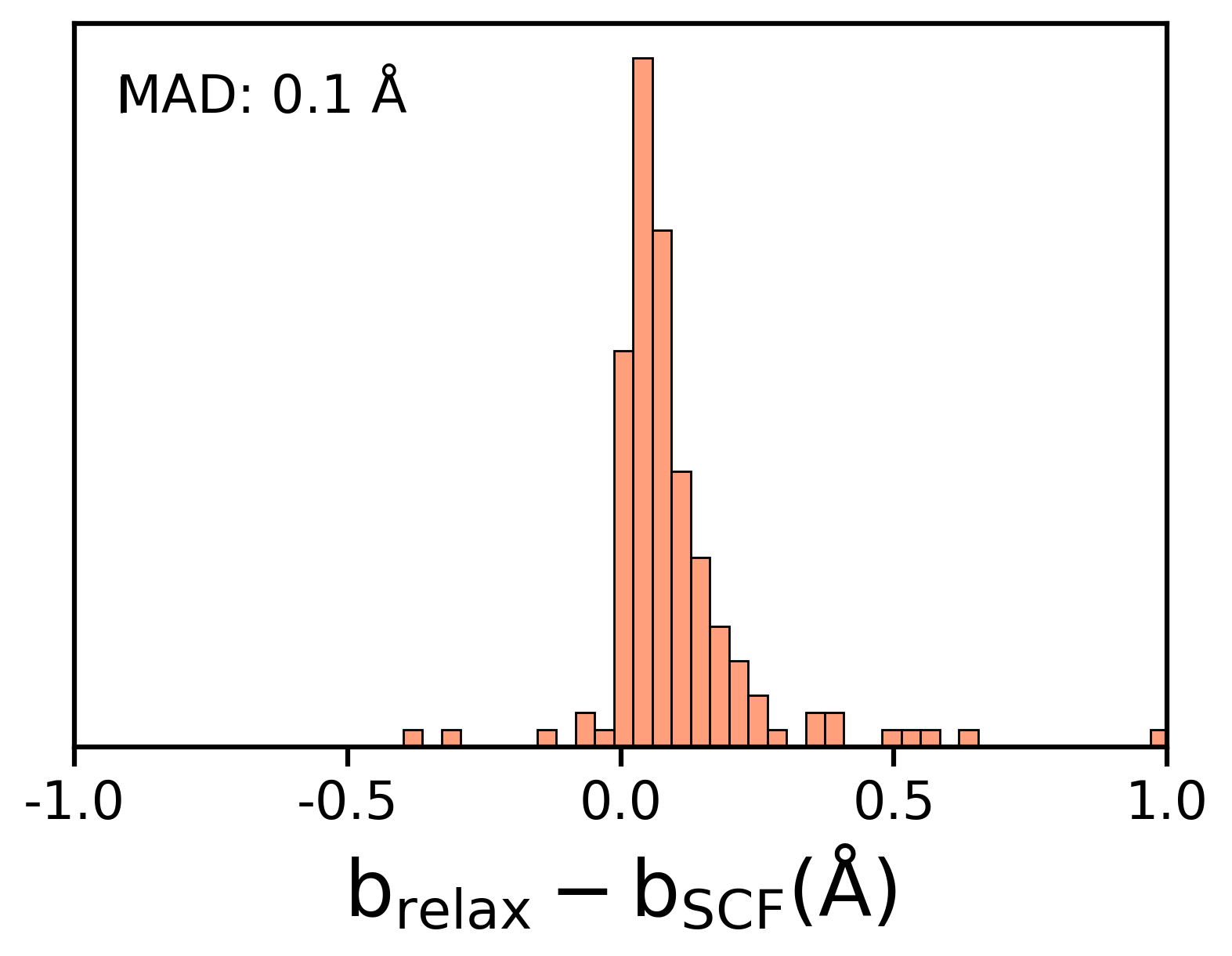}
    \caption*{}
  \end{subfigure}%
    \begin{subfigure}{0.33\textwidth}
    \includegraphics[width=\linewidth]{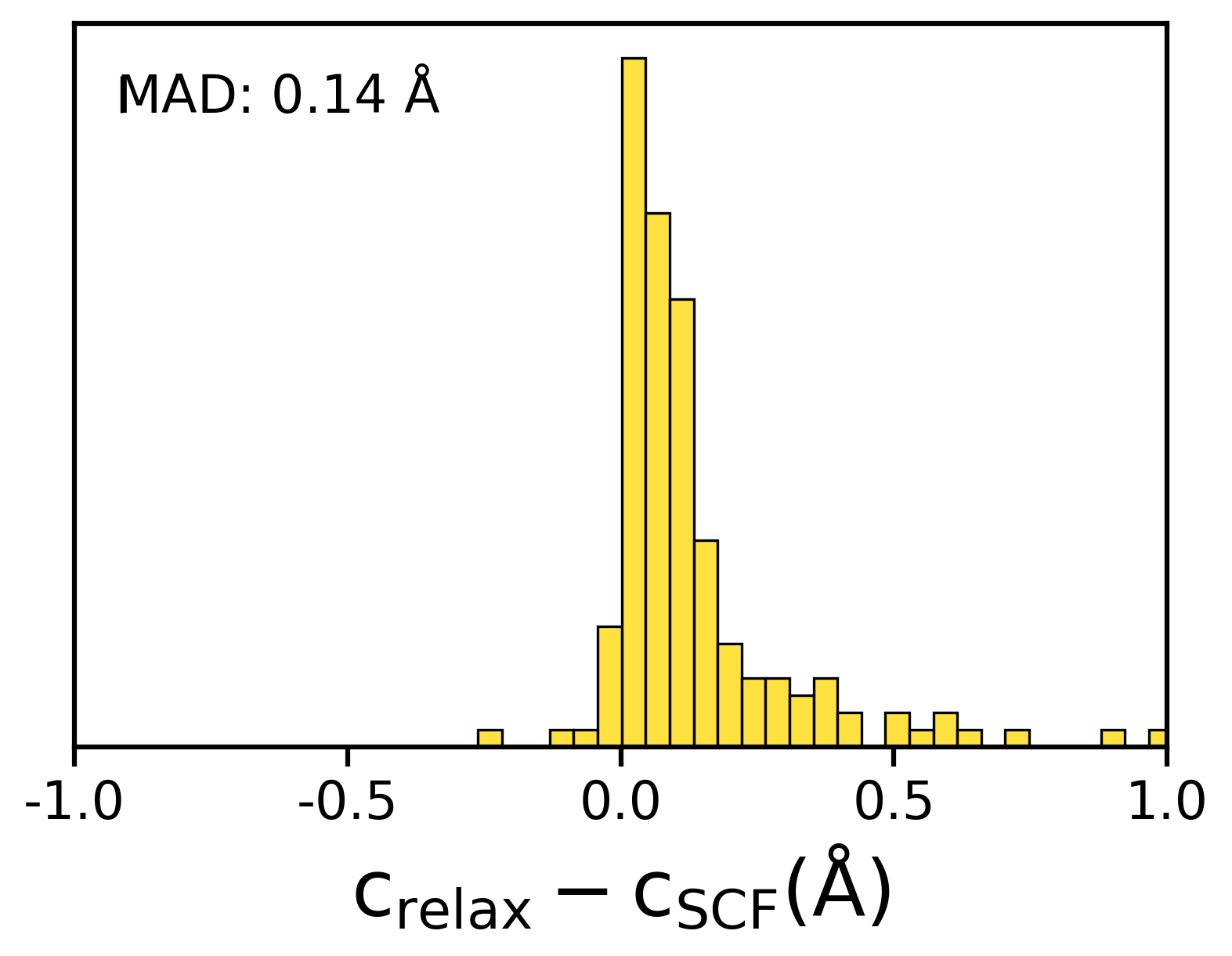}
    \caption*{}
  \end{subfigure}%
\caption*{\textbf{Figure S3.} Changes in the lattice parameters (a,b,c) when relaxed with HSE06 (labelled as relax) compared to a case where single-point HSE06 calculation (labelled as SCF) were carried out over PBE optimized structures for 153 oxides. The corresponding mean absolute deviation (MAD) between two cases is shown.}
\end{figure*}

\subsection{Effect of Geometry Relaxation at the HSE06 level for Calculating Formation Energy}
To evaluate the effect of the exchange-correlation functional on the equilibrium geometry of oxides, we have also optimized the structures of the 153 oxides used in our benchmark study with PBE and HSE06 methods. The distribution of difference in the lattice parameters obtained from PBE and HSE06 optimizations is shown in Figure S3. Then, we analyzed the impact of using geometries optimized with PBE vs HSE06 on the evaluation of the formation energies of these 153 oxides. We calculated formation energies i) with HSE06 for both geometry and energies and ii) with PBE for geometry and HSE06 for energies. In the latter case, we performed a single-point HSE06 calculation over a PBE geometry.  In Figure S4a, the formation energies obtained with both approaches are compared. This figure shows the  distribution of the deviation in formation energy values obtained with both approaches. Only minor difference are observed, with average deviations of 0.02 eV/atom. For this reason, the formation energy calculation of oxides by single point HSE06 calculations on a PBE geometry is adopted in the work. We have also evaluated the impact of PBE vs HSE06 geometries on band gaps for the 153 oxides using the full HSE06 approach and HSE06 single-point calculations on PBE geometries, as done for the formation energies. Figure S4b shows that the HSE06 geometry optimization has a significant impact on the band gaps, with a mean deviation of 0.26 eV.

\begin{figure*}[h]
\centering
\captionsetup{justification=justified, singlelinecheck=false}
  \begin{subfigure}{0.5\textwidth}
    \includegraphics[width=\linewidth]{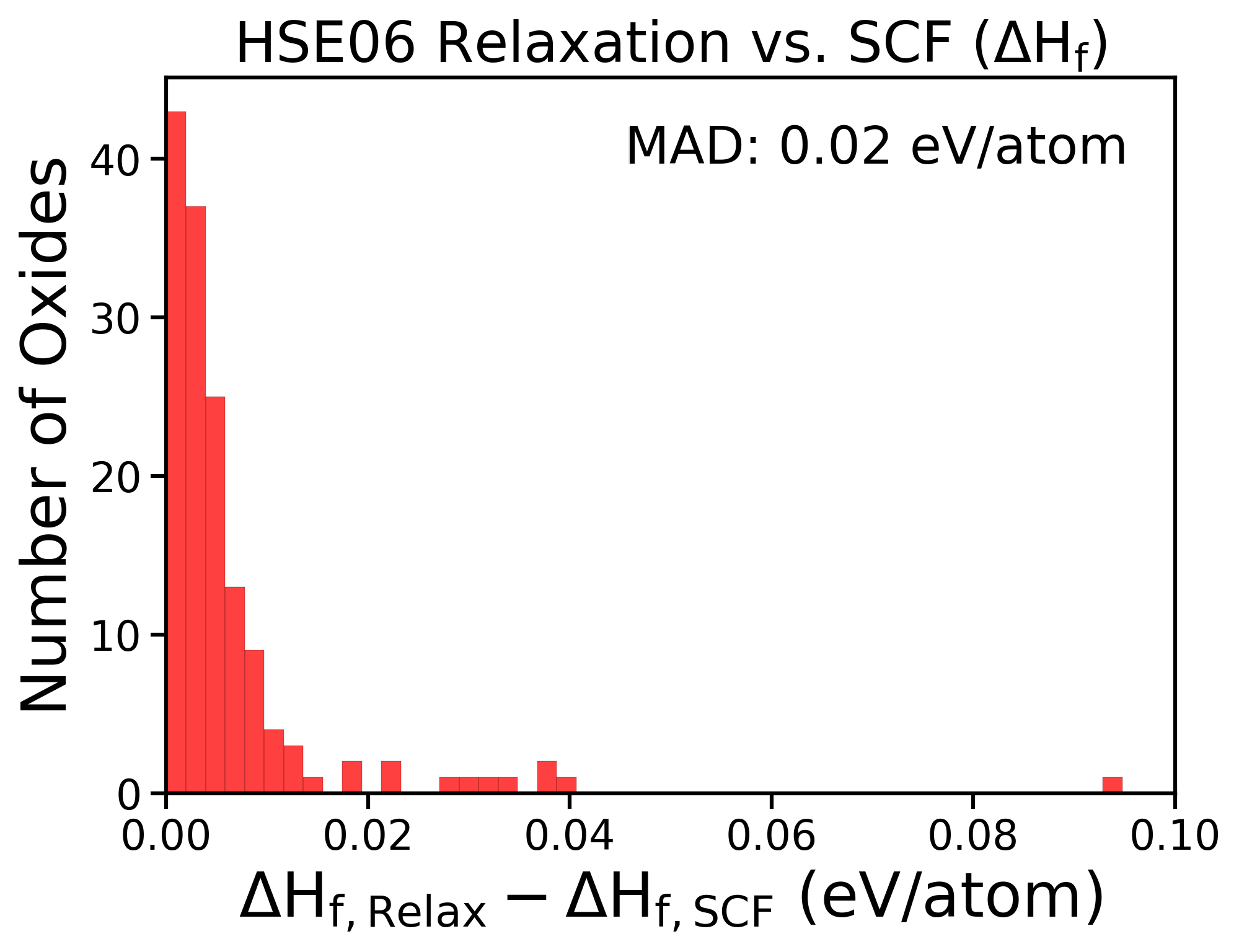}
    \caption*{(a)}
  \end{subfigure}%
  \hspace*{\fill}   
  \begin{subfigure}{0.5\textwidth}
    \includegraphics[width=\linewidth]{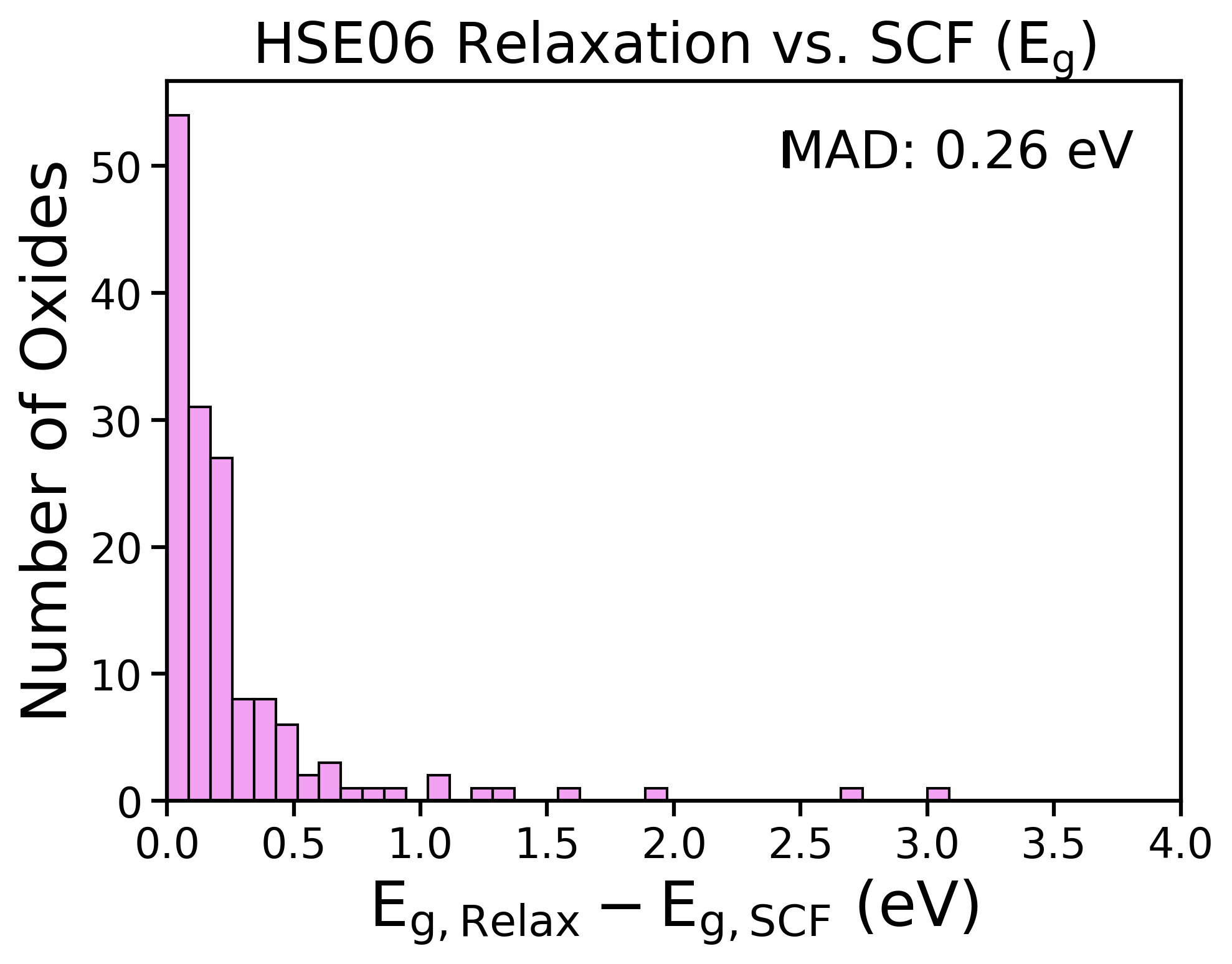}
    \caption*{(b)}
  \end{subfigure}%
\caption*{\textbf{Figure S4.} Effect of geometry optimization with HSE06 on (a) formation energies and (b) band gaps for 153 oxides. The formation energies ($\mathrm{\Delta H_f}$) and band gaps ($\mathrm{E_g}$) obtained from HSE06 optimized geometries ($\mathrm{\Delta H_{f, Relax}}$ and $\mathrm{E_{g, Relax}}$) and HSE06 single-point calculations on PBE optimized geometries ($\mathrm{\Delta H_{f, SCF}}$  and $\mathrm{E_{g, SCF}}$) are compared  with corresponding mean absolute deviation (MAD).}
\end{figure*}

\subsection{Effect of Fraction of Hartree-Fock Exchange}
The fraction of Hartree-Fock exchange in the Heyd-Scuseria-Ernzerhof (HSE) functional, termed $\mathrm{\alpha}$,  significantly influences the evaluation of materials properties such as band gaps \cite{zhang2005theoretical,vydrov2006importance}. Here, we have analyzed the effect of $\mathrm{\alpha}$ in the range of 0.1 to 0.6 (10 to 60 \%) on the formation energies of 70 oxides. These 70 oxides are randomly chosen among the 153 oxides considered in our benchmark study. In all the cases, the screening parameter value is fixed to the standard HSE06 value of 0.11 $\mathrm{Bohr^{-1}}$ and PBE-optimized geometries were used. We note that the formation energy calculation depends on the energy of the oxide material, of the \ce{O2} molecule, and of the reference elemental solid(s) - see Eqn. 1. We use the same $\mathrm{\alpha}$ for evaluating all energies entering in eqn 5. of the manuscript. The distribution of relative absolute errors in formation energies with respect to experiment \cite{wang2021framework} is shown in Figure S5. The error distribution for PBE and for HSE06 (with $\mathrm{\alpha}$ value of 0.25) are also included in the figure. The agreement between experimental and calculated formation energies changes significantly with $\mathrm{\alpha}$. HSE06 shows a MAE lower than PBE in computing the formation energies which is consistent with the analysis on 153 oxides shown in Figure S2. Formation energies tend to be less accurate beyond $\mathrm{\alpha}=0.40$. Nevertheless, no consistent and significant improvement in formation energies was observed with any $\mathrm{\alpha}$ values from that of HSE06. Hence we considered HSE06 with $\mathrm{\alpha} =0.25$ in this study. 

\begin{figure*}[ht]
\centering
\captionsetup{justification=raggedright, singlelinecheck=false}
    \includegraphics[width=0.8\linewidth]{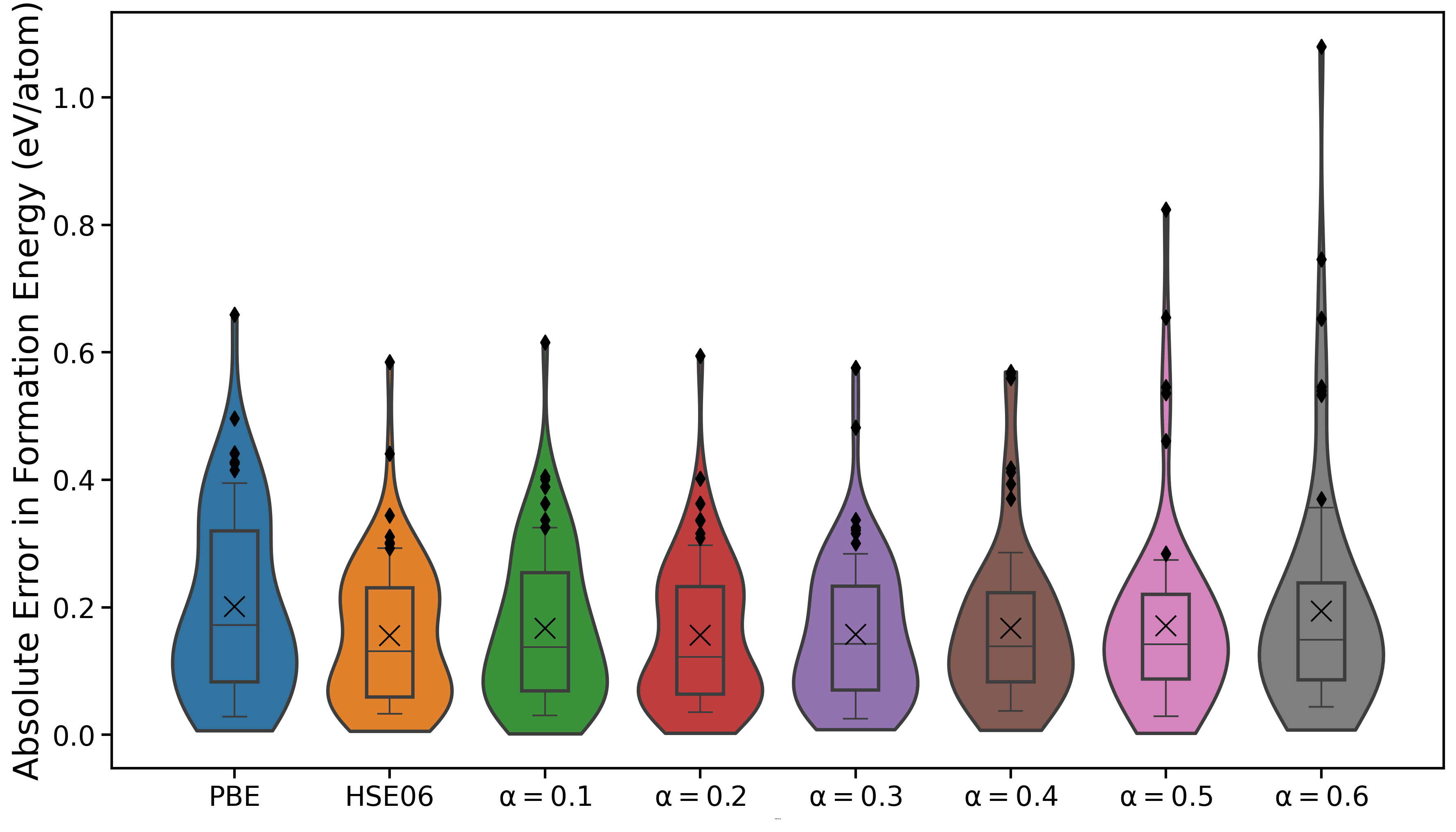}
    \caption*{\textbf{Figure S5} Relative absolute error distribution (calculated vs. experiment) for the formation energy of 70 oxides for different fractions of Hartree-Fock exact exchange ($\mathrm{\alpha}$). All geometries were optimized at the DFT-PBE level. The area between the upper and lower edges of the black box represents the interquartile range of errors and the whiskers from bottom to top correspond to 10 \%, median and 90 \% errors, respectively. The materials with errors higher than 90\% are indicated with black diamond symbols in respective error distribution violin plots.}
    \label{Supplementary Figure 4}
\end{figure*}

\subsection{Comparison of Transition metal vs. Non-transition Metal Oxide Formation Energies}
Previous studies have demonstrated that the HSE06 method leads to a more significant improvement in the accuracy of formation energy predictions for transition metal oxides compared to non-transition metal oxides.\cite{chevrier2010hybrid}. We have also investigated the formation energy errors separately for oxides containing transition metals as shown in Figure S6a. It can be seen that HSE06 provides a higher accuracy with respect to PBE for transition metal oxides. It has been proposed that the underestimation of PBE formation energies can be corrected by shifting them by a constant value, determined so as to minimize the errors in formation energies for oxides containing non-transition metals with respect to experiment \cite{wang2006oxidation}. The underlying assumption is that the self-interaction error is negligible for those systems and that most of the deviations originate from the poor description of $\mathrm{O_2}$ binding energy. Within the oxides based on non-transition-metal elements that we consider, we do not observe a constant difference between PBE and experimental formation energies. In fact, the formation energies are rather spread with respect to the values that would be obtained by applying the mentioned correction, with a standard deviation of 0.09 eV/atom (Figure S6b). Because of such a deviation and due to the fact that HSE06 provides a better description of $\mathrm{O_2}$ binding energy (-5.16 eV) compared to PBE, we abstain from utilizing such a correction on PBE formation energies.

\begin{figure*}[ht]
\centering
\captionsetup{justification=justified, singlelinecheck=false}
  \begin{subfigure}{0.33\textwidth}
    \includegraphics[width=\linewidth]{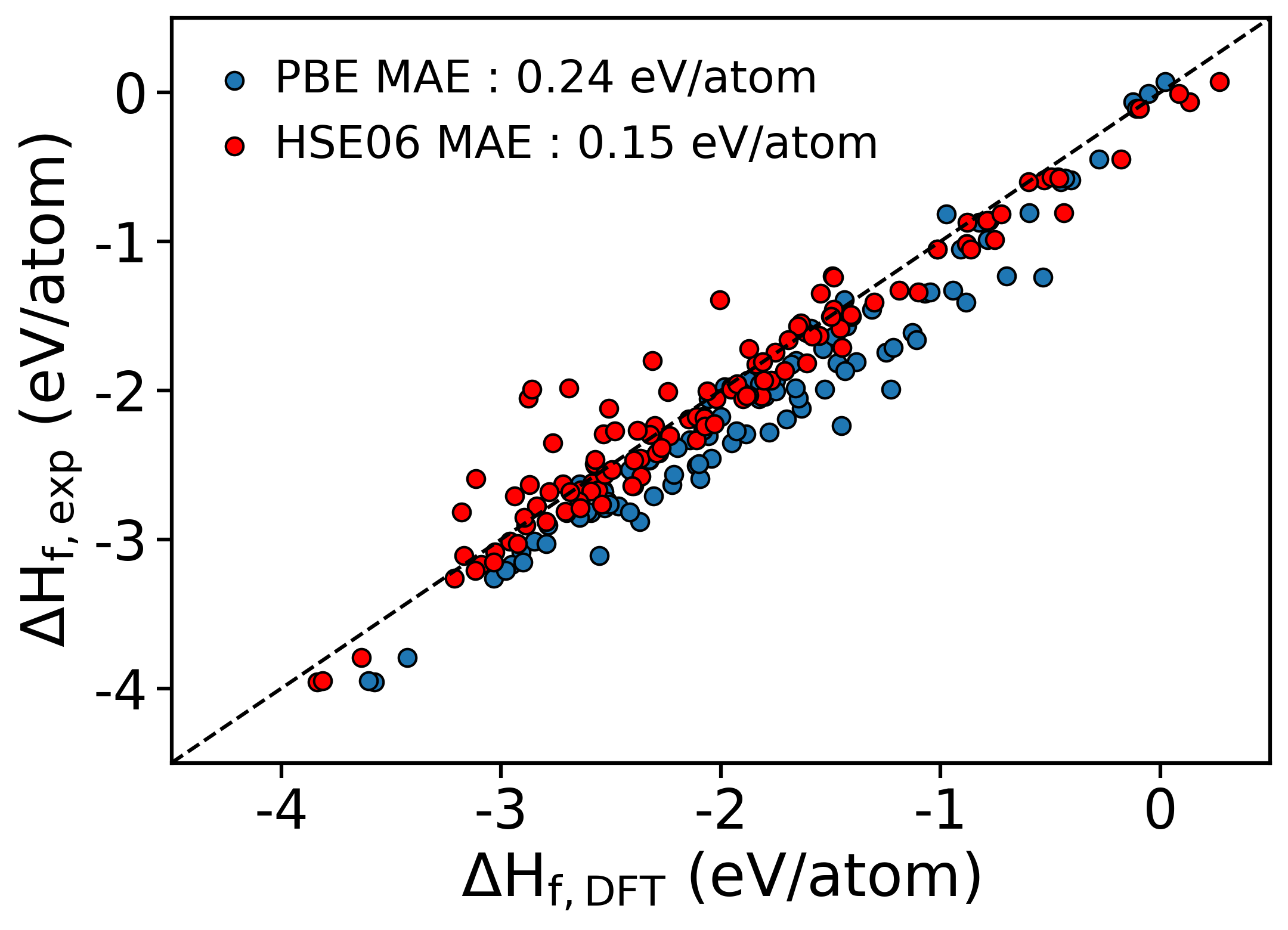}
    \caption*{(a)}
  \end{subfigure}%
  \hspace*{\fill}   
  \begin{subfigure}{0.33\textwidth}
    \includegraphics[width=\linewidth]{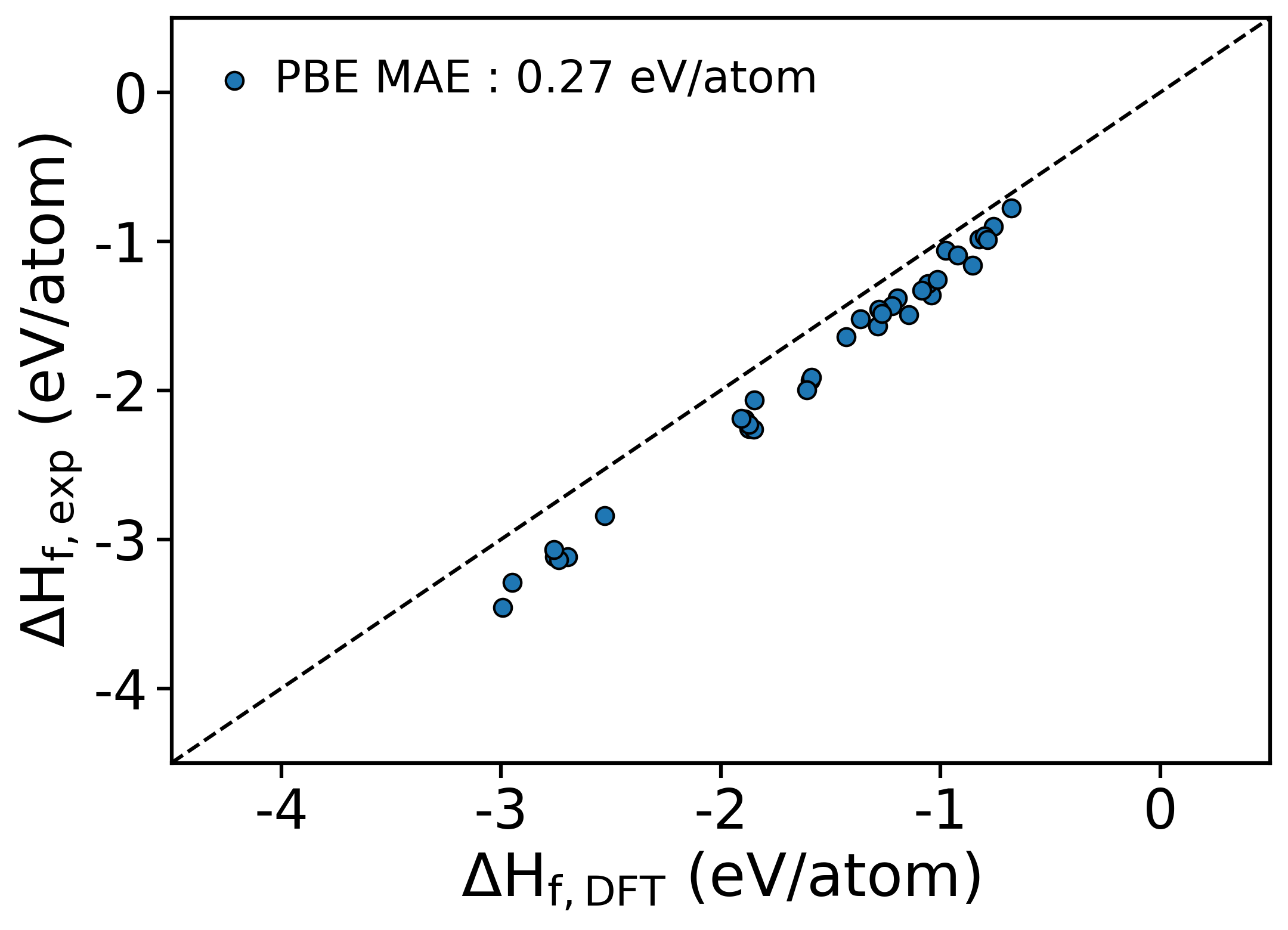}
    \caption*{(b)}
  \end{subfigure}%
    \begin{subfigure}{0.33\textwidth}
    \includegraphics[width=\linewidth]{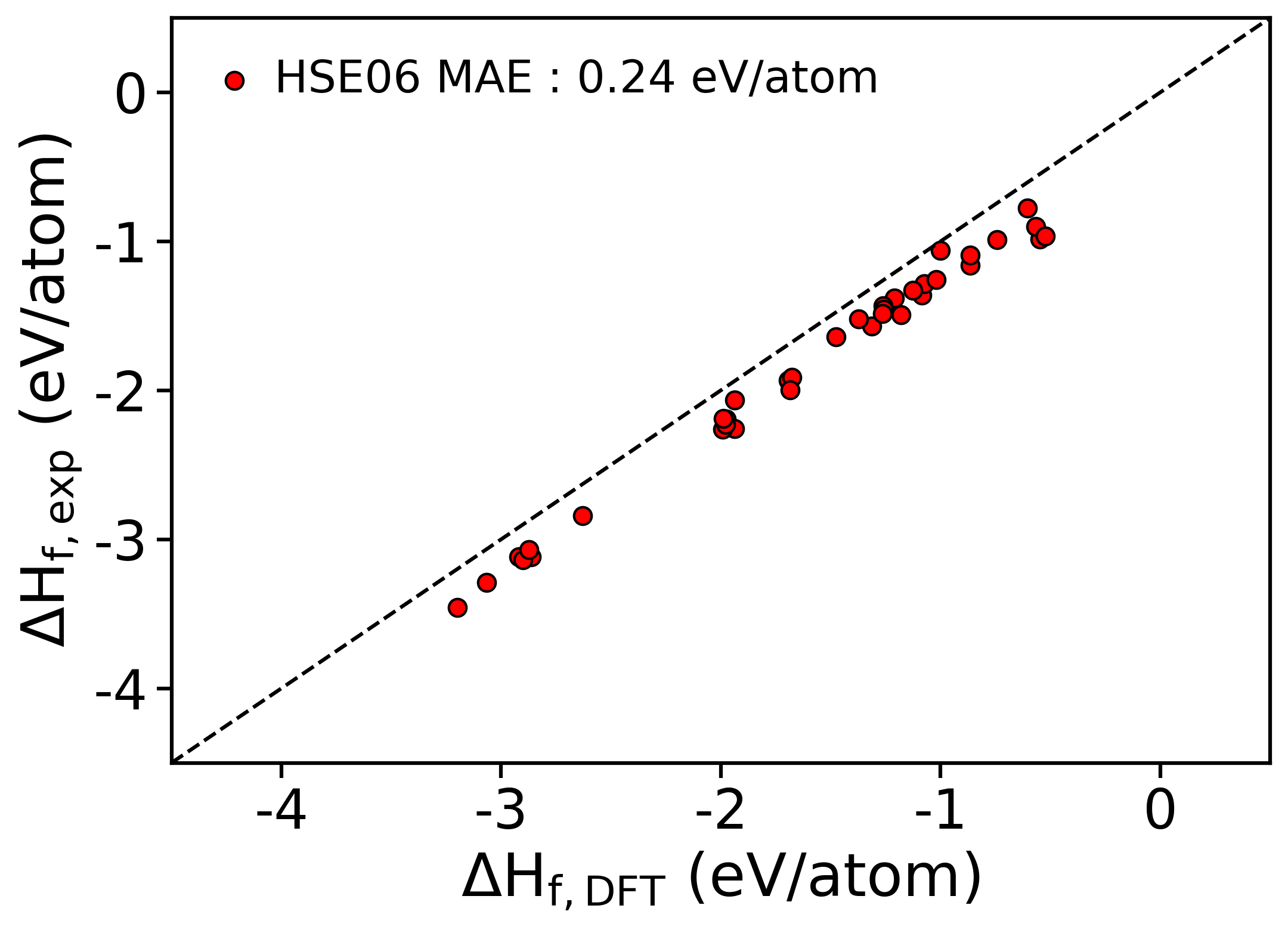}
    \caption*{(b)}
  \end{subfigure}%
  \caption*{\textbf{Figure S6.}  Comparison of formation energies of (a) transition-metal oxides computed using PBE and HSE06 (b) non-transition metal oxides computed using PBE and (c)  non-transition metal oxides computed using HSE06, with respect to experiment \cite{wang2021framework}. For the non-transition metal oxides, we do not observe a constant shift from experimental values for PBE. The standard deviation of the distribution of differences between PBE formation energies and experiment is equal to 0.09 eV/atom.}
\end{figure*}

\section{Pourbaix Diagram Analysis}
 An electrode material (X) undergoes redox reaction to form another phase (Y) in an aqueous medium under a given pH and applied potential (U)  following the equation;
\begin{equation}
X + n_w{H}_2O \rightleftharpoons Y + n_hH^+ + n_ee^-
\label{eqn:Nerst_1}
\end{equation}
Considering standard hydrogen electrode as the counter electrode, the equilibrium potential for the eqn. \ref{eqn:Nerst_1} can be written according to Nernst equation as;
\begin{equation}
U_{eq} = U^{0} + \frac{k_BT \ln10}{n_e}\left(\prod_{i \neq H^+} a_i^{n_i} - n_hpH \right)
\label{eqn:Nerst_2}
\end{equation}
where $k_B$ is the Boltzmann's constant, $T$ is the temperature (298.15 K), $a_i$ is the activity of species $i$ and $n_i$ is the number of species, respectively. Since the elemental solids are reference states, their chemical potentials is equal to zero. Thus, the equilibrium potential under standard state ($U^0$) can be computed from the formation energy ($\Delta H_{f}$) of the phases as;

\begin{equation}
U^{0} \approx \frac{\Delta H_{f}}{n_e}
\label{eqn:Nerst_3}
\end{equation}

By calculating the formation energies of all the relevant solid and solvated phases, eqn. \ref{eqn:Nerst_3} is derived for the electrochemical reactions the target material undergoes, resulting in a series of equilibrium lines which can be plotted on a pH-\textit{U} diagram and stable domains at the coordinates of the pH-\textit{U} space can be identified. We use a recent implementation of Pourbaix utility which is now available in ASE \cite{americo2023electronic}. 

\section{Benchmarking acid-stability of oxides}
From the formation energy benchmarking study, we have observed that HSE06 provides more accurate formation energies compared to PBE with respect to the experiment for a dataset of 153 oxides. Now, we have investigated whether such an accuracy improvement for HSE06 exists for the stability prediction of oxides in aqueous media by constructing their Pourbaix diagrams For this, Pourbaix diagrams are constructed. These diagrams characterize the stability of a material as a function of pH and electrochemical potential ($U$) \cite{pourbaix1966atlas}. For the construction of PBE and HSE06 Pourbaix diagrams, we utilize the formation energies of oxides and other competing phases evaluated by these methods. For experimental Pourbaix diagrams, formation energies from experiment \cite{wang2021framework} are utilized. In both cases, the formation energies of ions are taken from experiment \cite{pourbaix1966atlas,johnson1992supcrt92}. For the comparison to be consistent, during the benchmarking stage, the Pourbaix diagrams were constructed only by considering the competing phases for which formation energies are available at all three levels PBE, HSE06, and experiment. 

A comparison of experimental, PBE and HSE06 Pourbaix diagrams for the Ti-\ce{H2O} system is shown in Figure S7 as an example. The equilibrium solid black lines define the pH-\textit{U} domains at which different phases are the most stable. In this work, we are interested in the stability of oxide materials. Thus, we include, in the Pourbaix diagrams, a color scale that reflects the relative stability of a given target oxide, which is computed with respect to the most stable phase at different pH-\textit{U} coordinates. We refer to such energy difference as the Pourbaix decomposition energy of the oxide and denoted it $\Delta G_{pbx}$. In Figure S7, \ce{TiO2} is such target oxide and the color scale represents the $\Delta G_{pbx}$ associated with the decomposition of \ce{TiO2} to different phases across the pH-\textit{U} grid. In the Pourbaix diagram obtained with experimental formation energies (Figure S7a), \ce{TiO2} is stable between the equilibrium potentials of hydrogen evolution and water decomposition, highlighted by the blue dashed lines of the figure. \ce{TiO2} tends to dissolve into \ce{TiO2^2+} and decompose to \ce{TiH2} at $U\geq$ 2.2V and  $U\leq$ -0.6V, respectively. At extremely acidic conditions (pH $\leq$ -1.2), \ce{TiO2} can reduce to form \ce{Ti^2+} for U$\approx$-0.5V. The PBE diagram (Figure S7b) reproduces some of the qualitative characteristics of the experimental diagram such as the stability of \ce{TiO2^2+} and \ce{TiH2} at high and low potentials, respectively. However, an additional phase \ce{Ti^3+} is present in the PBE diagram compared to the experimental one. Furthermore, the stability domains for \ce{TiO2^2+} and \ce{Ti^2+} in the PBE diagram are larger compared to the experimental diagram. As a result, PBE provides a stability domain of \ce{TiO2} that is considerably smaller than it should. In the HSE06 diagram (Figure S7c), the stability domains of \ce{TiO2^2+}, \ce{Ti^2+} and \ce{Ti^3+} are reduced compared to PBE. As a result, the HSE06 diagram aligns more closely to experiment, with a similar stability region for the target oxide \ce{TiO2}. The significant disagreement between PBE and experimental Pourbaix diagrams can be attributed to the underestimation of the stability of Ti oxides, which is notably corrected by HSE06.


\begin{figure*}[ht]
\captionsetup{justification=justified,singlelinecheck=false}
  \begin{subfigure}{0.32 \textwidth}
    \includegraphics[width=\linewidth]{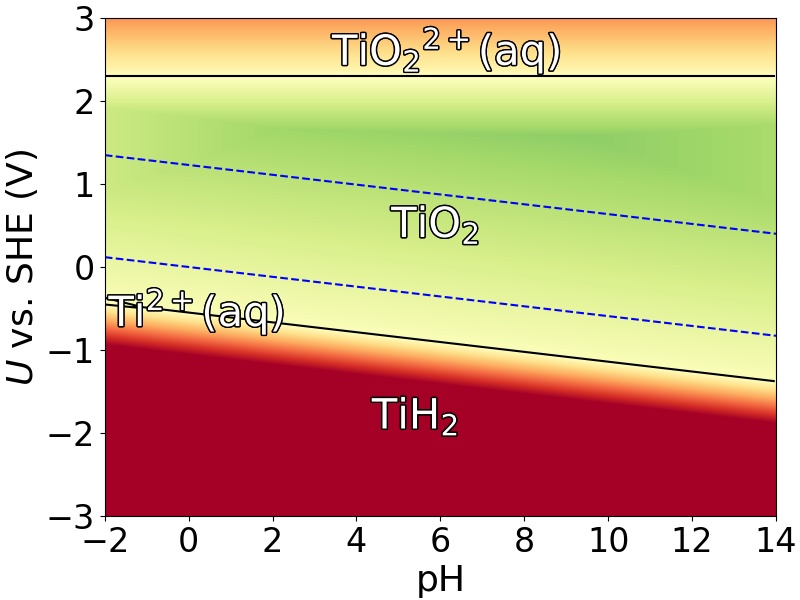}
    \caption{} \label{Fig:6a}
  \end{subfigure}%
  \hspace*{\fill}   
  \begin{subfigure}{0.32 \textwidth}
    \includegraphics[width=\linewidth]{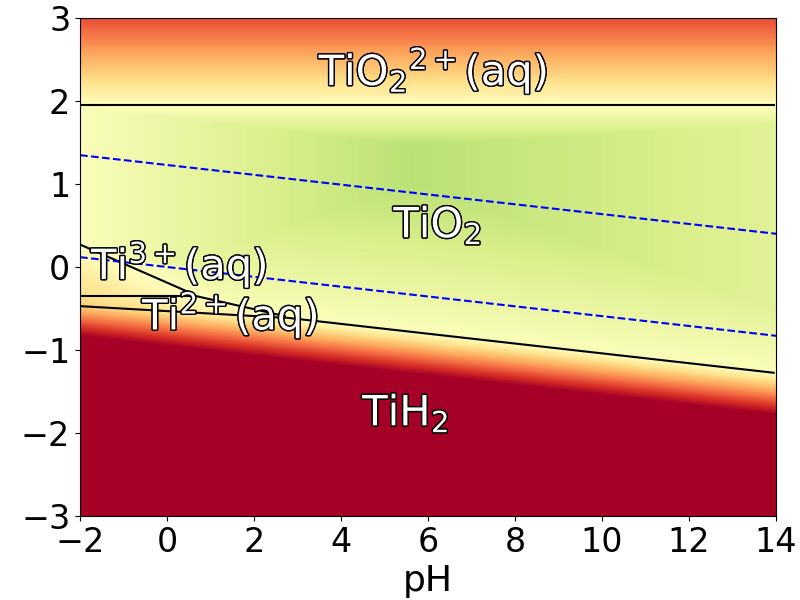}
    \caption{} \label{Fig:6b}
  \end{subfigure}%
  \hspace*{\fill}   
  \begin{subfigure}{0.357\textwidth}
    \includegraphics[width=\linewidth]{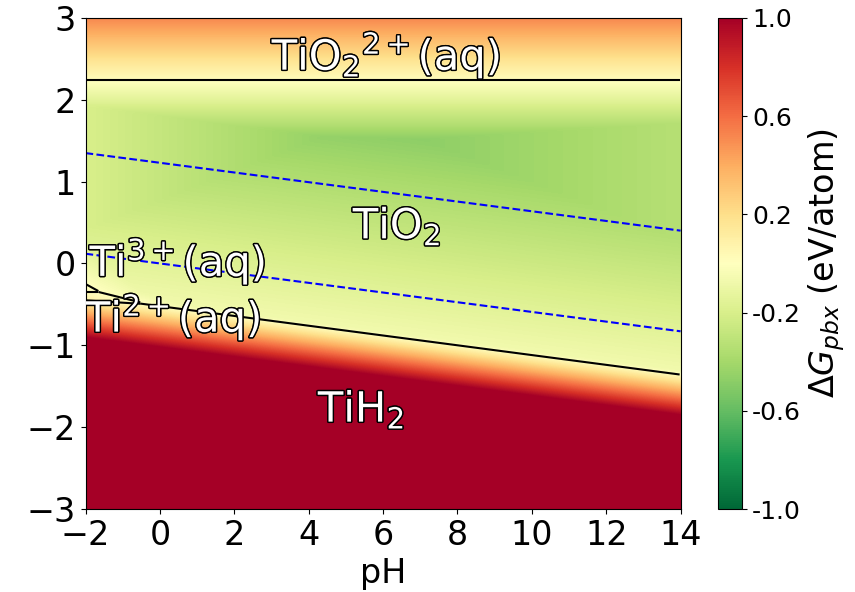}
    \caption{} \label{Fig:6c}
  \end{subfigure}
 \caption*{ \textbf{Figure S7.} (a) Experimental, (b) DFT-PBE and (c) DFT-HSE06 Pouraix diagrams of the Ti-\ce{H2O} system with \ce{TiO2} as the target oxide. An aqueous ion concentration of $10^{-6}$ M and temperature of 298.15 K are considered with the standard hydrogen electrode (SHE) as the counter electrode. The phases considered are \ce{TiO2}, \ce{Ti2O3}, \ce{Ti3O5}, \ce{Ti6O}, \ce{TiO}, \ce{Ti3O}, \ce{Ti4O7}, \ce{TiH2}, \ce{Ti(s)}, \ce{HTiO3-}, \ce{TiO2+}, $\mathrm{TiO_2^{2+}}$, $\ce{Ti^{2+}}$ and $\ce{Ti^{3+}}$ . The aqueous ionic phases are labelled with \qq{(aq)}. The formation energies of solids are obtained from DFT calculations whereas those of ionic aqueous phases are taken from refs \cite{pourbaix1966atlas,johnson1992supcrt92}. The stability domain of \ce{H2O} (pH$\in$[-2,14] and $U$$\in$[0,1.23] V) is indicated with blue dashed lines.}
 \label{Fig:6}
\end{figure*}

As a quantitative measure of acid stability, we consider the minimum Pourbaix decomposition energy across all the possible decomposition reactions the target oxide can undergo in a Pourbaix diagram at  pH=0 and $U$=1.23 V, as they represent typical acidic OER conditions. This quantity is denoted as $\Delta G_{pbx}^{\mathrm{OER}}$. It is important to note that the OER can occur in both acidic and alkaline environments, and catalysts are not limited to oxides \cite{yang2021non}. In this study, we focus on acidic conditions and oxide-based materials. However, the approach outlined here can be easily adapted for alkaline OER and applied to other types of materials.

First, $\Delta G_{pbx}^{\mathrm{OER}}$ for 153 oxides is computed from their Pourbaix diagrams constructed for the benchmarking study. The PBE and HSE06 calculated $\Delta G_{pbx}^{\mathrm{OER}}$ for 153 oxides show a MAE of 0.22 and 0.13 eV/atom with respect to the experiment. The distribution of errors for both methods is shown in Figure S2 (right panel). These results indicate the accuracy of HSE06 over PBE in reliably predicting the acid stability of oxides.

Following  this, we then evaluated $\Delta G_{pbx}^{\mathrm{OER}}$ for 250 oxides using both PBE and HSE06 methods. This materials set includes the 153 oxides considered in the benchmarking, plus about 100 additional oxides which share the same composition space. For the construction of these \textit{ab initio} Pourbaix diagrams, competing phases from the MP convex hull were retrieved using two criteria; i) energy of convex hull ($E_{hull}$) below 0.05 eV/atom and ii) less than 50 atoms per unit cell ($N_{atoms}$). When two materials with the same composition satisfy such criteria, we only consider the material with the lowest $E_{hull}$ according to MP data in our analysis. In addition to the materials selected from MP using these criteria, we also included in our analysis all the materials of the OQMD123 database that contain 1, 2 and 3 elements and are located in the convex hull \cite{landis2012computational}.  By using this approach for the selection of materials in the Pourbaix analysis, we aim to analyze the impact of non-local exchange-correlation via the HSE06 method in acid stability prediction rather than reconstructing the complete Pourbaix diagrams reported in databases like MP. 

The critical electrochemical decomposition reactions and the corresponding $\Delta G_{pbx}^{\mathrm{OER}}$ values for 250 oxides are provided in Table S1.  This dataset will be used for our SISSO analysis. In this dataset, only less than 50 oxides are found to be acid-stable. In general, ternary oxides are found to have lower stability than binary oxides. The $\Delta G_{pbx}^{\mathrm{OER}}$ distribution over the 250 materials has a mean value of 0.61 eV/atom and a standard deviation of 1.09 eV/atom (Figure S8). Such distribution highlights that there is a larger fraction of unstable oxides in such datasets. \ce{Ta2O5} possesses the highest acid stability ($\Delta G_{pbx}^{\mathrm{OER}}$ = -2.92 eV/atom). This is in line with the experimental observation of its high resistance for dissociation in the aqueous medium \cite{hwang2016electro}.

\begin{figure*}[h!]
\captionsetup{justification=centering, singlelinecheck=false}
  \begin{subfigure}{0.5\textwidth}
    \includegraphics[width=\linewidth]{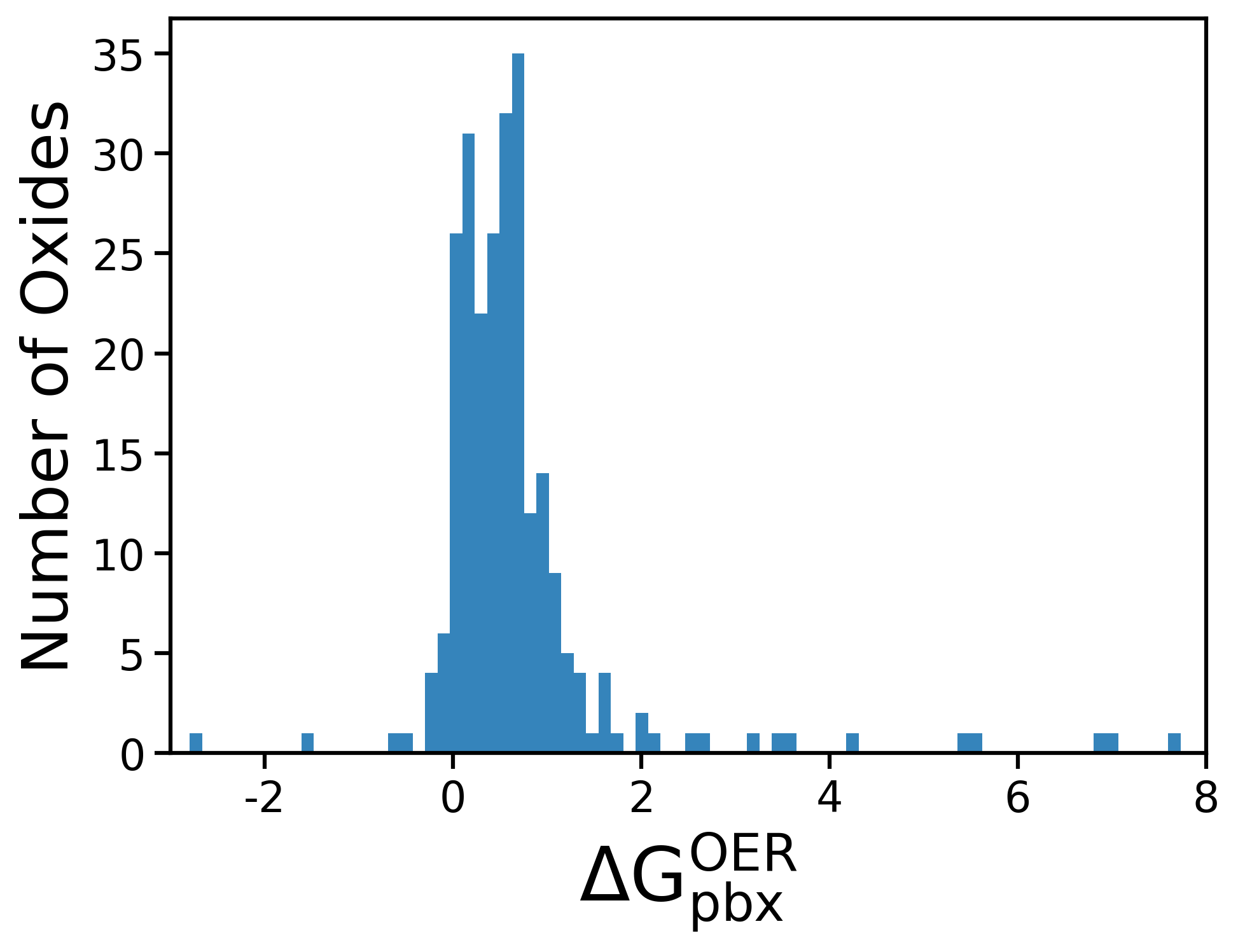}
    \caption*{(a)}
  \end{subfigure}%
  \hspace*{\fill}   
  \begin{subfigure}{0.5\textwidth}
    \includegraphics[width=\linewidth]{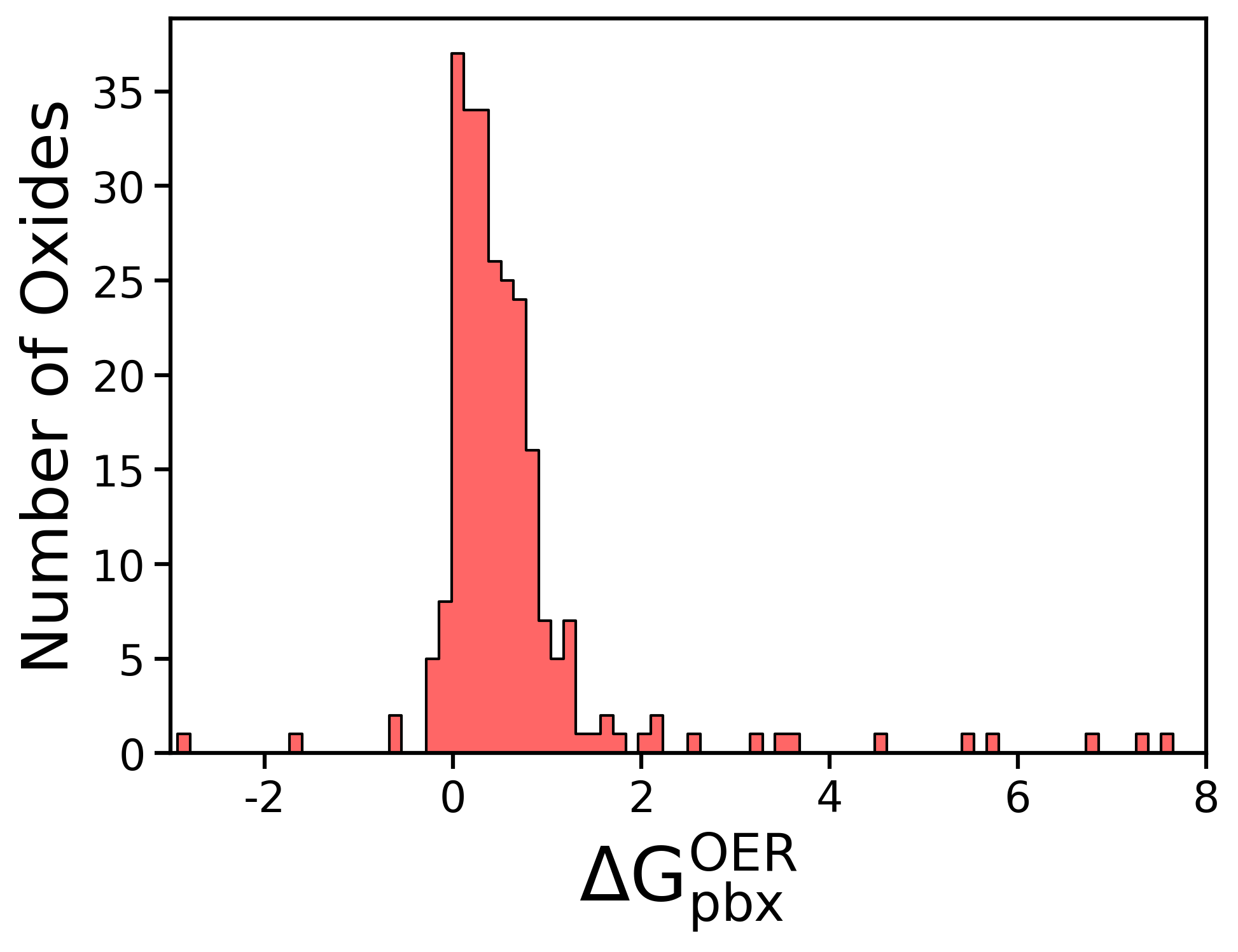}
    \caption*{(b)}
  \end{subfigure}%
\caption*{\textbf{Figure S8.} Distribution of $\Delta G_{pbx}^{\mathrm{OER}}$ for 250 oxides calculated with (a) PBE and (b) HSE06.}
\end{figure*}

\newpage
\begin{table*}[h]
\captionsetup{justification=raggedright, singlelinecheck=false}
    \caption*{\textbf{Table S1.} Critical electrochemical decomposition reaction and the minimum Pourbaix decomposition energy at OER-relevant conditions of $U$=1.23 V and pH=0  ($\Delta G_{pbx}^{\mathrm{OER}}$ in eV/atom) obtained from Pourbaix diagrams for 250 oxides calculated using PBE and HSE06 methods. The compositions of phases in the decomposition reaction are normalized per unit formula.}
    \centering
    \begin{adjustbox}{width=\textwidth}
    \small
    \begin{tabular}{p{2cm} *{4}{r}}
    \toprule
    \multirow{2}{*}{\textbf{Oxide}} & \multicolumn{2}{c}{PBE} & \multicolumn{2}{c}{HSE06} \\
    \cmidrule(lr){2-3} \cmidrule(lr){4-5}
     & \multicolumn{1}{c}{\textbf{Decomposition Reaction}} & \textbf{$\mathrm{\Delta G_{pbx}^{OER}}$ } & \multicolumn{1}{c}{\textbf{Decomposition Reaction}} & \textbf{$\mathrm{\Delta G_{pbx}^{OER}}$ } \\
    \midrule
\ce{AgO} & \ce{AgO + 2H+ + e-  ->  H2O + Ag+(aq)} & 0.268 & \ce{AgO + 2H+ + e-  ->  H2O + Ag+(aq)} & 0.526 \\
\ce{Al2O3} & \ce{Al2O3 + 6H+  ->  3H2O + 2Al3+(aq)} & 0.630 & \ce{Al2O3 + 6H+  ->  3H2O + 2Al3+(aq)} & 0.424 \\
\ce{AlCuO2} & \ce{AlCuO2 + 4H+  ->  2H2O + e- + Al3+(aq) + Cu2+(aq)} & 0.740 & \ce{AlCuO2 + 4H+  ->  2H2O + e- + Al3+(aq) + Cu2+(aq)} & 0.563 \\
\ce{As2O5} & \ce{As2O5 + 3H2O  ->  2H3AsO4(aq)} & 0.275 & \ce{As2O5 + 5/3H2O  ->  2/3As3H5O10} & 0.232 \\
\ce{Au2O3} & \ce{Au2O3 + 6H+ + 6e-  ->  3H2O + 2Au} & -0.055 & \ce{Au2O3 + 6H+ + 6e-  ->  3H2O + 2Au} & 0.083 \\
\ce{As2O3} & \ce{As2O3 + 5H2O  ->  4H+ + 4e- + 2H3AsO4(aq)} & 0.649 & \ce{As2O3 + 11/3H2O  ->  4H+ + 4e- + 2/3As3H5O10} & 0.635 \\
\ce{Ag2O} & \ce{Ag2O + 2H+  ->  H2O + 2Ag+(aq)} & 0.416 & \ce{Ag2O + 2H+  ->  H2O + 2Ag+(aq)} & 0.428 \\
\ce{Al2CoO4} & \ce{Al2CoO4 + 8H+  ->  4H2O + 2Al3+(aq) + Co2+(aq)} & 0.702 & \ce{Al2CoO4 + 5H+  ->  2H2O + e- + 2Al3+(aq) + CoHO2} & 0.363 \\
\ce{Ag3O4} & \ce{Ag3O4 + 8H+ + 5e-  ->  4H2O + 3Ag+(aq)} & 0.297 & \ce{Ag3O4 + 8H+ + 5e-  ->  4H2O + 3Ag+(aq)} & 0.570 \\
\ce{MnAl2O4} & \ce{Al2MnO4 + 8H+  ->  4H2O + 2Al3+(aq) + Mn2+(aq)} & 0.777 & \ce{Al2MnO4 + 4H+  ->  2H2O + 2e- + 2Al3+(aq) + MnO2} & 0.408 \\
\ce{Ba5V5O14} & \ce{Ba5V5O14 + 3H+  ->  3/2H2O + 7e- + 5Ba2+(aq) + 5/2V2O5} & 0.766 & \ce{Ba5V5O14 + 3H+  ->  3/2H2O + 7e- + 5Ba2+(aq) + 5/2V2O5} & 0.588 \\
\ce{BaMoO4} & \ce{BaMoO4 + 2H+  ->  H2O + Ba2+(aq) + MoO3} & 0.121 & \ce{BaMoO4 + 2H+  ->  H2O + Ba2+(aq) + MoO3} & 0.054 \\
\ce{BaO2} & \ce{BaO2 + 4H+ + 2e-  ->  2H2O + Ba2+(aq)} & 0.983 & \ce{BaO2 + 4H+ + 2e-  ->  2H2O + Ba2+(aq)} & 0.899 \\
\ce{BeO} & \ce{BeO + 2H+  ->  H2O + Be2+(aq)} & 0.613 & \ce{BeO + 2H+  ->  H2O + Be2+(aq)} & 0.452 \\
\ce{Ba2V3O9} & \ce{Ba2V3O9 + 3H+  ->  3/2H2O + e- + 2Ba2+(aq) + 3/2V2O5} & 0.212 & \ce{Ba2V3O9 + 3H+  ->  3/2H2O + e- + 2Ba2+(aq) + 3/2V2O5} & 0.169 \\
\ce{BaO} & \ce{BaO + 2H+  ->  H2O + Ba2+(aq)} & 1.783 & \ce{BaO + 2H+  ->  H2O + Ba2+(aq)} & 1.682 \\
\ce{BaV2O6} & \ce{BaV2O6 + 2H+  ->  H2O + Ba2+(aq) + V2O5} & 0.081 & \ce{BaV2O6 + 2H+  ->  H2O + Ba2+(aq) + V2O5} & 0.047 \\
\ce{Ba3V2O8} & \ce{Ba3V2O8 + 6H+  ->  3H2O + 3Ba2+(aq) + V2O5} & 0.303 & \ce{Ba3V2O8 + 6H+  ->  3H2O + 3Ba2+(aq) + V2O5} & 0.214 \\
\ce{Cr2O3} & \ce{Cr2O3 + 6H+  ->  3H2O + 2Cr3+(aq)} & 0.522 & \ce{Cr2O3 + H2O  ->  2CrHO2} & 0.213 \\
\ce{NaCr3O8} & \ce{Cr3NaO8 + 16H+ + 6e-  ->  8H2O + 3Cr3+(aq) + Na+(aq)} & 0.013 & \ce{Cr3NaO8 + 7H+ + 6e-  ->  2H2O + 3CrHO2 + Na+(aq)} & 0.258 \\
\ce{Cs(MoO3)3} & \ce{CsMo3O9  ->  e- + Cs+(aq) + 3MoO3} & 0.106 & \ce{CsMo3O9  ->  e- + Cs+(aq) + 3MoO3} & 0.082 \\
\ce{Cu2O} & \ce{Cu2O + 2H+  ->  H2O + 2e- + 2Cu2+(aq)} & 1.016 & \ce{Cu2O + 2H+  ->  H2O + 2e- + 2Cu2+(aq)} & 0.895 \\
\ce{CaO2} & \ce{CaO2 + 4H+ + 2e-  ->  2H2O + Ca2+(aq)} & 0.976 & \ce{CaO2 + 4H+ + 2e-  ->  2H2O + Ca2+(aq)} & 0.911 \\
\ce{CaCrO4} & \ce{CaCrO4 + 8H+ + 3e-  ->  4H2O + Ca2+(aq) + Cr3+(aq)} & 0.061 & \ce{CaCrO4 + 5H+ + 3e-  ->  2H2O + Ca2+(aq) + CrHO2} & 0.302 \\
\ce{Co3O4} & \ce{Co3O4 + 8H+ + 2e-  ->  4H2O + 3Co2+(aq)} & 0.375 & \ce{Co3O4 + 2H2O  ->  H+ + e- + 3CoHO2} & 0.164 \\
\ce{CaO} & \ce{CaO + 2H+  ->  H2O + Ca2+(aq)} & 1.323 & \ce{CaO + 2H+  ->  H2O + Ca2+(aq)} & 1.205 \\
\ce{CdO} & \ce{CdO + 2H+  ->  H2O + Cd2+(aq)} & 0.761 & \ce{CdO + 2H+  ->  H2O + Cd2+(aq)} & 0.708 \\
\ce{CrO2} & \ce{CrO2 + 4H+ + e-  ->  2H2O + Cr3+(aq)} & 0.148 & \ce{CrO2 + H+ + e-  ->  CrHO2} & 0.241 \\
\ce{Ca3V2O8} & \ce{Ca3V2O8 + 6H+  ->  3H2O + 3Ca2+(aq) + V2O5} & 0.346 & \ce{Ca3V2O8 + 6H+  ->  3H2O + 3Ca2+(aq) + V2O5} & 0.261 \\
\ce{Ca2V2O7} & \ce{Ca2V2O7 + 4H+  ->  2H2O + 2Ca2+(aq) + V2O5} & 0.211 & \ce{Ca2V2O7 + 4H+  ->  2H2O + 2Ca2+(aq) + V2O5} & 0.145 \\
\ce{CoO} & \ce{CoO + 2H+  ->  H2O + Co2+(aq)} & 0.989 & \ce{CoO + H2O  ->  H+ + e- + CoHO2} & 0.501 \\
\ce{Cs2Mo3O10} & \ce{Cs2Mo3O10 + 2H+  ->  H2O + 2Cs+(aq) + 3MoO3} & 0.075 & \ce{Cs2Mo3O10 + 2H+  ->  H2O + 2Cs+(aq) + 3MoO3} & 0.054 \\
\ce{MgCr2O4} & \ce{Cr2MgO4 + 8H+  ->  4H2O + 2Cr3+(aq) + Mg2+(aq)} & 0.619 & \ce{Cr2MgO4 + 2H+  ->  2CrHO2 + Mg2+(aq)} & 0.330 \\
\ce{Na4CrO4} & \ce{CrNa4O4 + 8H+ + e-  ->  4H2O + Cr3+(aq) + 4Na+(aq)} & 0.860 & \ce{CrNa4O4 + 5H+ + e-  ->  2H2O + CrHO2 + 4Na+(aq)} & 0.820 \\
\ce{KCrO2} & \ce{CrKO2 + 3/2H2O  ->  3H+ + 3e- + 1/2Cr2K2O7} & 0.929 & \ce{CrKO2 + H+  ->  CrHO2 + K+(aq)} & 0.671 \\
\ce{FeCuO2} & \ce{CuFeO2 + 4H+  ->  2H2O + e- + Cu2+(aq) + Fe3+(aq)} & 0.646 & \ce{CuFeO2 + 4H+  ->  2H2O + e- + Cu2+(aq) + Fe3+(aq)} & 0.401 \\
\ce{Fe2CoO4} & \ce{CoFe2O4 + 8H+  ->  4H2O + Co2+(aq) + 2Fe3+(aq)} & 0.560 & \ce{CoFe2O4 + 5H+  ->  2H2O + e- + CoHO2 + 2Fe3+(aq)} & 0.165 \\
\ce{ZnCr2O4} & \ce{Cr2ZnO4 + 8H+  ->  4H2O + 2Cr3+(aq) + Zn2+(aq)} & 0.500 & \ce{Cr2ZnO4 + 2H+  ->  2CrHO2 + Zn2+(aq)} & 0.214 \\
\ce{CrCuO2} & \ce{CrCuO2 + 4H+  ->  2H2O + e- + Cr3+(aq) + Cu2+(aq)} & 0.641 & \ce{CrCuO2 + H+  ->  e- + CrHO2 + Cu2+(aq)} & 0.389 \\
\ce{CrCuO4} & \ce{CrCuO4 + 8H+ + 3e-  ->  4H2O + Cr3+(aq) + Cu2+(aq)} & 0.071 & \ce{CrCuO4 + 5H+ + 3e-  ->  2H2O + CrHO2 + Cu2+(aq)} & 0.278 \\
\ce{Cs7O} & \ce{Cs7O + 2H+  ->  H2O + 5e- + 7Cs+(aq)} & 3.632 & \ce{Cs7O + 2H+  ->  H2O + 5e- + 7Cs+(aq)} & 3.632 \\
\ce{MgCrO4} & \ce{CrMgO4 + 8H+ + 3e-  ->  4H2O + Cr3+(aq) + Mg2+(aq)} & 0.141 & \ce{CrMgO4 + 5H+ + 3e-  ->  2H2O + CrHO2 + Mg2+(aq)} & 0.383 \\
\ce{K3CrO4} & \ce{CrK3O4 + H+  ->  1/2H2O + e- + 1/2Cr2K2O7 + 2K+(aq)} & 0.643 & \ce{CrK3O4 + 5H+ + 2e-  ->  2H2O + CrHO2 + 3K+(aq)} & 0.712 \\
\ce{KCr3O8} & \ce{Cr3KO8 + 9H+ + 3e-  ->  9/2H2O + 2Cr3+(aq) + 1/2Cr2K2O7} & 0.004 & \ce{Cr3KO8 + 7H+ + 6e-  ->  2H2O + 3CrHO2 + K+(aq)} & 0.240 \\
\ce{CrNiO4} & \ce{CrNiO4 + 8H+ + 3e-  ->  4H2O + Cr3+(aq) + Ni2+(aq)} & 0.130 & \ce{CrNiO4 + 5H+ + 3e-  ->  2H2O + CrHO2 + Ni2+(aq)} & 0.151 \\
\ce{Ca2Fe2O5} & \ce{Ca2Fe2O5 + 10H+  ->  5H2O + 2Ca2+(aq) + 2Fe3+(aq)} & 0.794 & \ce{Ca2Fe2O5 + 10H+  ->  5H2O + 2Ca2+(aq) + 2Fe3+(aq)} & 0.470 \\
\ce{TiCoO3} & \ce{CoTiO3 + 2H+  ->  H2O + Co2+(aq) + TiO2} & 0.385 & \ce{CoTiO3 + H2O  ->  H+ + e- + CoHO2 + TiO2} & 0.154 \\
\ce{K2Cr2O7} & \ce{Cr2K2O7 + 10/3H+ + 2e-  ->  5/3H2O + 2/3Cr3KO8 + 4/3K+(aq)} & -0.020 & \ce{Cr2K2O7 + 8H+ + 6e-  ->  3H2O + 2CrHO2 + 2K+(aq)} & 0.277 \\
\ce{CaCr2O4} & \ce{CaCr2O4 + 8H+  ->  4H2O + Ca2+(aq) + 2Cr3+(aq)} & 0.680 & \ce{CaCr2O4 + 2H+  ->  Ca2+(aq) + 2CrHO2} & 0.408 \\
\ce{Cr2CoO4} & \ce{CoCr2O4 + 8H+  ->  4H2O + Co2+(aq) + 2Cr3+(aq)} & 0.616 & \ce{CoCr2O4 + 2H2O  ->  H+ + e- + CoHO2 + 2CrHO2} & 0.186 \\
\ce{Na2CrO4} & \ce{CrNa2O4 + 8H+ + 3e-  ->  4H2O + Cr3+(aq) + 2Na+(aq)} & 0.127 & \ce{CrNa2O4 + 5H+ + 3e-  ->  2H2O + CrHO2 + 2Na+(aq)} & 0.338 \\
\ce{Ca4Fe9O17} & \ce{Ca4Fe9O17 + 34H+  ->  17H2O + e- + 4Ca2+(aq) + 9Fe3+(aq)} & 0.723 & \ce{Ca4Fe9O17 + 34H+  ->  17H2O + e- + 4Ca2+(aq) + 9Fe3+(aq)} & 0.286 \\
\ce{CuO} & \ce{CuO + 2H+  ->  H2O + Cu2+(aq)} & 0.470 & \ce{CuO + 2H+  ->  H2O + Cu2+(aq)} & 0.629 \\
\ce{K2CrO4} & \ce{CrK2O4 + H+  ->  1/2H2O + 1/2Cr2K2O7 + K+(aq)} & 0.135 & \ce{CrK2O4 + 5H+ + 3e-  ->  2H2O + CrHO2 + 2K+(aq)} & 0.330 \\
\ce{Ca(FeO2)2} & \ce{CaFe2O4 + 8H+  ->  4H2O + Ca2+(aq) + 2Fe3+(aq)} & 0.726 & \ce{CaFe2O4 + 8H+  ->  4H2O + Ca2+(aq) + 2Fe3+(aq)} & 0.281 \\
\ce{Ti2CoO5} & \ce{CoTi2O5 + 2H+  ->  H2O + Co2+(aq) + 2TiO2} & 0.230 & \ce{CoTi2O5 + H2O  ->  H+ + e- + CoHO2 + 2TiO2} & 0.070 \\
\ce{NaCrO2} & \ce{CrNaO2 + 4H+  ->  2H2O + Cr3+(aq) + Na+(aq)} & 0.692 & \ce{CrNaO2 + H+  ->  CrHO2 + Na+(aq)} & 0.453 \\
\ce{CaMoO4} & \ce{CaMoO4 + 2H+  ->  H2O + Ca2+(aq) + MoO3} & 0.141 & \ce{CaMoO4 + 2H+  ->  H2O + Ca2+(aq) + MoO3} & 0.067 \\
\ce{CaFeO3} & \ce{CaFeO3 + 6H+ + e-  ->  3H2O + Ca2+(aq) + Fe3+(aq)} & 0.534 & \ce{CaFeO3 + 6H+ + e-  ->  3H2O + Ca2+(aq) + Fe3+(aq)} & 0.431 \\
\ce{CsO3} & \ce{CsO3 + 6H+ + 5e-  ->  3H2O + Cs+(aq)} & 0.607 & \ce{CsO3 + 6H+ + 5e-  ->  3H2O + Cs+(aq)} & 0.804 \\
\ce{TiFe2O5} & \ce{Fe2TiO5 + 6H+  ->  3H2O + 2Fe3+(aq) + TiO2} & 0.360 & \ce{Fe2TiO5 + 6H+  ->  3H2O + 2Fe3+(aq) + TiO2} & -0.022 \\
\ce{Zn(FeO2)2} & \ce{Fe2ZnO4 + 8H+  ->  4H2O + 2Fe3+(aq) + Zn2+(aq)} & 0.577 & \ce{Fe2ZnO4 + 8H+  ->  4H2O + 2Fe3+(aq) + Zn2+(aq)} & 0.075 \\
\ce{FeO} & \ce{FeO + 2H+  ->  H2O + e- + Fe3+(aq)} & 1.228 & \ce{FeO + 2H+  ->  H2O + e- + Fe3+(aq)} & 0.810 \\
\ce{Ti2FeO5} & \ce{FeTi2O5 + 2H+  ->  H2O + e- + Fe3+(aq) + 2TiO2} & 0.284 & \ce{FeTi2O5 + 1/2H2O  ->  H+ + e- + 1/2Fe2TiO5 + 3/2TiO2} & 0.129 \\
\ce{Li2FeO3} & \ce{FeLi2O3 + 6H+ + e-  ->  3H2O + Fe3+(aq) + 2Li+(aq)} & 0.493 & \ce{FeLi2O3 + 6H+ + e-  ->  3H2O + Fe3+(aq) + 2Li+(aq)} & 0.470 \\
\ce{Na4FeO3} & \ce{FeNa4O3 + 6H+  ->  3H2O + e- + Fe3+(aq) + 4Na+(aq)} & 1.364 & \ce{FeNa4O3 + 6H+  ->  3H2O + e- + Fe3+(aq) + 4Na+(aq)} & 1.191 \\
\ce{Na3FeO3} & \ce{FeNa3O3 + 6H+  ->  3H2O + Fe3+(aq) + 3Na+(aq)} & 1.019 & \ce{FeNa3O3 + 6H+  ->  3H2O + Fe3+(aq) + 3Na+(aq)} & 0.797 \\
\ce{LiFeO2} & \ce{FeLiO2 + 4H+  ->  2H2O + Fe3+(aq) + Li+(aq)} & 0.682 & \ce{FeLiO2 + 4H+  ->  2H2O + Fe3+(aq) + Li+(aq)} & 0.256 \\
\ce{Na4FeO4} & \ce{FeNa4O4 + 8H+ + e-  ->  4H2O + Fe3+(aq) + 4Na+(aq)} & 0.900 & \ce{FeNa4O4 + 8H+ + e-  ->  4H2O + Fe3+(aq) + 4Na+(aq)} & 0.827 \\
\ce{Li5FeO4} & \ce{FeLi5O4 + 8H+(b  ->  4H2O + Fe3+(aq) + 5Li+(aq)} & 0.933 & \ce{FeLi5O4 + 8H+  ->  4H2O + Fe3+(aq) + 5Li+(aq)} & 0.915 \\
\ce{Ga2O3} & \ce{Ga2O3 + 6H+  ->  3H2O + 2Ga3+(aq)} & 0.426 & \ce{Ga2O3 + 6H+  ->  3H2O + 2Ga3+(aq)} & 0.284 \\
\bottomrule
\end{tabular}
\end{adjustbox}
\end{table*}

\newpage
\begin{table}[h]
\captionsetup{justification=centering, singlelinecheck=false}
    \caption*{\textit{Table S1 – Continued from previous page}}
    \centering
    \begin{adjustbox}{width=\textwidth}
    \small
    \begin{tabular}{p{2cm} *{4}{r}}
    \toprule
    \multirow{2}{*}{\textbf{Oxide}} & \multicolumn{2}{c}{PBE} & \multicolumn{2}{c}{HSE06} \\
    \cmidrule(lr){2-3} \cmidrule(lr){4-5}
     & \multicolumn{1}{c}{\textbf{Decomposition Reaction}} & \textbf{$\Delta G_{pbx}^{\mathrm{OER}}$} & \multicolumn{1}{c}{\textbf{Decomposition Reaction}} & \textbf{$\Delta G_{pbx}^{\mathrm{OER}}$} \\
    \midrule
\ce{GeO2} & \ce{GeO2 + 4H+ + 4e-  ->  2H2O + Ge} & -1.593 & \ce{GeO2 + 4H+ + 4e-  ->  2H2O + Ge} & -1.693 \\
\ce{In2O3} & \ce{In2O3 + 6H+  ->  3H2O + 2In3+(aq)} & 0.437 & \ce{In2O3 + 6H+  ->  3H2O + 2In3+(aq)} & 0.347 \\
\ce{IrO2} & \ce{IrO2 + H2O  ->  2H+ + 2e- + IrO3} & -0.176 & \ce{IrO2 + H2O  ->  2H+ + 2e- + IrO3} & -0.612 \\
\ce{KO3} & \ce{KO3 + 6H+ + 5e-  ->  3H2O + K+(aq)} & 0.529 & \ce{KO3 + 6H+ + 5e-  ->  3H2O + K+(aq)} & 0.737 \\
\ce{K2O2} & \ce{KO + 2H+ + e-  ->  H2O + K+(aq)} & 0.599 & \ce{KO + 2H+ + e-  ->  H2O + K+(aq)} & 0.590 \\
\ce{K2O} & \ce{K2O + 2H+  ->  H2O + 2K+(aq)} & 1.993 & \ce{K2O + 2H+  ->  H2O + 2K+(aq)} & 1.988 \\
\ce{KO2} & \ce{KO2 + 4H+ + 3e-  ->  2H2O + K+(aq)} & 0.678 & \ce{KO2 + 4H+ + 3e-  ->  2H2O + K+(aq)} & 0.955 \\
\ce{LiO2} & \ce{LiO2 + 4H+ + 3e-  ->  2H2O + Li+(aq)} & 0.597 & \ce{LiO2 + 4H+ + 3e-  ->  2H2O + Li+(aq)} & 0.747 \\
\ce{Li2O2} & \ce{LiO + 2H+ + e-  ->  H2O + Li+(aq)} & 0.439 & \ce{LiO + 2H+ + e-  ->  H2O + Li+(aq)} & 0.416 \\
\ce{Li2O} & \ce{Li2O + 2H+  ->  H2O + 2Li+(aq)} & 1.229 & \ce{Li2O + 2H+  ->  H2O + 2Li+(aq)} & 1.141 \\
\ce{LiO3} & \ce{LiO3 + 6H+ + 5e-  ->  3H2O + Li+(aq)} & 0.518 & \ce{LiO3 + 6H+ + 5e-  ->  3H2O + Li+(aq)} & 0.684 \\
\ce{MgMo2O7} & \ce{MgMo2O7 + 2H+  ->  H2O + Mg2+(aq) + 2MoO3} & 0.118 & \ce{MgMo2O7 + 2H+  ->  H2O + Mg2+(aq) + 2MoO3} & 0.095 \\
\ce{Mg2V2O7} & \ce{Mg2V2O7 + 4H+  ->  2H2O + 2Mg2+(aq) + V2O5} & 0.271 & \ce{Mg2V2O7 + 4H+  ->  2H2O + 2Mg2+(aq) + V2O5} & 0.209 \\
\ce{Mn2Mo3O8} & \ce{Mn2Mo3O8 + H2O  ->  2H+ + 6e- + 2Mn2+(aq) + 3MoO3} & 0.744 & \ce{Mn2Mo3O8 + 5H2O  ->  10H+ + 10e- + 2MnO2 + 3MoO3} & 0.433 \\
\ce{Na2MoO4} & \ce{MoNa2O4 + 2H+  ->  H2O + MoO3 + 2Na+(aq)} & 0.222 & \ce{MoNa2O4 + 2H+  ->  H2O + MoO3 + 2Na+(aq)} & 0.177 \\
\ce{MgO2} & \ce{MgO2 + 4H+ + 2e-  ->  2H2O + Mg2+(aq)} & 0.811 & \ce{MgO2 + 4H+ + 2e-  ->  2H2O + Mg2+(aq)} & 0.717 \\
\ce{MgMoO4} & \ce{MgMoO4 + 2H+  ->  H2O + Mg2+(aq) + MoO3} & 0.222 & \ce{MgMoO4 + 2H+  ->  H2O + Mg2+(aq) + MoO3} & 0.164 \\
\ce{TiMnO3} & \ce{MnTiO3 + 2H+  ->  H2O + Mn2+(aq) + TiO2} & 0.446 & \ce{MnTiO3 + H2O  ->  2H+ + 2e- + MnO2 + TiO2} & 0.134 \\
\ce{MoO3} & \ce{MoO3 + H2O  ->  H+ + HMoO4-(aq)} & -0.165 & \ce{MoO3 + H2O  ->  H+ + HMoO4-(aq)} & -0.092 \\
\ce{Mn3O4} & \ce{Mn3O4 + 8H+ + 2e-  ->  4H2O + 3Mn2+(aq)} & 0.583 & \ce{Mn3O4 + 2H2O  ->  4H+ + 4e- + 3MnO2} & 0.097 \\
\ce{MnO} & \ce{MnO + 2H+  ->  H2O + Mn2+(aq)} & 1.375 & \ce{MnO + H2O  ->  2H+ + 2e- + MnO2} & 0.608 \\
\ce{Mn2O3} & \ce{Mn2O3 + 6H+ + 2e-  ->  3H2O + 2Mn2+(aq)} & 0.421 & \ce{Mn2O3 + H2O  ->  2H+ + 2e- + 2MnO2} & 0.081 \\
\ce{Na(Mo2O3)2} & \ce{Mo4NaO6 + 6H2O  ->  12H+ + 13e- + 4MoO3 + Na+(aq)} & 1.397 & \ce{Mo4NaO6 + 6H2O  ->  12H+ + 13e- + 4MoO3 + Na+(aq)} & 1.305 \\
\ce{Mg3V2O8} & \ce{Mg3V2O8 + 6H+  ->  3H2O + 3Mg2+(aq) + V2O5} & 0.367 & \ce{Mg3V2O8 + 6H+  ->  3H2O + 3Mg2+(aq) + V2O5} & 0.285 \\
\ce{Na4MoO5} & \ce{MoNa4O5 + 4H+  ->  2H2O + MoO3 + 4Na+(aq)} & 0.574 & \ce{MoNa4O5 + 4H+  ->  2H2O + MoO3 + 4Na+(aq)} & 0.519 \\
\ce{MoO2} & \ce{MoO2 + H2O  ->  2H+ + 2e- + MoO3} & 0.606 & \ce{MoO2 + H2O  ->  2H+ + 2e- + MoO3} & 0.533 \\
\ce{TiMn2O4} & \ce{Mn2TiO4 + 4H+  ->  2H2O + 2Mn2+(aq) + TiO2} & 0.693 & \ce{Mn2TiO4 + 2H2O  ->  4H+ + 4e- + 2MnO2 + TiO2} & 0.244 \\
\ce{MgVO3} & \ce{MgVO3 + H+  ->  1/2H2O + e- + Mg2+(aq) + 1/2V2O5} & 0.562 & \ce{MgVO3 + H+  ->  1/2H2O + e- + Mg2+(aq) + 1/2V2O5} & 0.349 \\
\ce{MnMoO4} & \ce{MnMoO4 + 2H+  ->  H2O + Mn2+(aq) + MoO3} & 0.292 & \ce{MnMoO4 + H2O  ->  2H+ + 2e- + MnO2 + MoO3} & 0.458 \\
\ce{Mg2Mo3O8} & \ce{Mg2Mo3O8 + H2O  ->  2H+ + 6e- + 2Mg2+(aq) + 3MoO3} & 0.680 & \ce{Mg2Mo3O8 + H2O  ->  2H+ + 6e- + 2Mg2+(aq) + 3MoO3} & 0.548 \\
\ce{MgO} & \ce{MgO + 2H+  ->  H2O + Mg2+(aq)} & 1.062 & \ce{MgO + 2H+  ->  H2O + Mg2+(aq)} & 0.898 \\
\ce{MnO2} & \ce{MnO2 + 4H+ + 2e-  ->  2H2O + Mn2+(aq)} & 0.076 & \ce{MnO2 + 4/5H+ + 4/5e-  ->  2/5H2O + 1/5Mn5O8} & -0.050 \\
\ce{SrMoO4} & \ce{MoSrO4 + 2H+  ->  H2O + MoO3 + Sr2+(aq)} & 0.122 & \ce{MoSrO4 + 2H+  ->  H2O + MoO3 + Sr2+(aq)} & 0.054 \\
\ce{NbO} & \ce{NbO + 2H2O  ->  3H+ + 3e- + HNbO3(aq)} & 2.698 & \ce{NbO + 3/2H2O  ->  3H+ + 3e- + 1/2Nb2O5} & 2.627 \\
\ce{NbO2} & \ce{NbO2 + H2O  ->  H+ + e- + HNbO3(aq)} & 0.617 & \ce{NbO2 + 1/2H2O  ->  H+ + e- + 1/2Nb2O5} & 0.511 \\
\ce{NiO} & \ce{NiO + 2H+  ->  H2O + Ni2+(aq)} & 1.108 & \ce{NiO + 2H+  ->  H2O + Ni2+(aq)} & 0.156 \\
\ce{TiNiO3} & \ce{NiTiO3 + 2H+  ->  H2O + Ni2+(aq) + TiO2} & 0.378 & \ce{NiTiO3 + 2H+  ->  H2O + Ni2+(aq) + TiO2} & 0.010 \\
\ce{NaO2} & \ce{NaO2 + 4H+ + 3e-  ->  2H2O + Na+(aq)} & 0.673 & \ce{NaO2 + 4H+ + 3e-  ->  2H2O + Na+(aq)} & 0.866 \\
\ce{Na2O} & \ce{Na2O + 2H+  ->  H2O + 2Na+(aq)} & 1.646 & \ce{Na2O + 2H+  ->  H2O + 2Na+(aq)} & 1.604 \\
\ce{Na3VO4} & \ce{Na3VO4 + 3H+  ->  3/2H2O + 3Na+(aq) + 1/2V2O5} & 0.409 & \ce{Na3VO4 + 3H+  ->  3/2H2O + 3Na+(aq) + 1/2V2O5} & 0.358 \\
\ce{Nb2O5} & \ce{Nb2O5 + H2O  ->  2HNbO3(aq)} & 0.076 & \ce{Nb2O5 + H2O  ->  2HNbO3(aq)} & -0.022 \\
\ce{NaO3} & \ce{NaO3 + 6H+ + 5e-  ->  3H2O + Na+(aq)} & 0.500 & \ce{NaO3 + 6H+ + 5e-  ->  3H2O + Na+(aq)} & 0.684 \\
\ce{Na2O2} & \ce{NaO + 2H+ + e-  ->  H2O + Na+(aq)} & 0.532 & \ce{NaO + 2H+ + e-  ->  H2O + Na+(aq)} & 0.512 \\
\ce{OsO4} & \ce{OsO4 + 4H+ + 4e-  ->  2H2O + OsO2} & -0.446 & \ce{OsO4 + 4H+ + 4e-  ->  2H2O + OsO2} & -0.194 \\
\ce{OsO2} & \ce{OsO2 + 2H2O  ->  4H+ + 4e- + OsO4} & 0.743 & \ce{OsO2 + 2H2O  ->  4H+ + 4e- + OsO4} & 0.324 \\
\ce{PtO2} & \ce{PtO2 + 4/3H+ + 4/3e-  ->  2/3H2O + 1/3Pt3O4} & -0.130 & \ce{PtO2 + 4/3H+ + 4/3e-  ->  2/3H2O + 1/3Pt3O4} & -0.138 \\
\ce{PdO} & \ce{PdO + 2H+  ->  H2O + Pd2+(aq)} & 0.040 & \ce{PdO + 2H+  ->  H2O + Pd2+(aq)} & -0.107 \\
\ce{P2O5} & \ce{O5P2 + 3H2O  ->  2H3PO4(aq)} & 0.560 & \ce{O5P2 + 3H2O  ->  2H3O4P} & 0.584 \\
\ce{Pt3O4} & \ce{Pt3O4 + 2H2O  ->  4H+ + 4e- + 3PtO2} & 0.168 & \ce{Pt3O4 + 2H2O  ->  4H+ + 4e- + 3PtO2} & 0.177 \\
\ce{RbO2} & \ce{RbO2 + 4H+ + 3e-  ->  2H2O + Rb+(aq)} & 0.708 & \ce{RbO2 + 4H+ + 3e-  ->  2H2O + Rb+(aq)} & 0.985 \\
\ce{RbO3} & \ce{RbO3 + 6H+ + 5e-  ->  3H2O + Rb+(aq)} & 0.538 & \ce{RbO3 + 6H+ + 5e-  ->  3H2O + Rb+(aq)} & 0.722 \\
\ce{RuO4} & \ce{RuO4 + 4H+ + 4e-  ->  2H2O + RuO2} & -0.049 & \ce{RuO4 + 4H+ + 4e-  ->  2H2O + RuO2} & 0.269 \\
\ce{RhO2} & \ce{RhO2 + 4H+ + e-  ->  2H2O + Rh3+(aq)} & -0.196 & \ce{RhO2 + 4H+ + e-  ->  2H2O + Rh3+(aq)} & -0.164 \\
\ce{Rb2O2} & \ce{RbO + 2H+ + e-  ->  H2O + Rb+(aq)} & 0.643 & \ce{RbO + 2H+ + e-  ->  H2O + Rb+(aq)} & 0.631 \\
\ce{Rb2O3} & \ce{Rb2O3 + 6H+ + 4e-  ->  3H2O + 2Rb+(aq)} & 0.916 & \ce{Rb2O3 + 6H+ + 4e-  ->  3H2O + 2Rb+(aq)} & 1.086 \\
\ce{ReO2} & \ce{ReO2 + H2O  ->  2H+ + 2e- + ReO3} & 0.693 & \ce{ReO2 + 2H2O  ->  4H+ + 3e- + ReO4-(aq)} & 0.700 \\
\ce{Rb2O} & \ce{Rb2O + 2H+  ->  H2O + 2Rb+(aq)} & 2.163 & \ce{Rb2O + 2H+  ->  H2O + 2Rb+(aq)} & 2.150 \\
\ce{RuO2} & \ce{RuO2 + 2H2O  ->  4H+ + 4e- + RuO4} & 0.082 & \ce{RuO2 + 4H+ + e-  ->  2H2O + Ru3+(aq)} & -0.266 \\
\ce{Rb9O2} & \ce{Rb9O2 + 4H+  ->  2H2O + 5e- + 9Rb+(aq)} & 3.191 & \ce{Rb9O2 + 4H+  ->  2H2O + 5e- + 9Rb+(aq)} & 3.198 \\
\ce{ReO3} & \ce{ReO3 + H2O  ->  2H+ + e- + ReO4-(aq)} & -0.009 & \ce{ReO3 + H2O  ->  2H+ + e- + ReO4-(aq)} & 0.124 \\
\ce{SiO2} & \ce{O2Si  ->  SiO2(aq)} & 0.261 & \ce{O2Si  ->  SiO2(aq)} & 0.098 \\
\ce{SnO} & \ce{SnO + H2O  ->  2H+ + 2e- + SnO2} & 1.135 & \ce{SnO + H2O  ->  2H+ + 2e- + SnO2} & 1.269 \\
\ce{SrO2} & \ce{SrO2 + 4H+ + 2e-  ->  2H2O + Sr2+(aq)} & 0.978 & \ce{SrO2 + 4H+ + 2e-  ->  2H2O + Sr2+(aq)} & 0.898 \\
\ce{Sc2O3} & \ce{Sc2O3 + 6H+  ->  3H2O + 2Sc3+(aq)} & 0.472 & \ce{Sc2O3 + 6H+  ->  3H2O + 2Sc3+(aq)} & 0.213 \\
\ce{SrO} & \ce{SrO + 2H+  ->  H2O + Sr2+(aq)} & 1.570 & \ce{SrO + 2H+  ->  H2O + Sr2+(aq)} & 1.455 \\
\ce{SbO2} & \ce{O2Sb + 1/2H2O  ->  H+ + e- + 1/2O5Sb2} & 0.049 & \ce{O2Sb + 1/2H2O  ->  H+ + e- + 1/2O5Sb2} & 0.065 \\
\ce{Sb2O5} & \ce{O5Sb2 + 2H+ + 2e-  ->  H2O + 2O2Sb} & -0.042 & \ce{O5Sb2 + 2H+ + 2e-  ->  H2O + 2O2Sb} & -0.055 \\
\ce{Sb2O3} & \ce{O3Sb2 + 2H2O  ->  4H+ + 4e- + O5Sb2} & 0.336 & \ce{O3Sb2 + 2H2O  ->  4H+ + 4e- + O5Sb2} & 0.389 \\
\ce{SeO2} & \ce{O2Se + 6H2O  ->  2H+ + 2e- + H10O8Se} & 1.182 & \ce{O2Se + 6H2O  ->  2H+ + 2e- + H10O8Se} & 1.392 \\
\ce{SnO2} & \ce{SnO2 + 4H+ + 2e-  ->  2H2O + Sn2+(aq)} & -0.577 & \ce{SnO2 + 4H+ + 2e-  ->  2H2O + Sn2+(aq)} & -0.652 \\
\ce{TiO2} & \ce{TiO2 + 2H+  ->  H2O + TiO2+(aq)} & -0.101 & \ce{TiO2 + 1/4H+ + 1/4e-  ->  1/8H2O + 1/8Ti8O15} & -0.183 \\

    \bottomrule
    \end{tabular}
    \end{adjustbox}
    
\end{table}

\begin{table*}[h]
\captionsetup{justification=centering, singlelinecheck=false}
    \caption*{\textit{Table S1 – Continued from previous page}}
    \centering
    \begin{adjustbox}{width=\textwidth}
    \small
    \begin{tabular}{p{2cm} *{4}{r}}
    \toprule
    \multirow{2}{*}{\textbf{Oxide}} & \multicolumn{2}{c}{PBE} & \multicolumn{2}{c}{HSE06} \\
    \cmidrule(lr){2-3} \cmidrule(lr){4-5}
     & \multicolumn{1}{c}{\textbf{Decomposition Reaction}} & \textbf{$\Delta G_{pbx}^{\mathrm{OER}}$} & \multicolumn{1}{c}{\textbf{Decomposition Reaction}} & \textbf{$\Delta G_{pbx}^{\mathrm{OER}}$} \\
    \midrule
\ce{TeO2} & \ce{O2Te + 1/4H2O  ->  1/2H+ + 1/2e- + 1/4O9Te4} & -0.002 & \ce{O2Te + 1/4H2O  ->  1/2H+ + 1/2e- + 1/4O9Te4} & 0.007 \\
\ce{Ti3O5} & \ce{Ti3O5 + H2O  ->  2H+ + 2e- + 3TiO2} & 0.461 & \ce{Ti3O5 + H2O  ->  2H+ + 2e- + 3TiO2} & 0.525 \\
\ce{TeO3} & \ce{O3Te + 2H+ + 2e-  ->  H2O + O2Te} & 0.045 & \ce{O3Te + 3/2H+ + 3/2e-  ->  3/4H2O + 1/4O9Te4} & 0.037 \\
\ce{Ti2O3} & \ce{Ti2O3 + H2O  ->  2H+ + 2e- + 2TiO2} & 0.740 & \ce{Ti2O3 + H2O  ->  2H+ + 2e- + 2TiO2} & 0.824 \\
\ce{Ta2O5} & \ce{Ta2O5 + 10H+ + 10e-  ->  5H2O + 2Ta} & -2.795 & \ce{Ta2O5 + 10H+ + 10e-  ->  5H2O + 2Ta} & -2.925 \\
\ce{Te2O5} & \ce{O5Te2 + 2H+ + 2e-  ->  H2O + 2O2Te} & 0.004 & \ce{O5Te2 + H+ + e-  ->  1/2H2O + 1/2O9Te4} & 0.002 \\
\ce{Ti6O} & \ce{Ti6O + 11H2O  ->  22H+ + 22e- + 6TiO2} & 6.978 & \ce{Ti6O + 11H2O  ->  22H+ + 22e- + 6TiO2} & 7.387 \\
\ce{TiO} & \ce{TiO + H2O  ->  2H+ + 2e- + TiO2} & 2.022 & \ce{TiO + H2O  ->  2H+ + 2e- + TiO2} & 2.181 \\
\ce{Ti2O} & \ce{Ti2O + 3H2O  ->  6H+ + 6e- + 2TiO2} & 4.245 & \ce{Ti2O + 3H2O  ->  6H+ + 6e- + 2TiO2} & 4.501 \\
\ce{Ti3O} & \ce{Ti3O + 5H2O  ->  10H+ + 10e- + 3TiO2} & 5.417 & \ce{Ti3O + 5H2O  ->  10H+ + 10e- + 3TiO2} & 5.732 \\
\ce{VO} & \ce{VO + 3/2H2O  ->  3H+ + 3e- + 1/2V2O5} & 2.524 & \ce{VO + 3/2H2O  ->  3H+ + 3e- + 1/2V2O5} & 1.767 \\
\ce{V2O5} & \ce{V2O5 + 2H+  ->  H2O + 2VO2+(aq)} & -0.080 & \ce{V2O5 + 4/3H+ + 4/3e-  ->  2/3H2O + 1/3V6O13} & -0.093 \\
\ce{WO3} & \ce{WO3 + H2O  ->  H+ + HWO4-(aq)} & -0.214 & \ce{WO3 + H2O  ->  H+ + HWO4-(aq)} & -0.180 \\
\ce{WO2} & \ce{WO2 + H2O  ->  2H+ + 2e- + WO3} & 0.930 & \ce{WO2 + H2O  ->  2H+ + 2e- + WO3} & 0.887 \\
\ce{Y2O3} & \ce{Y2O3 + 6H+  ->  3H2O + 2Y3+(aq)} & 0.856 & \ce{Y2O3 + 6H+  ->  3H2O + 2Y3+(aq)} & 0.646 \\
\ce{ZrO2} & \ce{ZrO2 + 3H+  ->  H2O + ZrOH3+(aq)} & 0.264 & \ce{ZrO2 + 3H+  ->  H2O + ZrOH3+(aq)} & 0.054 \\
\ce{ZnO} & \ce{ZnO + 2H+  ->  H2O + Zn2+(aq)} & 0.698 & \ce{ZnO + 2H+  ->  H2O + Zn2+(aq)} & 0.560 \\
\ce{Zr3O} & \ce{Zr3O + 2H2O  ->  H+ + 10e- + 3ZrOH3+(aq)} & 6.900 & \ce{Zr3O + 2H2O  ->  H+ + 10e- + 3ZrOH3+(aq)} & 6.808 \\
\ce{Ag2O3} & \ce{Ag2O3 + 6H+ + 4e-  ->  3H2O + 2Ag+(aq)} & 0.337 & \ce{Ag2O3 + 6H+ + 4e-  ->  3H2O + 2Ag+(aq)} & 0.583 \\
\ce{Ag3O} & \ce{Ag3O + 2H+  ->  H2O + e- + 3Ag+(aq)} & 0.522 & \ce{Ag3O + 2H+  ->  H2O + e- + 3Ag+(aq)} & 0.560 \\
\ce{AsO2} & \ce{AsO2 + 2H2O  ->  H+ + e- + H3AsO4(aq)} & 0.423 & \ce{AsO2 + 4/3H2O  ->  H+ + e- + 1/3As3H5O10} & 0.391 \\
\ce{Ba2V2O7} & \ce{Ba2V2O7 + 4H+  ->  2H2O + 2Ba2+(aq) + V2O5} & 0.172 & \ce{Ba2V2O7 + 4H+  ->  2H2O + 2Ba2+(aq) + V2O5} & 0.107 \\
\ce{BaO10} & \ce{BaO10 + 20H+ + 18e-  ->  10H2O + Ba2+(aq)} & 0.367 & \ce{BaO10 + 20H+ + 18e-  ->  10H2O + Ba2+(aq)} & 0.755 \\
\ce{Ba3VO5} & \ce{Ba3VO5 + 5H+  ->  5/2H2O + e- + 3Ba2+(aq) + 1/2V2O5} & 1.069 & \ce{Ba3VO5 + 5H+  ->  5/2H2O + e- + 3Ba2+(aq) + 1/2V2O5} & 0.906 \\
\ce{Ba2VO4} & \ce{Ba2VO4 + 3H+  ->  3/2H2O + e- + 2Ba2+(aq) + 1/2V2O5} & 0.862 & \ce{Ba2VO4 + 3H+  ->  3/2H2O + e- + 2Ba2+(aq) + 1/2V2O5} & 0.687 \\
\ce{BaMoO3} & \ce{BaMoO3  ->  2e- + Ba2+(aq) + MoO3} & 0.830 & \ce{BaMoO3  ->  2e- + Ba2+(aq) + MoO3} & 0.765 \\
\ce{BaV3O8} & \ce{BaV3O8 + H+  ->  1/2H2O + e- + Ba2+(aq) + 3/2V2O5} & 0.136 & \ce{BaV3O8 + H+  ->  1/2H2O + e- + Ba2+(aq) + 3/2V2O5} & 0.048 \\
\ce{CaV2O6} & \ce{CaV2O6 + 2H+  ->  H2O + Ca2+(aq) + V2O5} & 0.108 & \ce{CaV2O6 + 2H+  ->  H2O + Ca2+(aq) + V2O5} & 0.071 \\
\ce{CrO} & \ce{CrO + 2H+  ->  H2O + e- + Cr3+(aq)} & 1.565 & \ce{CrO + H2O  ->  H+ + e- + CrHO2} & 1.182 \\
\ce{Ca(MoO2)2} & \ce{CaMo2O4 + 2H2O  ->  4H+ + 6e- + Ca2+(aq) + 2MoO3} & 1.616 & \ce{CaMo2O4 + 2H2O  ->  4H+ + 6e- + Ca2+(aq) + 2MoO3} & 1.221 \\
\ce{Cs2Mo5O16} & \ce{Cs2Mo5O16 + 2H+  ->  H2O + 2Cs+(aq) + 5MoO3} & 0.052 & \ce{Cs2Mo5O16 + 2H+  ->  H2O + 2Cs+(aq) + 5MoO3} & 0.031 \\
\ce{ZnCrO4} & \ce{CrZnO4 + 8H+ + 3e-  ->  4H2O + Cr3+(aq) + Zn2+(aq)} & 0.085 & \ce{CrZnO4 + 5H+ + 3e-  ->  2H2O + CrHO2 + Zn2+(aq)} & 0.331 \\
\ce{Cu2O3} & \ce{Cu2O3 + 6H+ + 2e-  ->  3H2O + 2Cu2+(aq)} & 0.454 & \ce{Cu2O3 + 6H+ + 2e-  ->  3H2O + 2Cu2+(aq)} & 0.587 \\
\ce{Ca8V6O23} & \ce{Ca8V6O23 + 16H+  ->  8H2O + 8Ca2+(aq) + 3V2O5} & 0.436 & \ce{Ca8V6O23 + 16H+  ->  8H2O + 8Ca2+(aq) + 3V2O5} & 0.386 \\
\ce{CaVO3} & \ce{CaVO3 + H+  ->  1/2H2O + e- + Ca2+(aq) + 1/2V2O5} & 0.573 & \ce{CaVO3 + H+  ->  1/2H2O + e- + Ca2+(aq) + 1/2V2O5} & 0.496 \\
\ce{Cs2Mo4O13} & \ce{Cs2Mo4O13 + 2H+  ->  H2O + 2Cs+(aq) + 4MoO3} & 0.063 & \ce{Cs2Mo4O13 + 2H+  ->  H2O + 2Cs+(aq) + 4MoO3} & 0.039 \\
\ce{CaV2O4} & \ce{CaV2O4 + H2O  ->  2H+ + 4e- + Ca2+(aq) + V2O5} & 1.059 & \ce{CaV2O4 + H2O  ->  2H+ + 4e- + Ca2+(aq) + V2O5} & 0.786 \\
\ce{Na2Cr2O7} & \ce{Cr2Na2O7 + 14H+ + 6e-  ->  7H2O + 2Cr3+(aq) + 2Na+(aq)} & 0.005 & \ce{Cr2Na2O7 + 8H+ + 6e-  ->  3H2O + 2CrHO2 + 2Na+(aq)} & 0.304 \\
\ce{CoO2} & \ce{CoO2 + 4H+ + 2e-  ->  2H2O + Co2+(aq)} & 0.216 & \ce{CoO2 + H+ + e-  ->  CoHO2} & 0.468 \\
\ce{CaV2O5} & \ce{CaV2O5  ->  2e- + Ca2+(aq) + V2O5} & 0.435 & \ce{CaV2O5  ->  2e- + Ca2+(aq) + V2O5} & 0.330 \\
\ce{Ca3Fe2O7} & \ce{Ca3Fe2O7 + 14H+ + 2e-  ->  7H2O + 3Ca2+(aq) + 2Fe3+(aq)} & 0.662 & \ce{Ca3Fe2O7 + 14H+ + 2e-  ->  7H2O + 3Ca2+(aq) + 2Fe3+(aq)} & 0.557 \\
\ce{Ca2V3O8} & \ce{Ca2V3O8 + H+  ->  1/2H2O + 3e- + 2Ca2+(aq) + 3/2V2O5} & 0.506 & \ce{Ca2V3O8 + H+  ->  1/2H2O + 3e- + 2Ca2+(aq) + 3/2V2O5} & 0.296 \\
\ce{CaV3O7} & \ce{CaV3O7 + 1/2H2O  ->  H+ + 3e- + Ca2+(aq) + 3/2V2O5} & 0.392 & \ce{CaV3O7 + 1/2H2O  ->  H+ + 3e- + Ca2+(aq) + 3/2V2O5} & 0.156 \\
\ce{Cr6O11} & \ce{Cr6O11 + 22H+ + 4e-  ->  11H2O + 6Cr3+(aq)} & 0.247 & \ce{Cr6O11 + H2O + 4H+ + 4e-  ->  6CrHO2} & 0.257 \\
\ce{Ca10V6O25} & \ce{Ca10V6O25 + 20H+  ->  10H2O + 10Ca2+(aq) + 3V2O5} & 0.377 & \ce{Ca10V6O25 + 20H+  ->  10H2O + 10Ca2+(aq) + 3V2O5} & 0.300 \\
\ce{Cs2MoO4} & \ce{Cs2MoO4 + 2H+  ->  H2O + 2Cs+(aq) + MoO3} & 0.309 & \ce{Cs2MoO4 + 2H+  ->  H2O + 2Cs+(aq) + MoO3} & 0.274 \\
\ce{K4CrO4} & \ce{CrK4O4 + H+  ->  1/2H2O + 2e- + 1/2Cr2K2O7 + 3K+(aq)} & 1.027 & \ce{CrK4O4 + 5H+ + e-  ->  2H2O + CrHO2 + 4K+(aq)} & 0.992 \\
\ce{Cu4O3} & \ce{Cu4O3 + 6H+  ->  3H2O + 2e- + 4Cu2+(aq)} & 0.712 & \ce{Cu4O3 + 6H+  ->  3H2O + 2e- + 4Cu2+(aq)} & 0.771 \\
\ce{Na3CrO4} & \ce{CrNa3O4 + 8H+ + 2e-  ->  4H2O + Cr3+(aq) + 3Na+(aq)} & 0.502 & \ce{CrNa3O4 + 5H+ + 2e-  ->  2H2O + CrHO2 + 3Na+(aq)} & 0.644 \\
\ce{Li3FeO4} & \ce{FeLi3O4 + 8H+ + 2e-  ->  4H2O + Fe3+(aq) + 3Li+(aq)} & 0.527 & \ce{FeLi3O4 + 8H+ + 2e-  ->  4H2O + Fe3+(aq) + 3Li+(aq)} & 0.541 \\
\ce{ZnFe5O8} & \ce{Fe5ZnO8 + 16H+  ->  8H2O + e- + 5Fe3+(aq) + Zn2+(aq)} & 0.640 & \ce{Fe5ZnO8 + 16H+  ->  8H2O + e- + 5Fe3+(aq) + Zn2+(aq)} & 0.111 \\
\ce{Li2FeO2} & \ce{FeLi2O2 + 4H+  ->  2H2O + e- + Fe3+(aq) + 2Li+(aq)} & 1.157 & \ce{FeLi2O2 + 4H+  ->  2H2O + e- + Fe3+(aq) + 2Li+(aq)} & 0.818 \\
\ce{Fe2O3} & \ce{Fe2O3 + 6H+  ->  3H2O + 2Fe3+(aq)} & 0.473 & \ce{Fe2O3 + 6H+  ->  3H2O + 2Fe3+(aq)} & 0.240 \\
\ce{Na7Fe3O8} & \ce{Fe3Na7O8 + 16H+  ->  8H2O + 3Fe3+(aq) + 7Na+(aq)} & 0.947 & \ce{Fe3Na7O8 + 16H+  ->  8H2O + 3Fe3+(aq) + 7Na+(aq)} & 0.695 \\
\ce{Na2FeO3} & \ce{FeNa2O3 + 6H+ + e-  ->  3H2O + Fe3+(aq) + 2Na+(aq)} & 0.649 & \ce{FeNa2O3 + 6H+ + e-  ->  3H2O + Fe3+(aq) + 2Na+(aq)} & 0.552 \\
\ce{NaFeO2} & \ce{FeNaO2 + 4H+  ->  2H2O + Fe3+(aq) + Na+(aq)} & 0.746 & \ce{FeNaO2 + 4H+  ->  2H2O + Fe3+(aq) + Na+(aq)} & 0.320 \\
\ce{TiFeO3} & \ce{FeTiO3 + 2H+  ->  H2O + e- + Fe3+(aq) + TiO2} & 0.451 & \ce{FeTiO3 + 1/2H2O  ->  H+ + e- + 1/2Fe2TiO5 + 1/2TiO2} & 0.262 \\
\ce{Na4FeO5} & \ce{FeNa4O5 + 10H+ + 3e-  ->  5H2O + Fe3+(aq) + 4Na+(aq)} & 0.767 & \ce{FeNa4O5 + 10H+ + 3e-  ->  5H2O + Fe3+(aq) + 4Na+(aq)} & 0.823 \\
\ce{Na3FeO4} & \ce{FeNa3O4 + 8H+ + 2e-  ->  4H2O + Fe3+(aq) + 3Na+(aq)} & 0.671 & \ce{FeNa3O4 + 8H+ + 2e-  ->  4H2O + Fe3+(aq) + 3Na+(aq)} & 0.694 \\
\ce{Na8FeO6} & \ce{FeNa8O6 + 12H+ + e-  ->  6H2O + Fe3+(aq) + 8Na+(aq)} & 1.226 & \ce{FeNa8O6 + 12H+ + e-  ->  6H2O + Fe3+(aq) + 8Na+(aq)} & 1.168 \\
\ce{Fe3O4} & \ce{Fe3O4 + 8H+  ->  4H2O + e- + 3Fe3+(aq)} & 0.700 & \ce{Fe3O4 + 8H+  ->  4H2O + e- + 3Fe3+(aq)} & 0.118 \\
\ce{IrO3} & \ce{IrO3 + 2H+ + 2e-  ->  H2O + IrO2} & 0.132 & \ce{IrO3 + 2H+ + 2e-  ->  H2O + IrO2} & 0.459 \\
\ce{LiO8} & \ce{LiO8 + 16H+ + 15e-  ->  8H2O + Li+(aq)} & 0.205 & \ce{LiO8 + 16H+ + 15e-  ->  8H2O + Li+(aq)} & 0.378 \\
\ce{NaMo6O17} & \ce{Mo6NaO17 + H2O  ->  2H+ + 3e- + 6MoO3 + Na+(aq)} & 0.089 & \ce{Mo6NaO17 + H2O  ->  2H+ + 3e- + 6MoO3 + Na+(aq)} & 0.091 \\
\ce{SrMoO3} & \ce{MoSrO3  ->  2e- + MoO3 + Sr2+(aq)} & 0.830 & \ce{MoSrO3  ->  2e- + MoO3 + Sr2+(aq)} & 0.761 \\
\ce{MgV3O8} & \ce{MgV3O8 + H+  ->  1/2H2O + e- + Mg2+(aq) + 3/2V2O5} & 0.191 & \ce{MgV3O8 + H+  ->  1/2H2O + e- + Mg2+(aq) + 3/2V2O5} & 0.112 \\
\ce{Na3MoO4} & \ce{MoNa3O4 + 2H+  ->  H2O + e- + MoO3 + 3Na+(aq)} & 0.712 & \ce{MoNa3O4 + 2H+  ->  H2O + e- + MoO3 + 3Na+(aq)} & 0.629 \\
\ce{MgV2O6} & \ce{MgV2O6 + 2H+  ->  H2O + Mg2+(aq) + V2O5} & 0.187 & \ce{MgV2O6 + 2H+  ->  H2O + Mg2+(aq) + V2O5} & 0.145 \\
\ce{Mn5O8} & \ce{Mn5O8 + 16H+ + 6e-  ->  8H2O + 5Mn2+(aq)} & 0.342 & \ce{Mn5O8 + 2H2O  ->  4H+ + 4e- + 5MnO2} & 0.058 \\
\ce{Sr2MoO4} & \ce{MoSr2O4 + 2H+  ->  H2O + 2e- + MoO3 + 2Sr2+(aq)} & 1.039 & \ce{MoSr2O4 + 2H+  ->  H2O + 2e- + MoO3 + 2Sr2+(aq)} & 0.962 \\
\ce{MgV4O10} & \ce{MgV4O10  ->  2e- + Mg2+(aq) + 2V2O5} & 0.238 & \ce{MgV4O10  ->  2e- + Mg2+(aq) + 2V2O5} & 0.134 \\
\ce{NaMoO2} & \ce{MoNaO2 + H2O  ->  2H+ + 3e- + MoO3 + Na+(aq)} & 1.516 & \ce{MoNaO2 + H2O  ->  2H+ + 3e- + MoO3 + Na+(aq)} & 1.171 \\
\ce{Na2Mo2O7} & \ce{Mo2Na2O7 + 2H+  ->  H2O + 2MoO3 + 2Na+(aq)} & 0.095 & \ce{Mo2Na2O7 + 2H+  ->  H2O + 2MoO3 + 2Na+(aq)} & 0.064 \\
\ce{NaV2O5} & \ce{NaV2O5  ->  e- + Na+(aq) + V2O5} & 0.166 & \ce{NaV2O5  ->  e- + Na+(aq) + V2O5} & 0.102 \\
\ce{NaVO3} & \ce{NaVO3 + H+  ->  1/2H2O + Na+(aq) + 1/2V2O5} & 0.124 & \ce{NaVO3 + H+  ->  1/2H2O + Na+(aq) + 1/2V2O5} & 0.090 \\
\ce{Ti4(Ni5O8)3} & \ce{Ni15Ti4O24 + 32H+ + 2e-  ->  16H2O + 15Ni2+(aq) + 4TiO2} & 0.666 & \ce{Ni15Ti4O24 + 32H+ + 2e-  ->  16H2O + 15Ni2+(aq) + 4TiO2} & 0.062 \\
\ce{NaVO2} & \ce{NaVO2 + 1/2H2O  ->  H+ + 2e- + Na+(aq) + 1/2V2O5} & 0.982 & \ce{NaVO2 + 1/2H2O  ->  H+ + 2e- + Na+(aq) + 1/2V2O5} & 0.729 \\
\ce{Nb12O29} & \ce{Nb12O29 + 7H2O  ->  2H+ + 2e- + 12HNbO3(aq)} & 0.133 & \ce{Nb12O29 + H2O  ->  2H+ + 2e- + 6Nb2O5} & 0.071 \\
\ce{NaV2O4} & \ce{NaV2O4 + H2O  ->  2H+ + 3e- + Na+(aq) + V2O5} & 0.609 & \ce{NaV2O4 + H2O  ->  2H+ + 3e- + Na+(aq) + V2O5} & 0.385 \\

    \bottomrule
    \end{tabular}
    \end{adjustbox}
    
\end{table*}

\begin{table}[h]
    \caption*{\textit{Table S1 – Continued from previous page}}
    \centering
    \begin{adjustbox}{width=1\textwidth}
    \small
    \begin{tabular}{p{2cm} *{4}{r}}
    \toprule
    \multirow{2}{*}{\textbf{Oxide}} & \multicolumn{2}{c}{PBE} & \multicolumn{2}{c}{HSE06} \\
    \cmidrule(lr){2-3} \cmidrule(lr){4-5}
     & \multicolumn{1}{c}{\textbf{Decomposition Reaction}} & \textbf{$\mathrm{\Delta G_{pbx}}$} & \multicolumn{1}{c}{\textbf{Decomposition Reaction}} & \textbf{$\mathrm{\Delta G_{pbx}}$} \\
    \midrule

\ce{NaV3O8} & \ce{NaV3O8 + H+  ->  1/2H2O + Na+(aq) + 3/2V2O5} & 0.034 & \ce{NaV3O8 + H+  ->  1/2H2O + Na+(aq) + 3/2V2O5} & 0.020 \\     
\ce{Ni3O4} & \ce{Ni3O4 + 8H+ + 2e-  ->  4H2O + 3Ni2+(aq)} & 0.705 & \ce{Ni3O4 + 8H+ + 2e-  ->  4H2O + 3Ni2+(aq)} & 0.172 \\
\ce{Na5VO5} & \ce{Na5VO5 + 5H+  ->  5/2H2O + 5Na+(aq) + 1/2V2O5} & 0.742 & \ce{Na5VO5 + 5H+  ->  5/2H2O + 5Na+(aq) + 1/2V2O5} & 0.695 \\
\ce{Na2O9} & \ce{Na2O9 + 18H+ + 16e-  ->  9H2O + 2Na+(aq)} & 0.356 & \ce{Na2O9 + 18H+ + 16e-  ->  9H2O + 2Na+(aq)} & 0.656 \\
\ce{Na4VO4} & \ce{Na4VO4 + 3H+  ->  3/2H2O + e- + 4Na+(aq) + 1/2V2O5} & 0.897 & \ce{Na4VO4 + 3H+  ->  3/2H2O + e- + 4Na+(aq) + 1/2V2O5} & 0.776 \\
\ce{Ti6O11} & \ce{Ti6O11 + H2O  ->  2H+ + 2e- + 6TiO2} & 0.214 & \ce{Ti6O11 + H2O  ->  2H+ + 2e- + 6TiO2} & 0.255 \\
\ce{Ti8O13} & \ce{Ti8O13 + 3H2O  ->  6H+ + 6e- + 8TiO2} & 0.541 & \ce{Ti8O13 + 3H2O  ->  6H+ + 6e- + 8TiO2} & 0.623 \\
\ce{Ti5O8} & \ce{Ti5O8 + 2H2O  ->  4H+ + 4e- + 5TiO2} & 0.577 & \ce{Ti5O8 + 2H2O  ->  4H+ + 4e- + 5TiO2} & 0.656 \\
\ce{Ti4O7} & \ce{Ti4O7 + H2O  ->  2H+ + 2e- + 4TiO2} & 0.330 & \ce{Ti4O7 + H2O  ->  2H+ + 2e- + 4TiO2} & 0.389 \\
\ce{V5O9} & \ce{V5O9 + 7/2H2O  ->  7H+ + 7e- + 5/2V2O5} & 0.489 & \ce{V5O9 + 7/2H2O  ->  7H+ + 7e- + 5/2V2O5} & 0.255 \\
\ce{Sb6O13} & \ce{O13Sb6 + 2H2O  ->  4H+ + 4e- + 3O5Sb2} & 0.060 & \ce{O13Sb6 + 2H2O  ->  4H+ + 4e- + 3O5Sb2} & 0.081 \\
\ce{Te4O9} & \ce{O9Te4 + 2H+ + 2e-  ->  H2O + 4O2Te} & 0.002 & \ce{O9Te4 + H2O  ->  2H+ + 2e- + 2O5Te2} & -0.003 \\
\ce{V6O13} & \ce{V6O13 + 2H2O  ->  4H+ + 4e- + 3V2O5} & 0.172 & \ce{V6O13 + 2H2O  ->  4H+ + 4e- + 3V2O5} & 0.103 \\
\ce{V7O13} & \ce{V7O13 + 9/2H2O  ->  9H+ + 9e- + 7/2V2O5} & 0.431 & \ce{V7O13 + 9/2H2O  ->  9H+ + 9e- + 7/2V2O5} & 0.237 \\
\ce{Zr4O} & \ce{Zr4O + 3H2O  ->  2H+ + 14e- + 4ZrOH3+(aq)} & 7.729 & \ce{Zr4O + 3H2O  ->  2H+ + 14e- + 4ZrOH3+(aq)} & 7.653 \\
\ce{PdO2} & \ce{PdO2 + 4H+ + 2e-  ->  2H2O + Pd2+(aq)} & 0.047 & \ce{PdO2 + 2H+ + 2e-  ->  H2O + PdO} & 0.220 \\
\ce{VO2} & \ce{VO2 + 1/2H2O  ->  H+ + e- + 1/2V2O5} & 0.321 & \ce{VO2 + 1/2H2O  ->  H+ + e- + 1/2V2O5} & 0.139 \\
\ce{Se2O5} & \ce{O5Se2 + 11H2O  ->  2H+ + 2e- + 2H10O8Se} & 1.077 & \ce{O5Se2 + 11H2O  ->  2H+ + 2e- + 2H10O8Se} & 1.276 \\
\ce{Ti11O18} & \ce{Ti11O18 + 4H2O  ->  8H+ + 8e- + 11TiO2} & 0.523 & \ce{Ti11O18 + 4H2O  ->  8H+ + 8e- + 11TiO2} & 0.600 \\
\ce{PO2} & \ce{O2P + 2H2O  ->  H+ + e- + H3PO4(aq)} & 1.091 & \ce{O2P + 2H2O  ->  H+ + e- + H3O4P} & 1.163 \\
\ce{Zr2O} & \ce{Zr2O + H2O  ->  6e- + 2ZrOH3+(aq)} & 5.578 & \ce{Zr2O + H2O  ->  6e- + 2ZrOH3+(aq)} & 5.471 \\
\ce{V3O5} & \ce{V3O5 + 5/2H2O  ->  5H+ + 5e- + 3/2V2O5} & 0.685 & \ce{V3O5 + 5/2H2O  ->  5H+ + 5e- + 3/2V2O5} & 0.334 \\
\ce{Ti8O15} & \ce{Ti8O15 + H2O  ->  2H+ + 2e- + 8TiO2} & 0.161 & \ce{Ti8O15 + H2O  ->  2H+ + 2e- + 8TiO2} & 0.191 \\
\ce{Sn5O6} & \ce{Sn5O6 + 4H2O  ->  8H+ + 8e- + 5SnO2} & 0.830 & \ce{Sn5O6 + 4H2O  ->  8H+ + 8e- + 5SnO2} & 0.925 \\
\ce{Ti7O13} & \ce{Ti7O13 + H2O  ->  2H+ + 2e- + 7TiO2} & 0.184 & \ce{Ti7O13 + H2O  ->  2H+ + 2e- + 7TiO2} & 0.218 \\
\ce{Ti5O9} & \ce{Ti5O9 + H2O  ->  2H+ + 2e- + 5TiO2} & 0.259 & \ce{Ti5O9 + H2O  ->  2H+ + 2e- + 5TiO2} & 0.308 \\
\ce{W3O8} & \ce{W3O8 + H2O  ->  2H+ + 2e- + 3WO3} & 0.287 & \ce{W3O8 + H2O  ->  2H+ + 2e- + 3WO3} & 0.272 \\
\ce{Rb6O} & \ce{Rb6O + 2H+  ->  H2O + 4e- + 6Rb+(aq)} & 3.477 & \ce{Rb6O + 2H+  ->  H2O + 4e- + 6Rb+(aq)} & 3.482 \\
\ce{Ti19O30} & \ce{Ti19O30 + 8H2O  ->  16H+ + 16e- + 19TiO2} & 0.612 & \ce{Ti19O30 + 8H2O  ->  16H+ + 16e- + 19TiO2} & 0.692 \\

   \bottomrule
    \end{tabular}
    \end{adjustbox}
    
\end{table}

\begin{table*}[h!]
\captionsetup{justification=raggedright, singlelinecheck=false}
\caption*{\textbf{Table S2.} List of primary features considered for SISSO models. The primary features computed as the composition average are denoted in angled brackets. The primary features obtained from atomicfeaturespackage \cite{afp} are calculated with FHI-aims using HSE06 functional.} 
\begin{tabular}{lrr}
\toprule
\centering
 Primary Feature  & Symbol & Source \\

 \midrule
 Radius of s orbital & $\langle R_S \rangle$ & \textit{atomicfeaturespackage  (DFT-HSE06)} \\
 Radius of valence orbital & $\langle R_{VAL} \rangle$ & \textit{atomicfeaturespackage  (DFT-HSE06)} \\
 Kohn-Sham eigenvalue of the highest occupied orbital & $\langle E_{H} \rangle$ & \textit{atomicfeaturespackage  (DFT-HSE06)} \\
 Kohn-Sham eigenvalue of the lowest unoccupied orbital & $\langle E_L \rangle$ & \textit{atomicfeaturespackage (DFT-HSE06)} \\
 Atomic number & $\langle AN \rangle$ & \textit{atomicfeaturespackage  (DFT-HSE06)}\\
 Ionization potential & $\langle IP \rangle$  & Pymatgen \\
 Electron Affinity & $\langle EA \rangle$ & Pymatgen \\
 Covalent radius & $\langle R_{COV} \rangle$ & Pymatgen \\
 Electronegativity & $\langle EN \rangle$ & Pymatgen \\
 Number of valence electrons & $\langle N_{VAL} \rangle$ & Matminer\\
  Number of vacant orbitals & $\langle N_{VAC} \rangle$ & Matminer\\
 Cohesive energy & $\langle CE \rangle$ & Kittel et al. \cite{kittel1955solid} \\
 Maximum oxidation state & $\max_{OS}$ & Matminer\\
 Standard deviation of oxidation states & $\mathrm{\sigma_{OS}}$ & Matminer\\
\bottomrule
\end{tabular}
\end{table*}

\section{Nested Cross Validation for Identifying ideal complexity of sisso models}
 The performance of SISSO model for $\Delta G_{pbx}^{\mathrm{OER}}$ is evaluated using a nested cross-validation (NCV) strategy. This employs a two-level CV approach, where hyperparameter optimization occurs in the inner loop, and model validation and selection are carried out in the outer loop. Initially, the dataset is randomly divided into five folds using a 5-fold CV. In the outer loop, four folds are designated as training data, while the remaining fold serves as held-out \textit{test set}. The test set is employed for assessing the model performance via test or prediction errros. The training data is further split into five subsets through an additional 5-fold CV which are utilized for hyperparameter optimization. In the inner loop, four folds are taken as training data, while the remaining fold serves as \textit{validation set}.

In the context of SISSO, the model's complexity is determined by two hyperparameters: the rung ($q$) and the model dimensionality ($D$). These hyperparameters are optimized by training SISSO models on the four subsets with varying $q$ and $D$ values, and evaluating the model performance on the remaining validation sets, quantified in terms of the average root mean squared error (RMSE) and the distribution of validation errors (Figure S9). Subsequently, in the outer loop, the SISSO model is trained and evaluated on each fold, employing the selected optimal hyperparameters, with model performance measured by the average RMSE across the initially held-out test data. By analyzing the performance across all these folds, an estimate of the model's generalization performance is obtained.

\begin{figure*}[ht]
    \centering
    \captionsetup{justification=raggedright, singlelinecheck=false}
    \includegraphics[width=\linewidth]{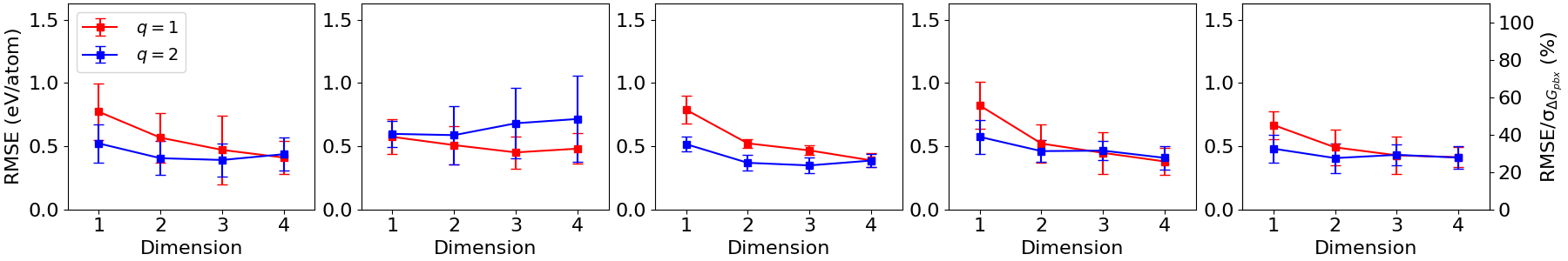}
    \caption*{\textbf{Figure S9.} Nested cross validation analysis for identifying the optimal complexity of SISSO model. The primary y-axis represents the average validation RMSE obtained for the validation sets in the inner loops of NCV for each dimension at both $q$=1 and $q$=2. Corresponding standard deviation is plotted as the error bar. The second y-axis represents the RMSE divided by the standard deviation of the distribution of $\mathrm{\Delta G_{pbx}^{OER}}$ values across the 250 materials of the initial training set for active learning, in percentage.}
    \label{Fig:8}
\end{figure*}

\begin{figure*}[ht]
    \centering
    \captionsetup{justification=raggedright, singlelinecheck=false}
    \includegraphics[width=0.5\linewidth]{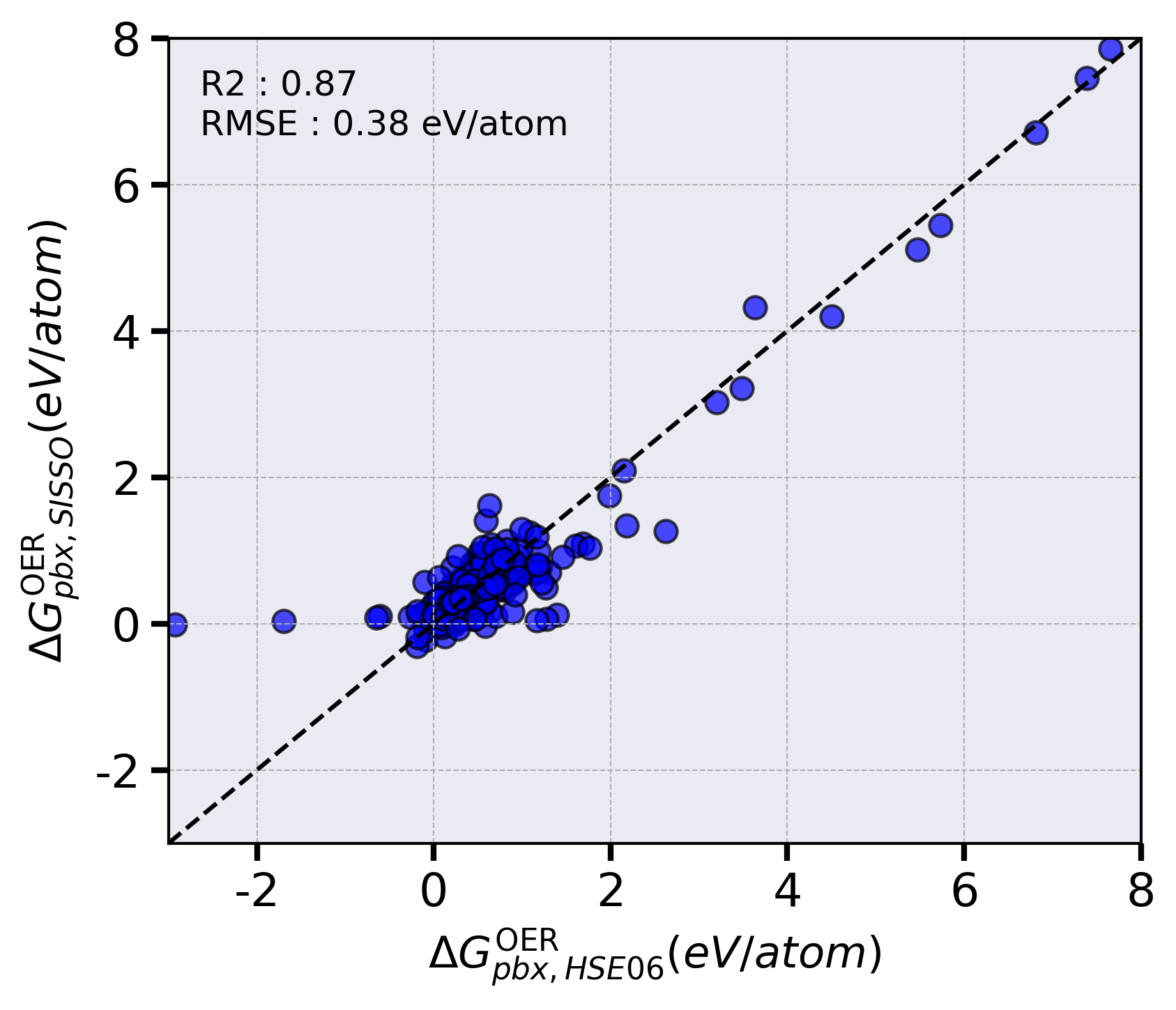}
    \caption*{\textbf{Figure S10.} Performance of the SISSO model with optimal complexity for the training data of 250 oxides.}
    \label{Fig:9}
\end{figure*}

The SISSO model with optimal complexity identified from NCV is further trained on the entire training dataset. The 2D model thus obtained is:
\begin{equation}
\begin{split}
\Delta G_{pbx,SISSO}^{\mathrm{OER}}  = 
&-2.23 \times 10^{-1} \mathrm{eV/atom} \\&+ 3.27 \times 10^{-1} \text{\AA}  * d_1 \\&+ 2.3 \times 10^{-1} \text{\AA}^3 * d_2
\label{SISSO_model}
\end{split}
\end{equation}

where the descriptors $d_1 = \bigl[\sigma_{OS}\langle R_{S} \rangle \ln(\langle N_{VAC} \rangle)\bigr]$ and $d_2 = \bigl[{\langle R_{COV} \rangle}^2\langle R_{S} \rangle\langle N_{VAC} \rangle \bigr]$. The key primary features identified by SISSO are are the standard deviation of oxidation state distribution ($\sigma_{OS}$), the composition-average number of vacant orbitals ($\langle N_{VAC} \rangle$), the covalent radii ($\langle R_{COV} \rangle$) and s-orbital radii ($\langle R_{S} \rangle$). These properties are linked with the chemical bonding in oxides and consequently, play a key role in determining the energetics of their decomposition reactions. 

\FloatBarrier
\begin{figure*}[h]
    \centering
    \captionsetup{justification=raggedright, singlelinecheck=false}
    \includegraphics[width=\linewidth]{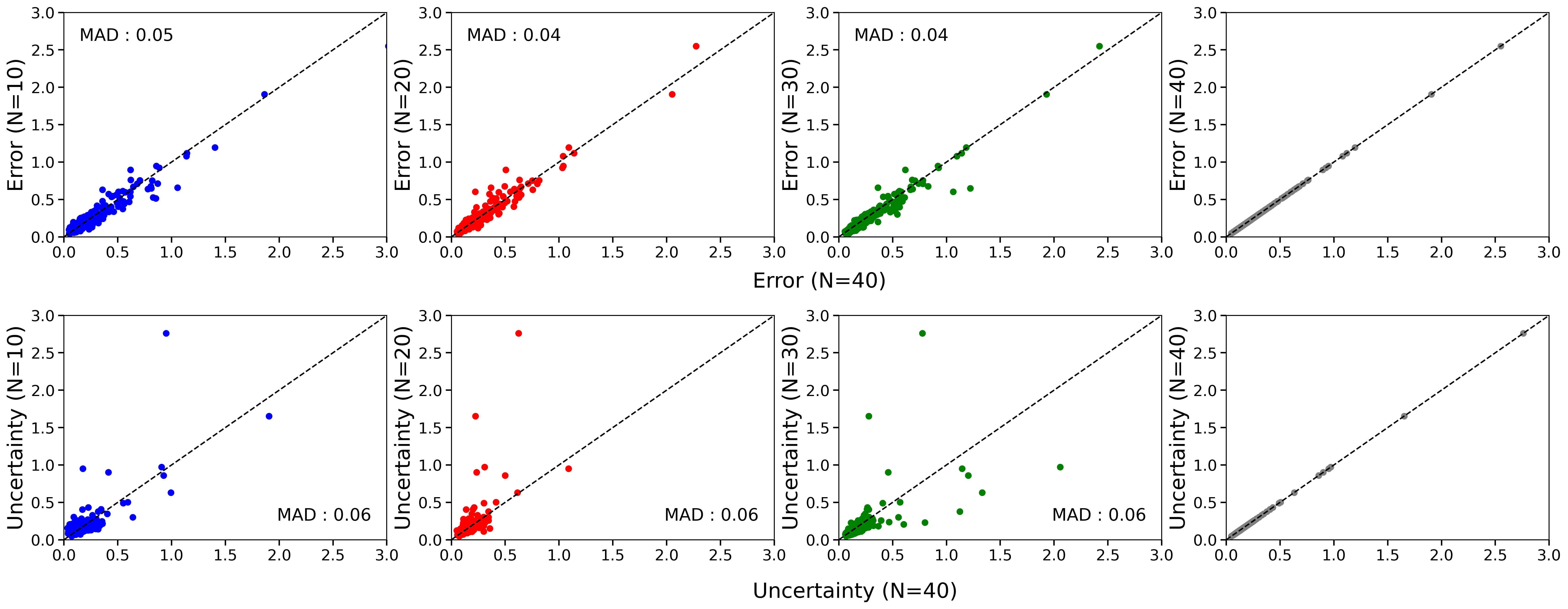}
    \caption*{\textbf{Figure S11.} Convergence test results for the size of SISSO ensembles (N) for employing them in AL framework. For this, the ensembles are created by randomly subsampling the training data (250 oxides) with replacement (bootstrapping) and conducted 10-fold cross validation (CV) at each ensemble size. The CV test errors and corresponding uncertainty estimates for each N, quantified as the standard deviation of prediction for each material across all the bootstrap samples, is compared against the largest N value (here 40). The corresponding mean absolute deviation (MAD) is labelled. The similarity in the distribution of test errors and uncertainty estimates suggest the convergence at ensemble size of 10. Hence this is considered for the SL campaigns.}
    \label{Fig:10}
\end{figure*}
\vspace{1cm}
\FloatBarrier

\section{Curation of Candidate Space Datasets for Sequential Learning}
For the materials in the candidate space, $\Delta G_{pbx}^{\mathrm{OER}}$ values are not known. The AL framework is used to efficiently identify promising oxides in the candidate space. The initial structures for these materials were obtained from MP. Binary and ternary oxides satisfying the $E_{hull}$ and $N_{atoms}$ criteria described in the above section are selected which contain, in addition to oxygen, elements from  groups 1, 2 of the periodic table, as well as main-group elements and transition metals. In order to keep the number of DFT calculations at a manageable level, we only included, in the candidate space, oxides associated with 15 or less competing phases. Finally, oxides containing the elements bisumuth, polonium, and lead were excluded from the candidate space because they were noted to have convergence issues within the default basis sets considered for DFT calculations. The resulting candidate space contains 1470 oxides with 693 unique elemental combinations. These oxides along with corresponding competing phases correspond to a total number of $\sim$12k materials.

\begin{figure*}[!ht]
\captionsetup{justification=justified, singlelinecheck=false}
  \begin{subfigure}{0.45\textwidth}
    \includegraphics[width=\linewidth]{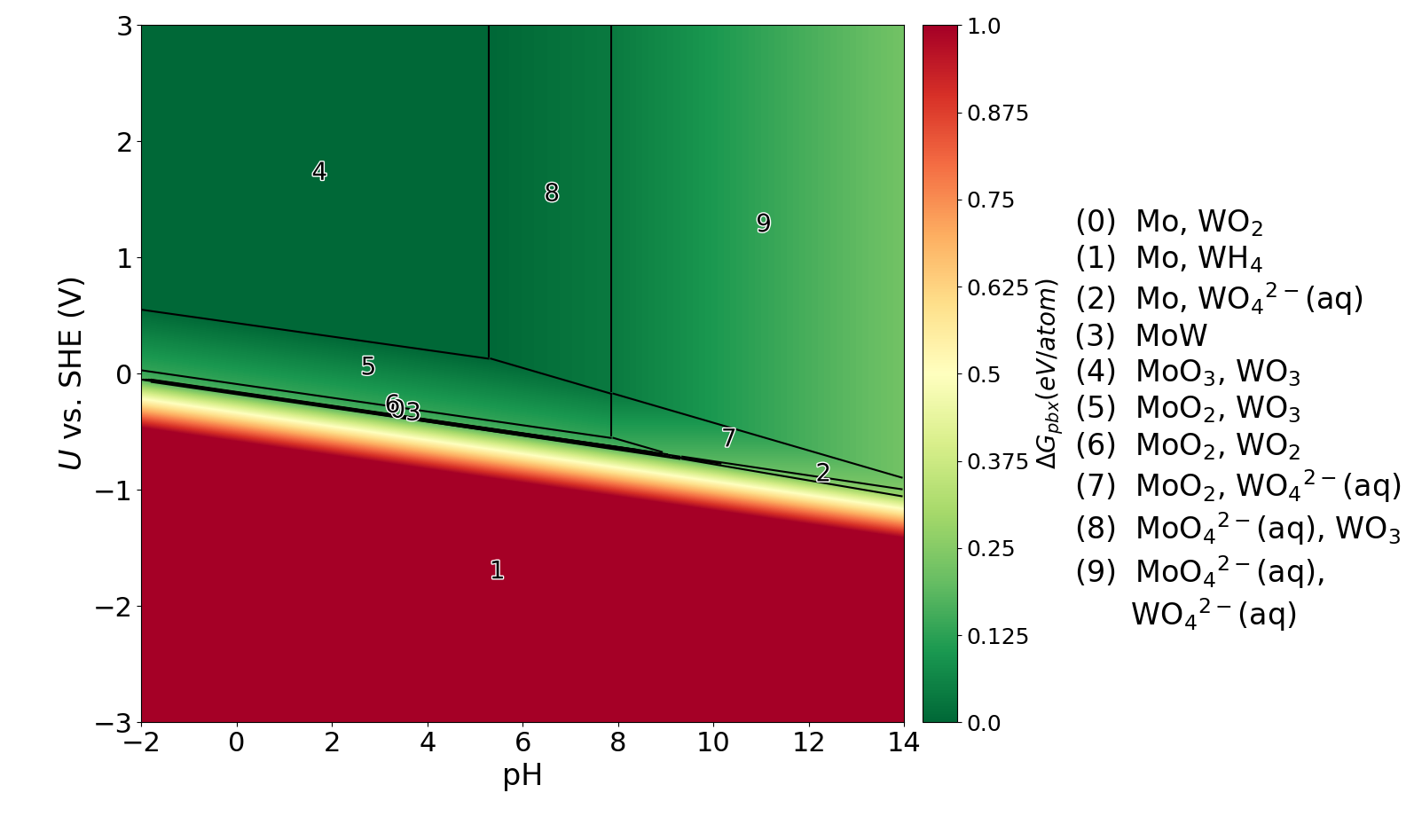}
    \caption*{\ce{MoWO6}}
  \end{subfigure}%
    \begin{subfigure}{0.45\textwidth}
    \includegraphics[width=\linewidth]{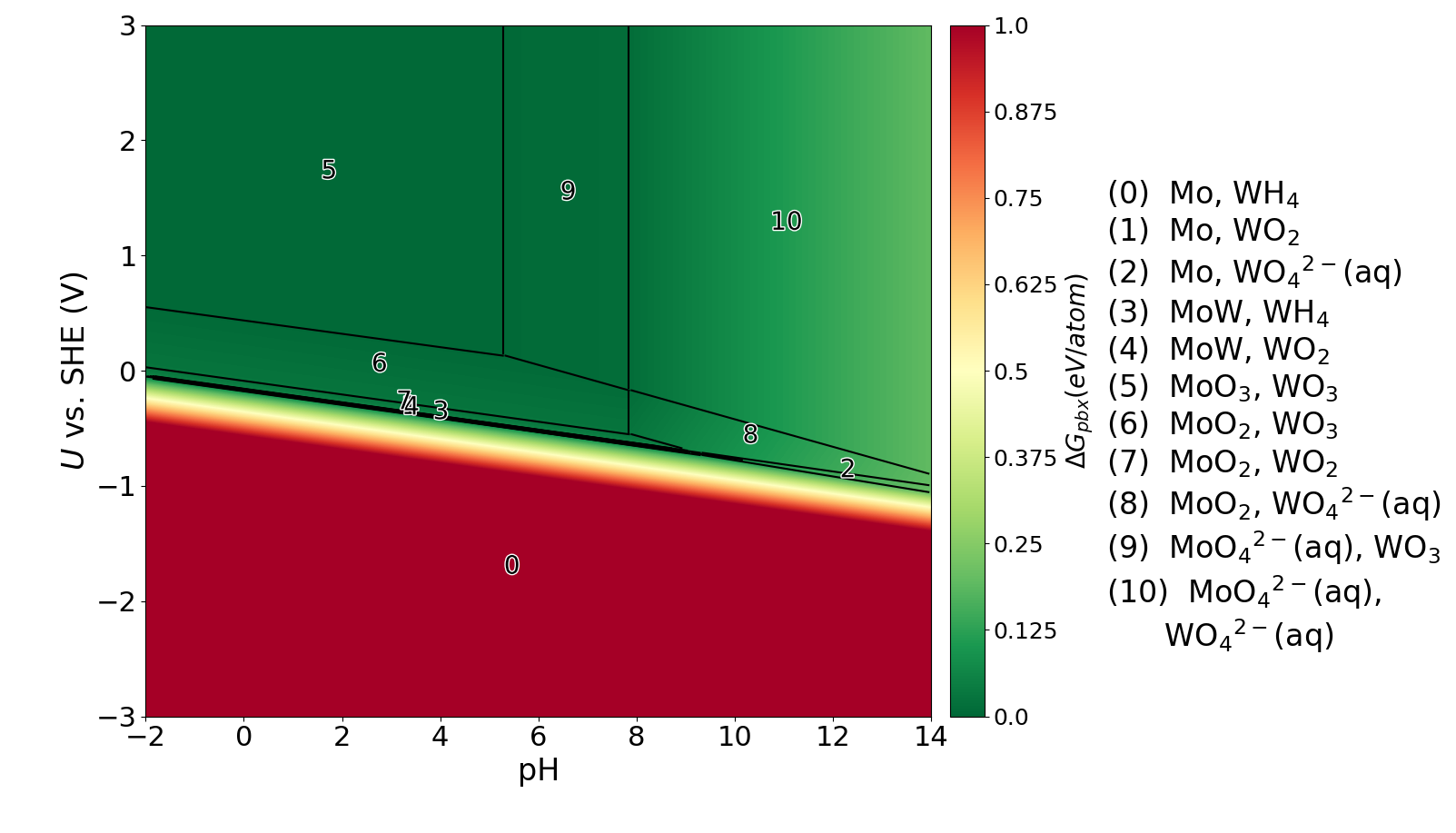}
    \caption*{\ce{MoW11O36}}
  \end{subfigure}%
  \\

\begin{subfigure}{0.45\textwidth}
    \includegraphics[width=\linewidth]{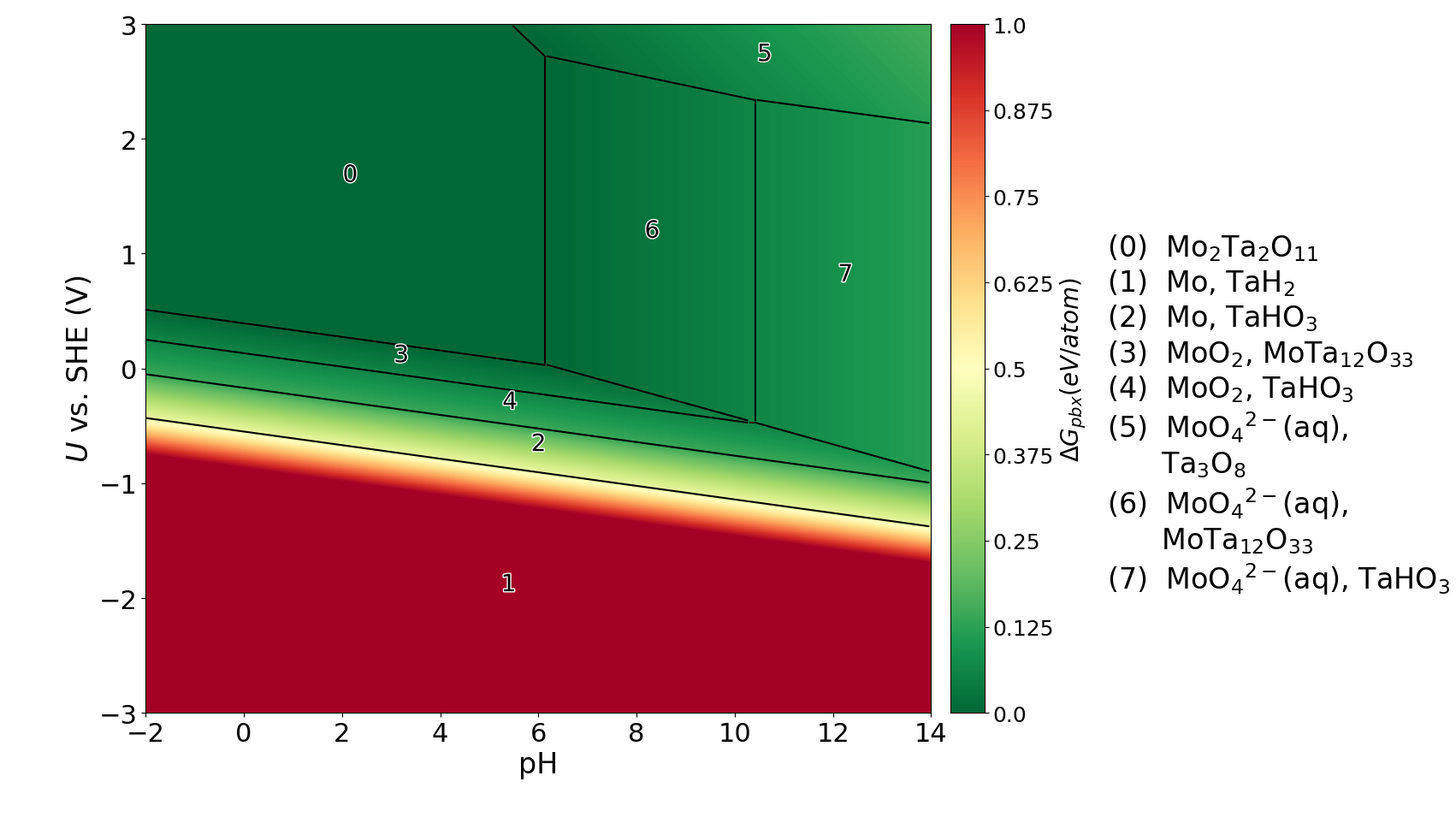}
    \caption*{\ce{Mo2Ta2O11}}
  \end{subfigure}%
      \begin{subfigure}{0.45\textwidth}
    \includegraphics[width=\linewidth]{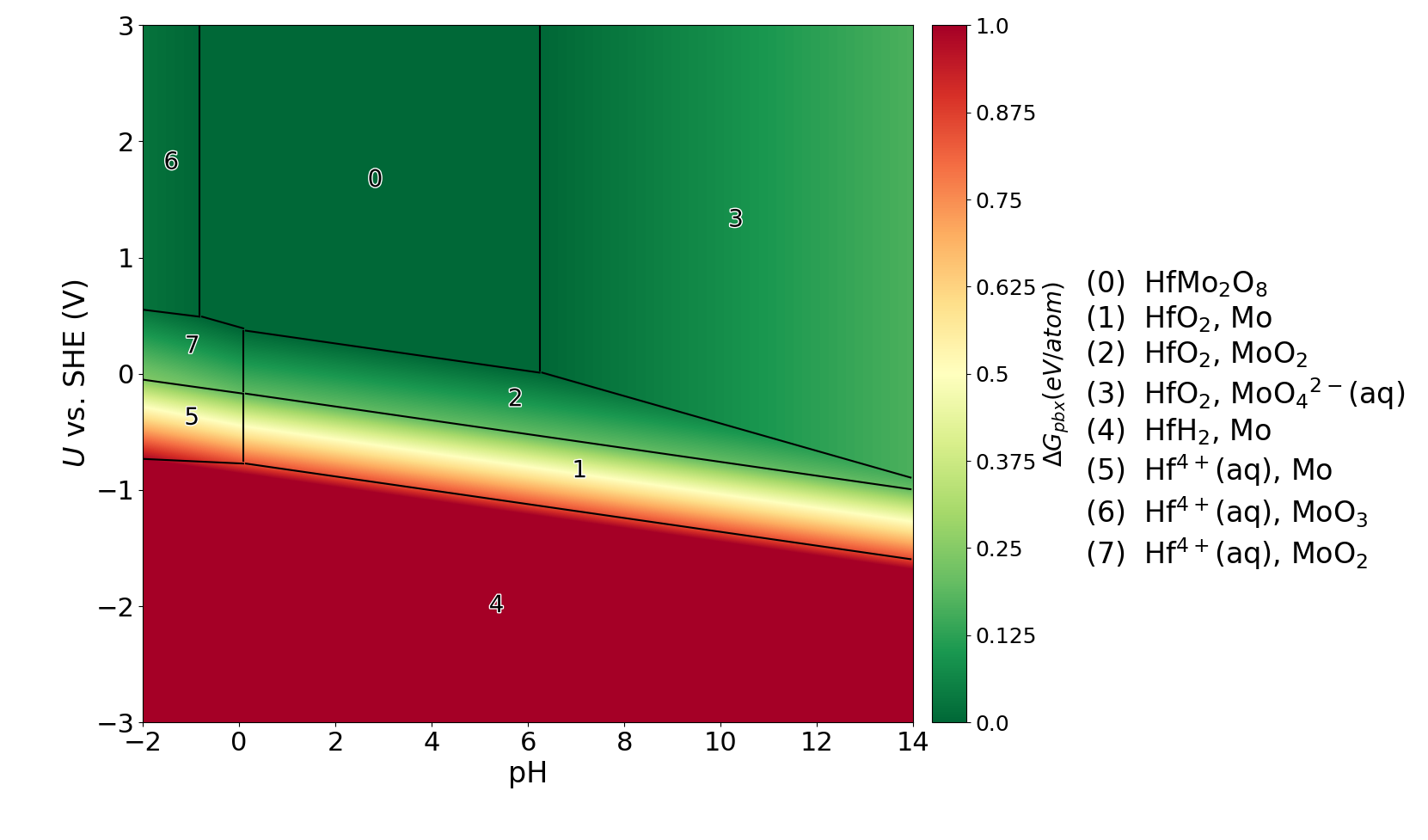}
    \caption*{\ce{Hf(MoO4)2}}
  \end{subfigure}%
  \\
  
 \begin{subfigure}{0.45\textwidth}
    \includegraphics[width=\linewidth]{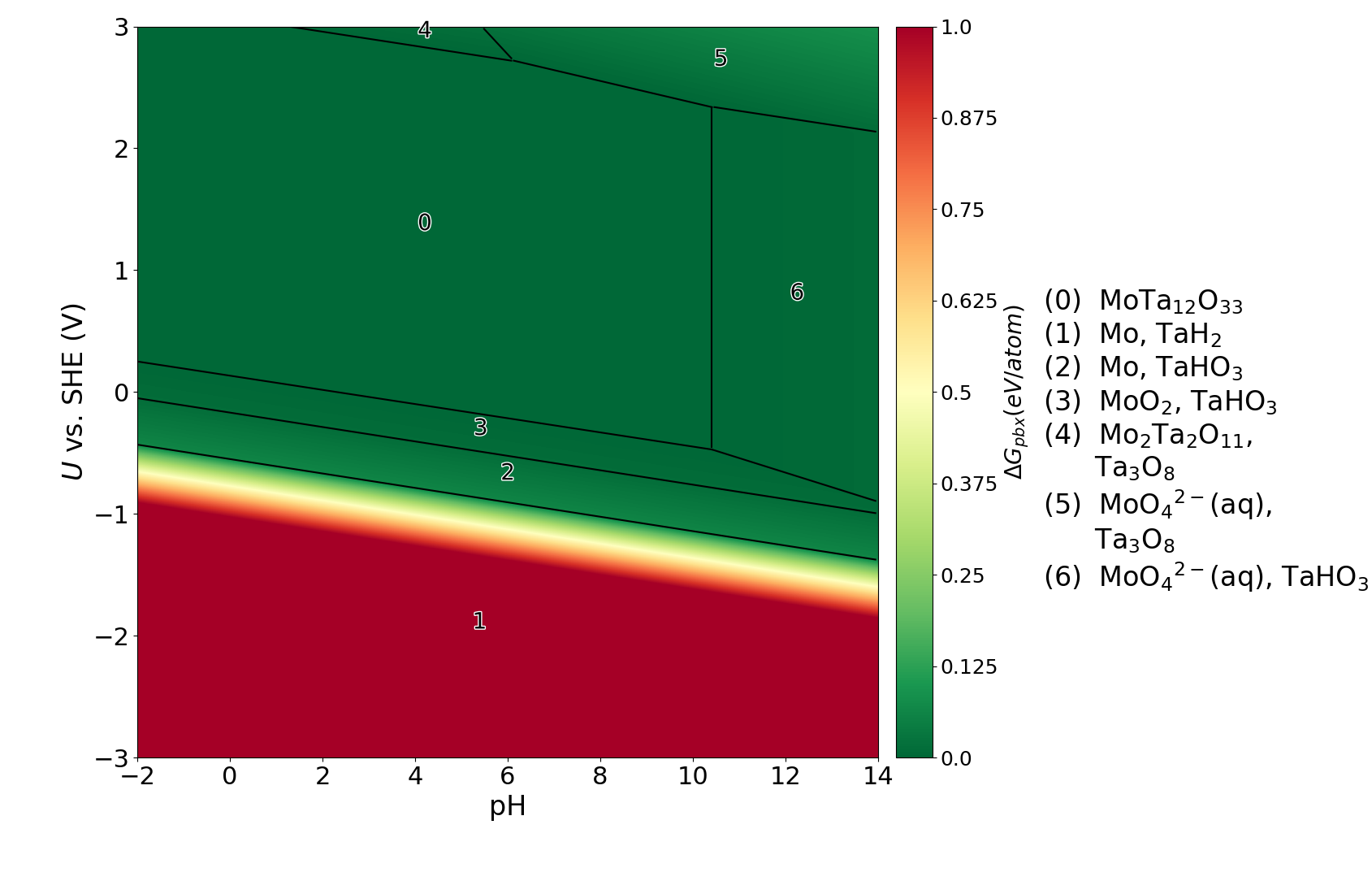}
    \caption*{\ce{MoTa12O33}}
  \end{subfigure}%
     \begin{subfigure}{0.45\textwidth}
    \includegraphics[width=\linewidth]{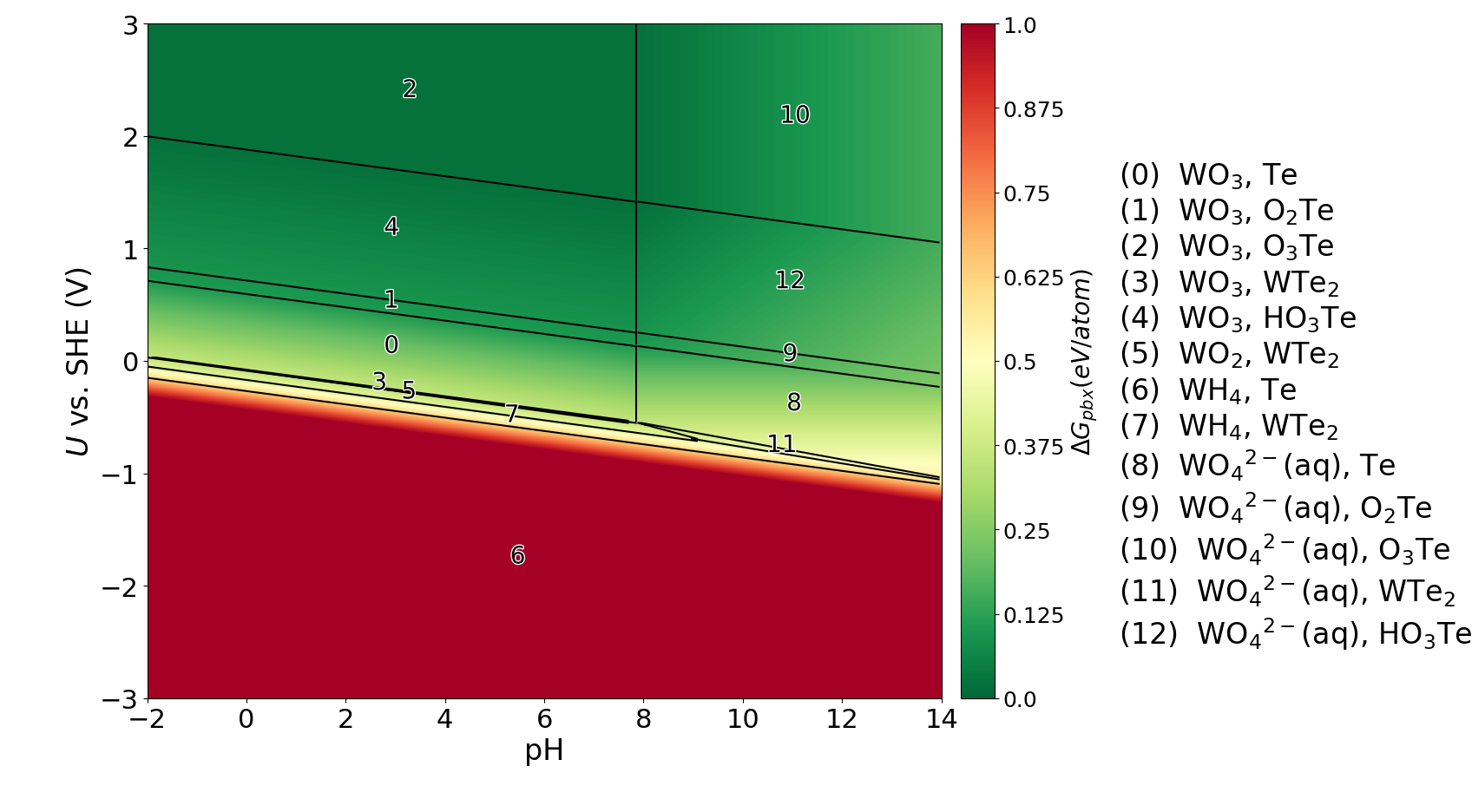}
    \caption*{\ce{Te(WO4)3}}
   \end{subfigure}%
   \\
   
\begin{subfigure}{0.45\textwidth}
    \includegraphics[width=\linewidth]{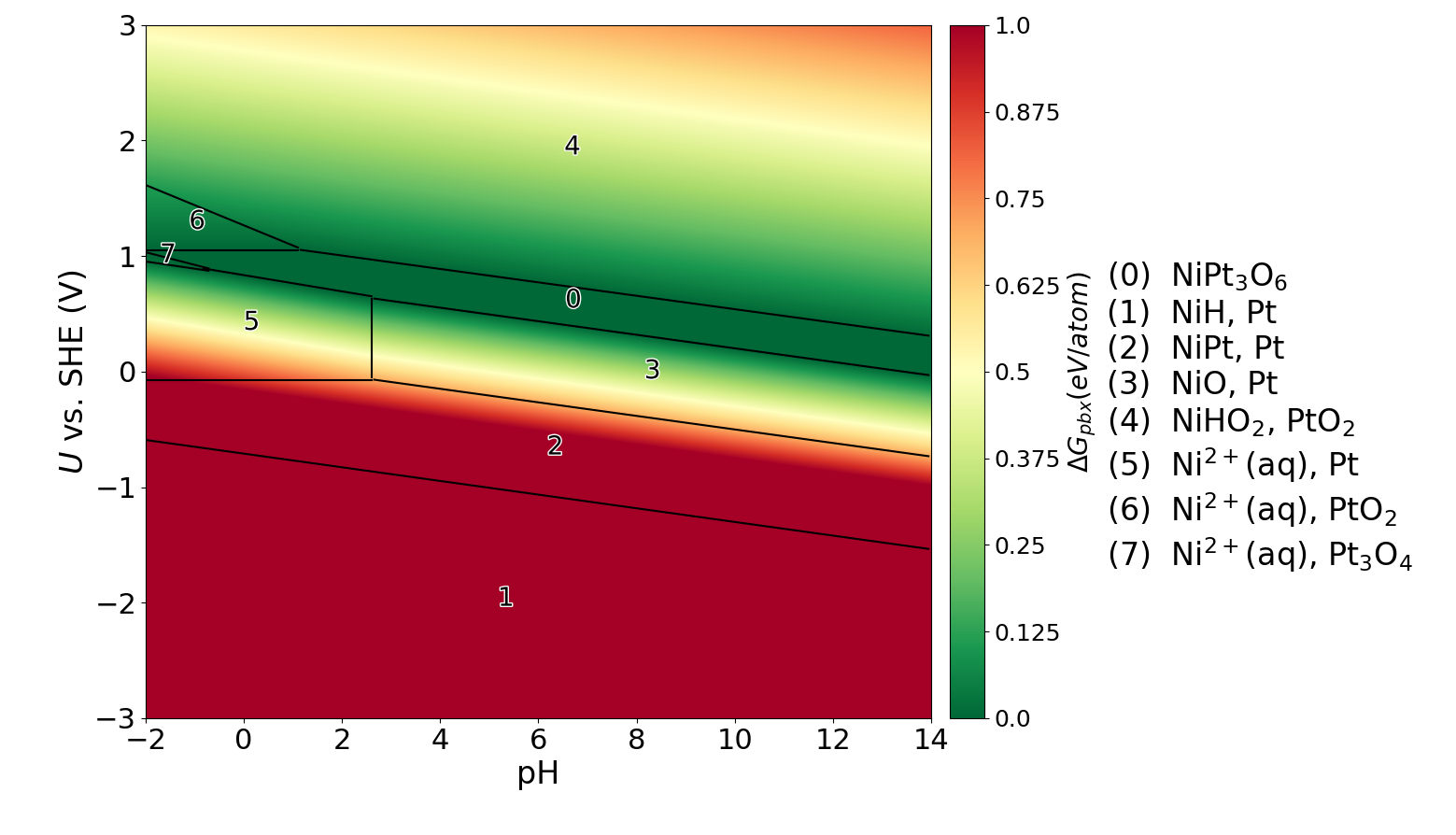}
    \caption*{\ce{Ni(PtO2)3}}
  \end{subfigure}%
   \begin{subfigure}{0.45\textwidth}
    \includegraphics[width=\linewidth]{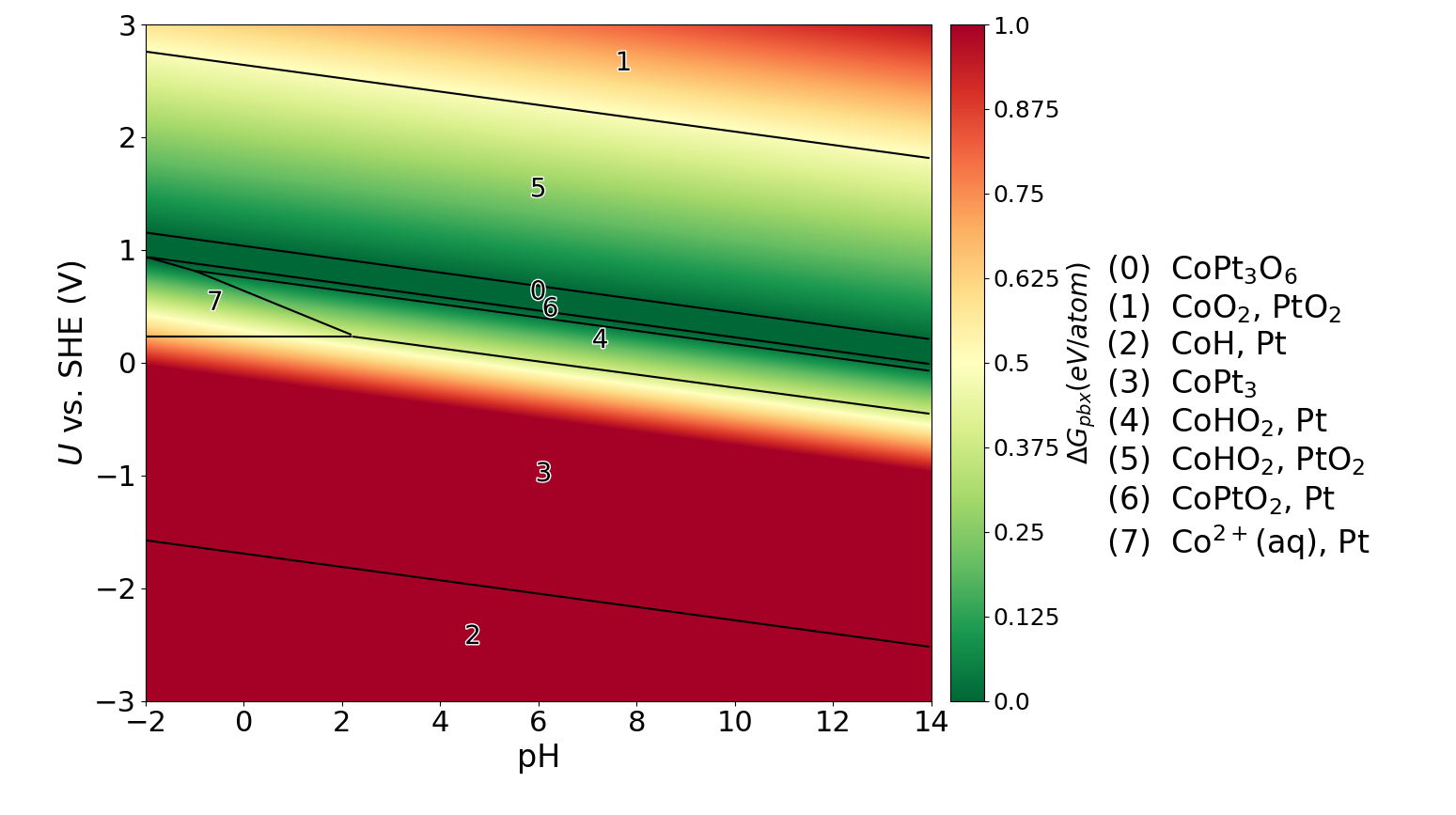}
    \caption*{\ce{Co(PtO2)3}}
  \end{subfigure}%
   \\ 
  
\textit{**Continued in next page}
 \end{figure*}
 
 \FloatBarrier
\begin{figure*}[!ht]
\captionsetup{justification=justified, singlelinecheck=false}
     \begin{subfigure}{0.5\textwidth}
    \includegraphics[width=\linewidth]{figures/Fe2(MoO4)3.png}
    \caption*{\ce{Fe2(MoO4)3}}
  \end{subfigure}
  \begin{subfigure}{0.45\textwidth}
    \includegraphics[width=\linewidth]{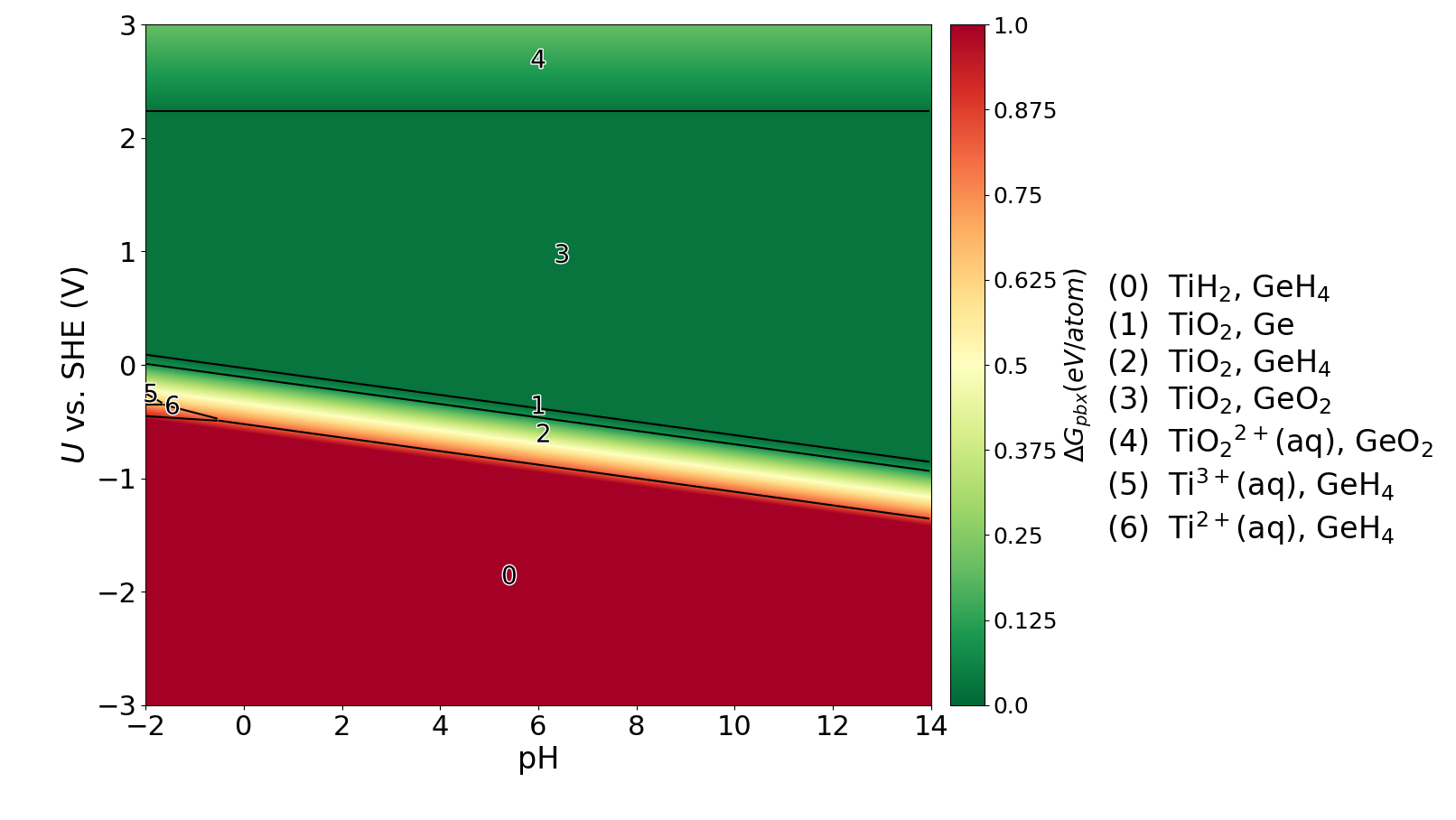}
    \caption*{\ce{Ti(GeO3)2}}
  \end{subfigure}
   \\
   
 \begin{subfigure}{0.45\textwidth}
    \includegraphics[width=\linewidth]{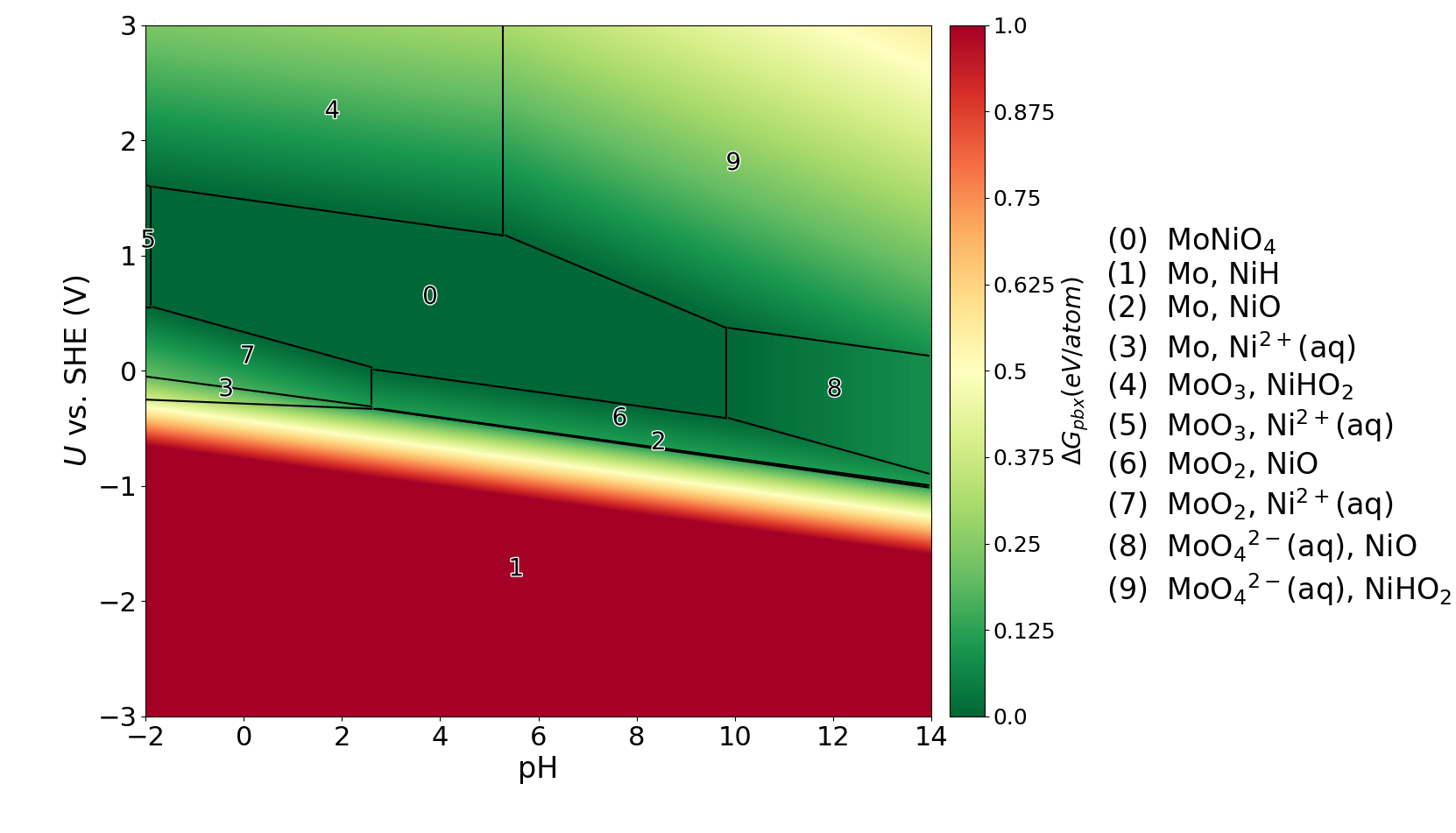}
    \caption*{\ce{NiMoO4}}
  \end{subfigure}
  \begin{subfigure}{0.45\textwidth}
    \includegraphics[width=\linewidth]{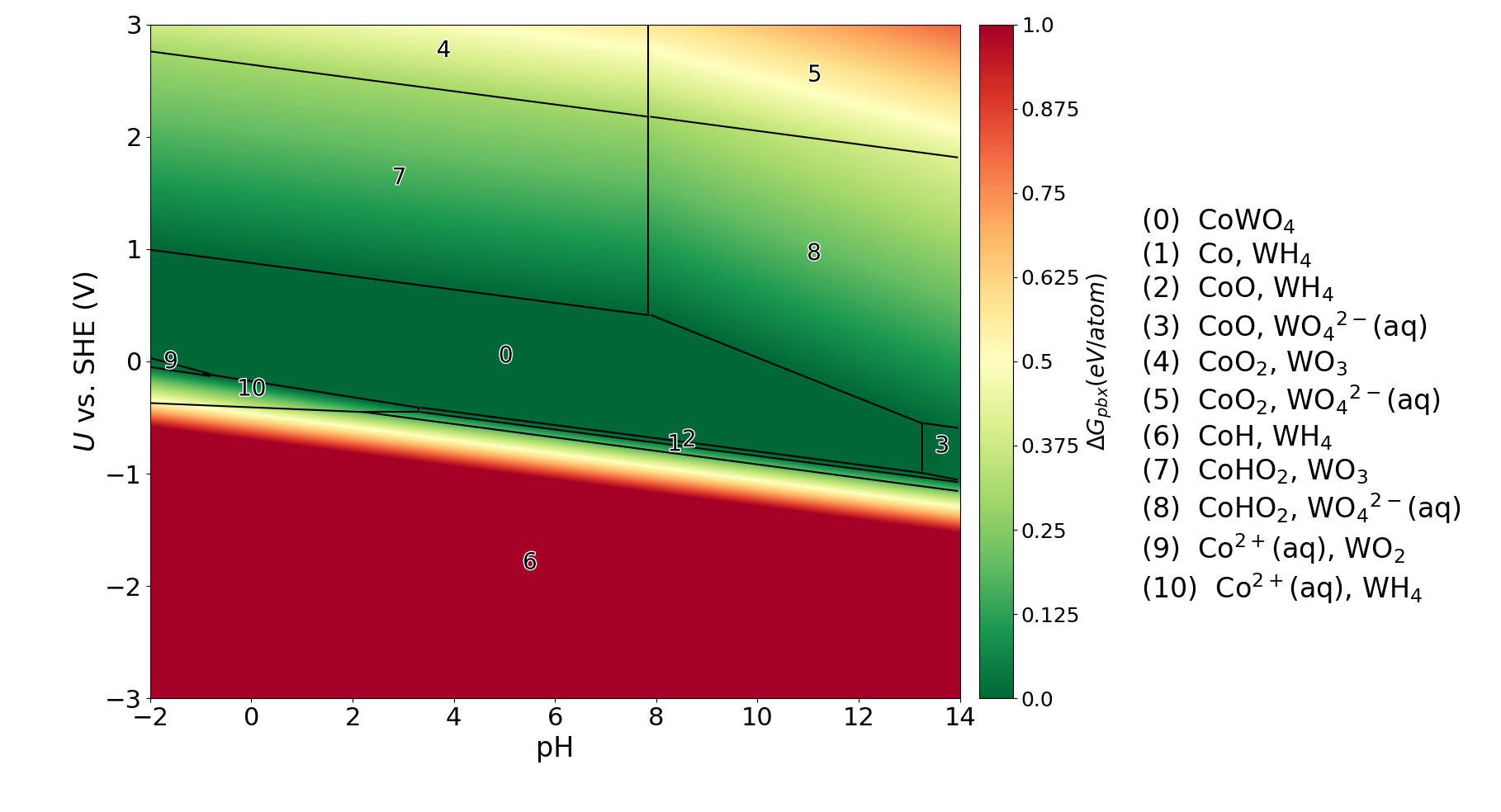}
    \caption*{\ce{CoWO4}}
  \end{subfigure}
  \\
  \caption*{\textbf{Figure S12.} Pourbaix diagrams of 12  selected from sequential learning with $\mathrm{\Delta G_{pbx}^{OER} \leq} 0.1$ eV/atom. The phases present in the diagrams are labelled on the right side.} 
\end{figure*}

\bibliography{apssamp}